\documentclass[aps,preprint,showpacs,superscriptaddress,groupedaddress,amsmath]{revtex4}

\usepackage{amsmath,amsfonts,graphicx,bm,amssymb,array}
\usepackage[usenames]{color}
\usepackage{array}
\usepackage{float}
\usepackage{sidecap}
\usepackage{graphicx}
\usepackage{bm}
\usepackage{axodraw4j}
\usepackage{color}
\usepackage[section]{placeins}
\usepackage{booktabs}
\usepackage{subfigure}

\def\etmiss{E\!\!\!\!\slash_{T}}

\def\ptmiss{p\!\!\!\slash_{T}}

\def\pslash{\not{\hbox{\kern-4pt $p$}}}
\def\qslash{\not{\hbox{\kern-4pt $q$}}}
\def\lv{\not{\hbox{\kern-4pt $L$}}}
\def\lsim{\mathrel{\raise.3ex\hbox{$<$\kern-.75em\lower1ex\hbox{$\sim$}}}}
\def\gsim{\mathrel{\raise.3ex\hbox{$>$\kern-.75em\lower1ex\hbox{$\sim$}}}}
\def\ifmath#1{\relax\ifmmode #1\else $#1$\fi}

\definecolor{DarkRed}{rgb}{0.55,0.00,0.00}
\def\leqn#1{(\ref{#1})}

\newcommand{\nc}{\newcommand}
\nc{\postscript}[2]{\setlength{\epsfxsize}{#2\hsize}\centerline{\epsfbox{#1}}}
\nc{\beq}{\begin{equation}}   \nc{\eeq}{\end{equation}}
\nc{\bea}{\begin{eqnarray}}   \nc{\eea}{\end{eqnarray}}
\nc{\baa}{\begin{array}}      \nc{\eaa}{\end{array}}
\nc{\bit}{\begin{itemize}}    \nc{\eit}{\end{itemize}}
\nc{\ben}{\begin{enumerate}}  \nc{\een}{\end{enumerate}}
\nc{\bce}{\begin{center}}     \nc{\ece}{\end{center}}
\nc{\non}{\nonumber}

\def\ttbar{t\bar{t}}
\def\atlas{ATLAS}
\def\cms{CMS}

\def\ttbar{t\bar{t}}

\expandafter\def\csname ttbarsr0 \endcsname{\ensuremath{144}}
\expandafter\def\csname ttbarsr1 \endcsname{\ensuremath{21}}
\expandafter\def\csname ttbarsr2 \endcsname{\ensuremath{11}}
\expandafter\def\csname pasr0 \endcsname{\ensuremath{21}}
\expandafter\def\csname pasr1 \endcsname{\ensuremath{13}}
\expandafter\def\csname pbsr0 \endcsname{\ensuremath{19}}
\expandafter\def\csname pbsr1 \endcsname{\ensuremath{5.5}}
\expandafter\def\csname pbsr2 \endcsname{\ensuremath{3.3}}
\expandafter\def\csname pesr0 \endcsname{\ensuremath{12}}
\expandafter\def\csname pesr2 \endcsname{\ensuremath{2.3}}
\expandafter\def\csname ttbarsr0t \endcsname{\ensuremath{35}}
\expandafter\def\csname ttbarsr1t \endcsname{\ensuremath{5.2}}
\expandafter\def\csname ttbarsr2t \endcsname{\ensuremath{2.6}}
\expandafter\def\csname pasr0t \endcsname{\ensuremath{16}}
\expandafter\def\csname pasr1t \endcsname{\ensuremath{10}}
\expandafter\def\csname pesr0t \endcsname{\ensuremath{9.4}}
\expandafter\def\csname pesr2t \endcsname{\ensuremath{1.9}}
\expandafter\def\csname pbsr0t \endcsname{\ensuremath{15}}
\expandafter\def\csname pbsr1t \endcsname{\ensuremath{4.4}}
\expandafter\def\csname pbsr2t \endcsname{\ensuremath{2.6}}

\def\mtop{m_{\mathrm{top}}}

\def\cphi{\cos{\phi}}

\begin{document}

\title{Deconstructed Transverse Mass Variables}

\author{Ahmed Ismail} \affiliation{Argonne National Laboratory, Argonne, IL 60439, USA} \affiliation{SLAC National Accelerator Laboratory, 2575 Sand Hill Road, Menlo Park, CA 94025, USA} \affiliation{University of Illinois, Chicago, IL 60607, USA}
\author{Reinhard~Schwienhorst} \affiliation{Department of Physics and Astronomy, Michigan State University, East Lansing MI 48824, USA}
\author{Joseph S.\! Virzi} \affiliation{Lawrence Berkeley National Laboratory, Physics Division, 1 Cyclotron Rd., Berkeley, CA 94720, USA}
\author{Devin G.\! E.\! Walker} \affiliation{SLAC National Accelerator Laboratory, 2575 Sand Hill Road, Menlo Park, CA 94025, USA}

\begin{abstract} 
\noindent
Traditional searches for R-parity conserving natural supersymmetry (SUSY) require large transverse mass and missing energy cuts to separate the signal from large backgrounds.  SUSY models with compressed spectra inherently produce signal events with small amounts of missing energy that are hard to explore.  We use this difficulty to motivate the construction of ``deconstructed'' transverse mass variables which are designed preserve information on both the norm and direction of the missing momentum. We demonstrate the effectiveness of these variables in searches for the pair production of supersymmetric top-quark partners which subsequently decay into a final state with an isolated lepton, jets and missing energy.  We show that the use of deconstructed transverse mass variables extends the accessible compressed spectra parameter space beyond the region probed by traditional methods.  The parameter space can further be expanded to neutralino masses that are larger than the difference between the stop and top masses.
In addition, we also discuss how these variables allow for novel
searches of single stop production, in order to directly probe unconstrained stealth stops in the small stop- and neutralino-mass regime. We also demonstrate the utility of these variables for generic gluino and stop searches in all-hadronic final states.
Overall, we demonstrate that deconstructed transverse variables are essential to any search wanting to maximize signal separation from the background when the signal has undetected particles in the final state. 
\end{abstract}

\pacs{14.65.Jk, 14.65.Ha, 12.60.-i, 12.60.Jv}

\preprint{SLAC-PUB-16080}

\maketitle



\section{Introduction}
\label{sec:intro}

For the first time in history, the TeV scale is being directly probed by the Large Hadron Collider (LHC). Already, a 125.5~GeV Higgs-boson-like particle has been discovered~\cite{Aad:2012tfa,Chatrchyan:2012ufa}. With the ongoing confirmation of this particle as the Standard Model (SM) Higgs boson, the experimentally successful Standard Model (SM) will be complete. However, a serious theoretical inconsistency remains:  It is well known that radiative corrections, generated dominantly by top quark loops, push the SM Higgs boson to have a mass of the order of the Planck scale ($10^{19}$~GeV). This implies that these corrections must be ``fine-tuned'' in order to recover the observed Higgs mass at the LHC.  Natural models of new physics~\cite{Dimopoulos:1981zb,Cohen:1996vb,Weinberg:1975gm,Susskind:1978ms,Hill:1980sq,Hill:1991at,Chivukula:1998wd,Dobrescu:1997nm,Hill:2002ap,Kaplan:1983sm,ArkaniHamed:2001nc,ArkaniHamed:2002qy,ArkaniHamed:1998rs,Randall:1999ee,Cheng:2004yc,Cheng:2003ju} ameliorate this problem by adding light top partners to the SM which cancel (some or all of) the top quark-induced radiative corrections to the Higgs boson mass. For example, in the Minimal Supersymmetric Standard Model (MSSM), the role of the top partner is fulfilled by the stops, at least one of which may be expected to be light from naturalness considerations. In this paper, we describe a new search technique for light top partners, concentrating on stops in R-parity conserving supersymmetric models.  Despite this focus, we note this search strategy is applicable to a wide range of natural models with light top partners.

A significant fraction of the parameter space for light stops has been ruled out by various LHC searches~\cite{Aad:2014kra,Chatrchyan:2013xna,Cahill-Rowley:2014twa}.  Most of these searches focus on the decay $\tilde{t} \to t + \chi$. The most important parameter in such searches is the mass difference between the stop and the lightest neutralino~\cite{Han:2008gy}, 
\begin{equation}
\Delta M_{\tilde{t}\,\chi} =  m_{\tilde{t}} - m_{\chi} \, .  \label{eq:massdifference}
\end{equation}
As $\Delta M_{\tilde{t}\,\chi}$ grows smaller and approaches $m_\mathrm{top}$, the decaying stops inherently generate only a small amount of missing energy, as the daughter neutralino of a stop is produced with little momentum. Traditional stop searches typically rely on large missing energy cuts to separate signal from background. Natural SUSY can thus still be inaccessible in the case of compressed mass spectra, i.e.,
\begin{equation} 
m_{\tilde{t}} \sim m_\chi + m_\mathrm{top}. \label{eq:cspectra}
\end{equation}
In these instances, signal events from pair-produced stops are similar to the overwhelming SM $\ttbar$ background unless one implements analysis techniques that exploit any residual differences. We provide such an analysis technique in this paper. We also apply our technique to the case of very compressed mass spectra, where $\Delta M_{\tilde{t}\,\chi} < m_\mathrm{top}$. For these spectra, in the absence of a light chargino, the typical decay pattern of a stop is $\tilde{t} \to b + W^+ + \chi^0$, again with limited missing energy. 

At the LHC, the most sensitive search channel for stops is the single lepton + jets + MET final state, in which pair-produced stops yield one top quark decaying to a lepton, neutrino and $b$~quark and another fully hadronic top quark~\cite{Aad:2012xqa,Chatrchyan:2012uea,ATLAS-CONF-2013-037,Aad:2014kra,Chatrchyan:2013xna}.
(Here the leptons are either electrons or muons.)  We focus on this lepton+jets final state.  The missing transverse momentum in the signal events receives contributions from the neutrino from the $W$~boson decay as well as from the two neutralinos.  However, the same final state is produced by lepton+jets $\ttbar$ decays in the SM, with the missing energy provided solely by the neutrino (as well as detector effects).  Moreover, the LHC is a top quark factory with a large SM production cross section~\cite{Czakon:2013goa,Chatrchyan:2012uea}, and $\ttbar$ is generally the dominant background for one-lepton stop searches.  In particular, for this study, top quark pair events with taus produced from $W$~bosons have additional missing transverse energy from the tau decay, and we will see that they generate an important additional background that can easily be rejected.

\textbf{Motivation for Deconstructed Transverse Masses:}  
The most important quantity in searches for new physics in final states with multiple undetected particles is the missing momentum, in particular its correlations with other objects in an event. Experimental searches use transverse mass cuts to separate the signal from SM backgrounds in these searches. These cuts, however, often destroy the information about the missing momentum and its correlations. 
Thus, we ``deconstruct'' the transverse mass variables to preserve the maximal amount of information about the missing momentum in order to maximize the separation of signal and background. By exploring magnitudes and angular correlations simultaneously, the deconstructed variables improve signal-background separation compared to transverse mass cuts.
Our study of stop pair production (in the compressed limit) provides a platform to show how these variables can increase the sensitivity of the current searches. Overall, the deconstruction can be generalized to other transverse masses and similar objects, improving the signal sensitivity in a wide array of analyses.
Before moving on, we note that other strategies have been proposed to access stop pair production in this region of compressed mass spectra: A variant on the traditional transverse mass variables~\cite{Alves:2012ft} as well as additional transverse mass variables~\cite{Han:2008gy,Nojiri:2008ir}.

This paper is organized as follows:  We first detail the signal and background processes for our stop pair analysis in Section~\ref{sec:prelims}. We also describe how we simulate detector effects, lepton and jet reconstruction and our basic acceptance cuts. In Section~\ref{sec:deconstruct}, we detail the deconstruction of the transverse mass variables.  A detailed discussion of our final event selection in several different signal regions is presented in Section~\ref{sec:analysis}. Stealth stops are addressed through a single stop search in Section~\ref{sec:singlestop}. Section~\ref{sec:additionalapps} contains additional applications for deconstruction techniques, including a discussion of all-hadronic final states. Section~\ref{sec:conclusion} summarizes our findings.

\section{Preliminaries}
\label{sec:prelims}

In this section, we describe the signal and major background processes for stop pair production in the lepton+jets mode.  We include details of the event generation and detector simulations.  We also discuss our basic selection cuts. We set the top mass to 173~GeV~\cite{ATLAS:2014wva}.

\subsection{Stop Pair Production Signal Processes}
\label{sec:signal}
We explore the pair production of the supersymmetric partner of the top quark, the stop, assuming that the stop decays to the lightest neutralino with a branching fraction of 100\%. The decay occurs either through
\begin{align}
\tilde{t} \to t + \chi && \tilde{t} \to b + W^+ + \chi\,,
\end{align}
where the latter process occurs for the case where $m_{\tilde{t}} < m_t + m_\chi$ and $\chi$ is the neutralino dark matter.  We focus on processes with one lepton in the final state.  The full partonic process is given by
\begin{eqnarray}
p + p &\to& \tilde{t} + \tilde{t}^* \to t + \bar{t} + \chi + \chi \to \ell + 2b + 2j + \etmiss \label{eq:signal}\\
p + p &\to& \tilde{t} + \tilde{t}^* \to b + \bar{b} + W^+ + W^- + \chi + \chi \to \ell + 2b + 2j + \etmiss\,, \nonumber 
\end{eqnarray}
where the second line accounts for off-shell top quarks. We include on- and off-shell top quarks
in the analysis to cover the difficult-to-access compressed region, see Section~\ref{sec:sr2}.
The cross section for this process is independent of the neutralino mass as long as the decay is kinematically possible.

While this paper focuses on a supersymmetric scenario, there are non-supersymmetric analogues to which our analysis procedure equally applies.  For example, in little Higgs models with $T$-parity~\cite{Cheng:2004yc}, heavy top quark partners can be produced analogously to stops and subsequently decay into a top quark and dark photon (dark matter candidate).  
The cross section for this process is typically about 50\% larger than that for stop pair production. 

We generate signal events for several combinations of $m_{\tilde{t}}$ and $m_\chi$ using
SHERPA 1.4.3 ~\cite{Gleisberg:2008ta} or
MadGraph 5~\cite{Alwall:2011uj} for the event generation.
MadGraph uses the CTEQ6 set of parton distribution functions (PDF)~\cite{Pumplin:2002vw};
Sherpa uses the CTEQ10 PDF set~\cite{Lai:2010vv}.
We scale the signal cross sections to next-to-leading order in the strong coupling with next-to-leading log resummation~\cite{Kramer:2012bx}.
Off-shell decays are included in our event generation. When generating stops that undergo a
three-body decay, we increase the Breit-Wigner cutoff parameter to a large value ($\sim 10^4$)
in the MadGraph run card to include intermediate particles (top quark and $W$~boson) that go
significantly off shell. The cross sections for the three signal mass points under consideration
are given in Table~\ref{tab:xs}.

\begin{table}[!h!tbp]
\begin{center}
\caption{Stop pair production cross section at NLO at a 8~TeV proton-proton collider~\cite{Kramer:2012bx}.}
\begin{tabular}{|c|c|}
\hline
stop, neutralino mass & cross section [pb] \\
\hline
$m_{\tilde{t}}$ = 350 GeV, $m_{\chi}$ = 200 GeV & 0.81 \\
$m_{\tilde{t}}$ = 400 GeV, $m_{\chi}$ = 200 GeV & 0.36 \\
$m_{\tilde{t}}$ = 500 GeV, $m_{\chi}$ = 200 GeV & 0.086 \\
\hline
\end{tabular}
\label{tab:xs}
\end{center}
\end{table}

\subsection{Background Processes}
\label{sec:process}

The main background to the lepton+jets signal signature is from SM top pair production ($t\bar{t}$), with smaller
backgrounds from $W$+jets and top pair production in association with a $W$~or $Z$~boson ($t\bar{t}V$) and other backgrounds from single top, diboson, and QCD multijet production~\cite{ATLAS:2012maq}. 

We focus on the dominant $t\bar{t}$ background here without loss of generality and use SHERPA~\cite{Gleisberg:2008ta} to simulate $t\bar{t}$+jets. All top quark decays are included except for fully hadronic decays, and are separated into lepton+jets ($lq$) and dilepton ($ll$) modes.
In particular, we include top decays where the $W$~boson decays to tau leptons.
We will see in Section~\ref{sec:analysis} that dilepton events with one tau that decays hadronically constitute
a significant background. We use the CTEQ6 set of parton distribution functions (PDF)~\cite{Pumplin:2002vw}.
The top pair background is normalized to the NNLO cross section~\cite{Czakon:2013goa}.

Showering for all signal and background processes is provided by Pythia~6~\cite{Sjostrand:2006za}.
All events are processed by a detector simulation described in Section~\ref{sec:detector_effects},
which in particular provides the smearing of $\etmiss$ that appropriately enhances the amount
of $t\bar{t}$ in the signal region.

\subsection{Jet and Lepton Reconstruction}
\label{sec:detector_effects}
We account for detector effects using a simplified detector simulation as follows.
Final state interacting particles (excluding muons, neutrinos and SUSY particles) with transverse momentum $ p_{T} \ge 100 $ MeV and $ | \eta | \le 3 $
are clustered into jets using the Anti-$k_T$ algorithm in the FastJet framework~\cite{Cacciari:2011ma}.
The jet energy $E$ is smeared according to $ \Delta E / E = 0.5 / \sqrt{E} \oplus 0.03 $.
Electron energies are smeared according to $ \Delta E / E = 0.1 / \sqrt{E} \oplus 0.007 $;
muon energies are smeared according to $ \Delta E / E = 0.04 $.

After the smearing step, electrons and muons are selected if they have $ p_{T} > 10$~GeV and
$|\eta|\le 2.5$. Jets are selected if they have $ p_{T} \ge 25 $~GeV, $ | \eta | \le 2.5 $
and minimum $ \Delta R(ej) \ge 0.2 $ to an electron,
where $\Delta R$ is the standard definition of the separation cone
$\Delta R_{ij} = \sqrt{\left( \eta_{i} - \eta_{j} \right)^2 + \left( \phi_{i} - \phi_{j} \right)^2}$.
Leptons must be separated from jets by $\Delta R(lj) > 0.4$.
$\etmiss$ is calculated using smeared electrons, muons and jets (with $ | \eta | \le 5 $) before any vetoes,
and by construction balances the transverse momentum of the entire event.
Jets are tagged as $b$-jets if the truth record shows a weakly-decaying $B$-hadron with $ p_{T} \ge 5 $~GeV and $ | \eta | \le 2.5 $ with $ \Delta R \left( j, B \right) \le 0.3 $.  We do not account for pileup and other detector effects.

\subsection{Event Selection}
\label{sec:selection}

Following the criteria used by \atlas{}~\cite{Aad:2014kra} and \cms{}~\cite{Chatrchyan:SUS14011}, events are selected if they have at least four jets, 
exactly one lepton (electron or muon) with $p_T>25$~GeV and $|\eta|<2.5$, 
and missing transverse energy $\etmiss>150$~GeV. 
At least one jet must be $b$-tagged.
The jets are required to pass cuts of
$p_T($jet 1$)>80$~GeV, $p_T($jet 2$)>60$~GeV, $p_T($jet 3$)>40$~GeV and $p_T($jet 4$)>25$~GeV.
Events with leptons with 10~GeV~$<p_{T}<$~25~GeV are rejected in order to reduce the dilepton
and tau backgrounds. 
We further impose requirements on $\Delta \phi$ between the $\etmiss$ and each of the two jets with the largest $p_{T}$,
namely $\Delta \phi ({\rm jet\,1},\etmiss) \ge 0.8$ and $\Delta \phi ({\rm jet\,2},\etmiss) \ge 0.8$.

\section{Deconstructed Transverse Masses}
\label{sec:deconstruct}

When searching for signal processes with multiple undetected particles in the final state, the missing momentum is the most important quantity needed to separate the signal from the background.  To maximize this separation, we emphasize that both the magnitude and direction of the missing momentum are needed.  The central idea of deconstruction is to consider both of these observables and, in particular, to not inadvertently destroy information with cuts that tie these observables together.  A common cut that does the latter is the traditional transverse mass cut. In essence, deconstructed transverse mass variables allow the transverse mass cut to vary on an event-by-event basis in order to maximize the signal.

The transverse mass variable was first defined as a way to reconstruct the $W$~boson mass even in the
presence of a neutrino in the final state~\cite{Barger:1983wf}.  Since our goal is to maximize the separation of signal from background, new transverse mass variables are mandatory.
To understand how to construct these variables, consider the signal and $\ttbar$ background processes.  All describe a final state with a leptonically decaying $W$ boson plus $n$-jets.  The $W$~boson decays into a high-$p_T$ electron or muon.
The $W$~transverse mass is defined as
\begin{equation}
m_T = \sqrt{2\, E_{T \ell}\, \etmiss - 2\,\vec{p}_{T \ell}\cdot \vec{\ptmiss} } \,, \label{eq:Wtransmass}
\end{equation}  
where $\vec{p}_{T \ell}$ is the transverse momentum of the lepton and $E_{T\ell} = |\vec{p}_{T\ell}|$.
The missing momentum and energy are defined as $\vec{\ptmiss} = - \sum_i \,\vec{p}_{T\,\mathrm{visible}}$ and $\etmiss = |\vec{\ptmiss}|$, respectively.  By convention, the $m_T$ equation makes the implicit assumption that only neutrinos are contributing to the observed missing energy.  To separate the signal containing additional invisible particles besides neutrinos from the background, a cut is often implemented,
\begin{equation}
m_T > m_T^0 > m_W.  \label{eq:examplecut}
\end{equation}
Here the $m_T^0$ value is chosen to maximize the  signal-background separation.  Given the uniform cut on $m_T$ and the structure of the transverse mass variable under the square root in equation~\leqn{eq:Wtransmass}, it is clear that some information about the missing energy and momentum vector is discarded.  To refine the separation between signal and background, we first rewrite equation~\leqn{eq:Wtransmass} as
\begin{equation}
1 - \frac{m_T^2}{2\, E_{T\ell}\, \etmiss} = \cphi.
\end{equation}
Here $\cphi$ is the transverse angle between the lepton $p_T$ and the missing transverse momentum,
\begin{equation}
\cphi = \frac{\vec{p}_{l\,T} \cdot \vec{\ptmiss}}{p_{l\,T}\, \ptmiss}\;.
\label{eq:cosphi}
\end{equation}
Then, rather than using $m_T^0$ as a strict cut, we define
\begin{equation}
Q \equiv 1 - \frac{{m_T^0}^2}{2\, E_{T\ell}\, \etmiss}.
\label{eq:Q}
\end{equation}
$Q$ and $\cphi$ are now independent variables.  We have therefore ``deconstructed'' the transverse mass from Eq.~\ref{eq:Wtransmass} into an angular component (Eq.~\ref{eq:cosphi}) and a dimensionless magnitude component (Eq.~\ref{eq:Q}).

In events where additional particles escape the detector together with the neutrino
from the $W$~boson decay, the total missing transverse energy vector is
\begin{equation}
\vec{\etmiss} = \vec{p}_{\nu\,T} + \sum_\chi{\vec{p}_{\chi\,T}}\;.
\end{equation}
With this, the transverse mass variable generalizes to
\begin{equation}
m^2_T = 2\,E_{l\,T}\,\etmiss\bigl(1 - \cphi_{l\etmiss} \bigr).
\label{eq:genmt}
\end{equation}

\begin{figure}[!h!tbp] 
\centering
\subfigure[]{\label{subfig:etmiss}\includegraphics[width=0.475\textwidth]{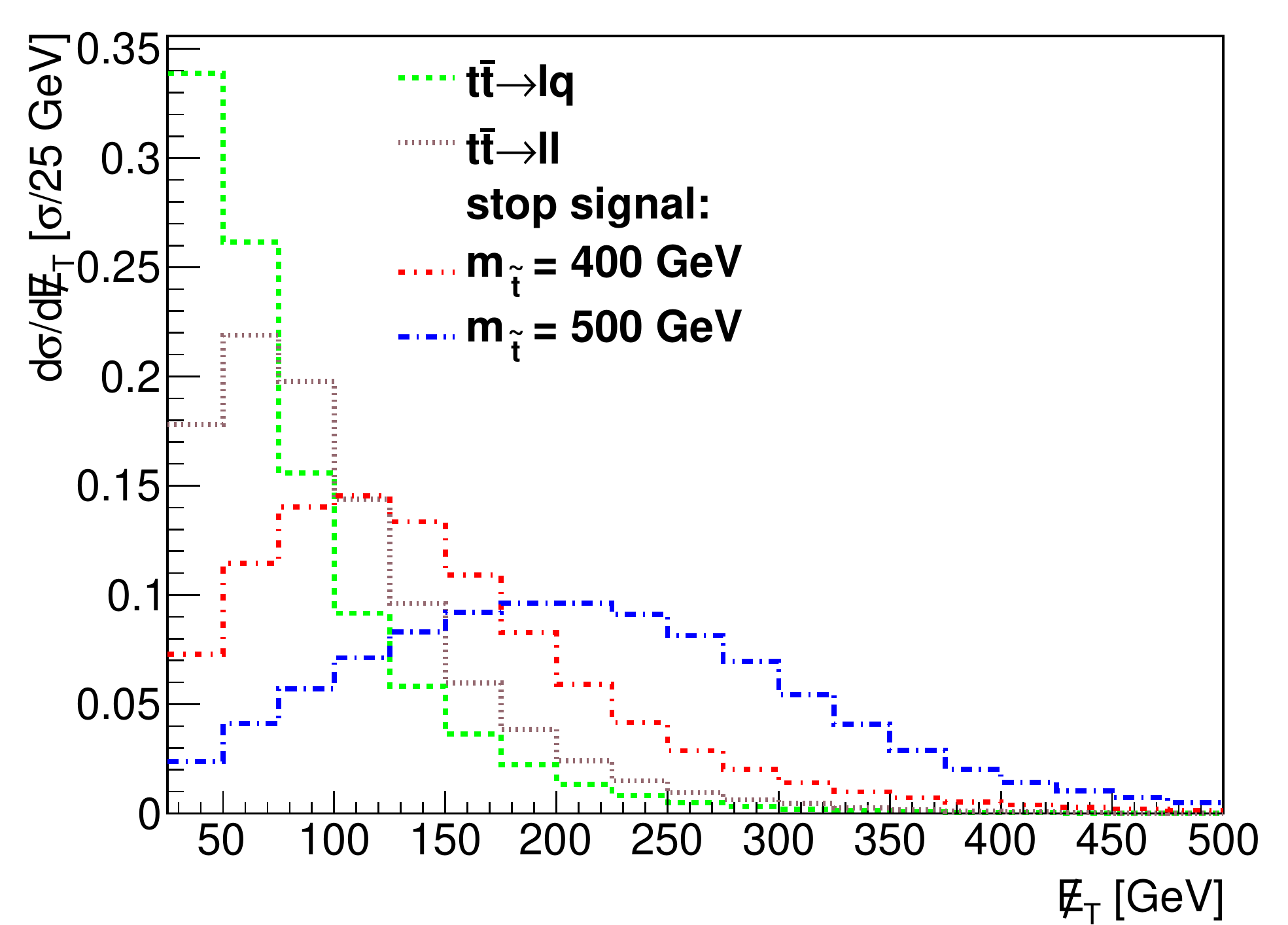}}
\subfigure[]{\label{subfig:mtw}\includegraphics[width=0.475\textwidth]{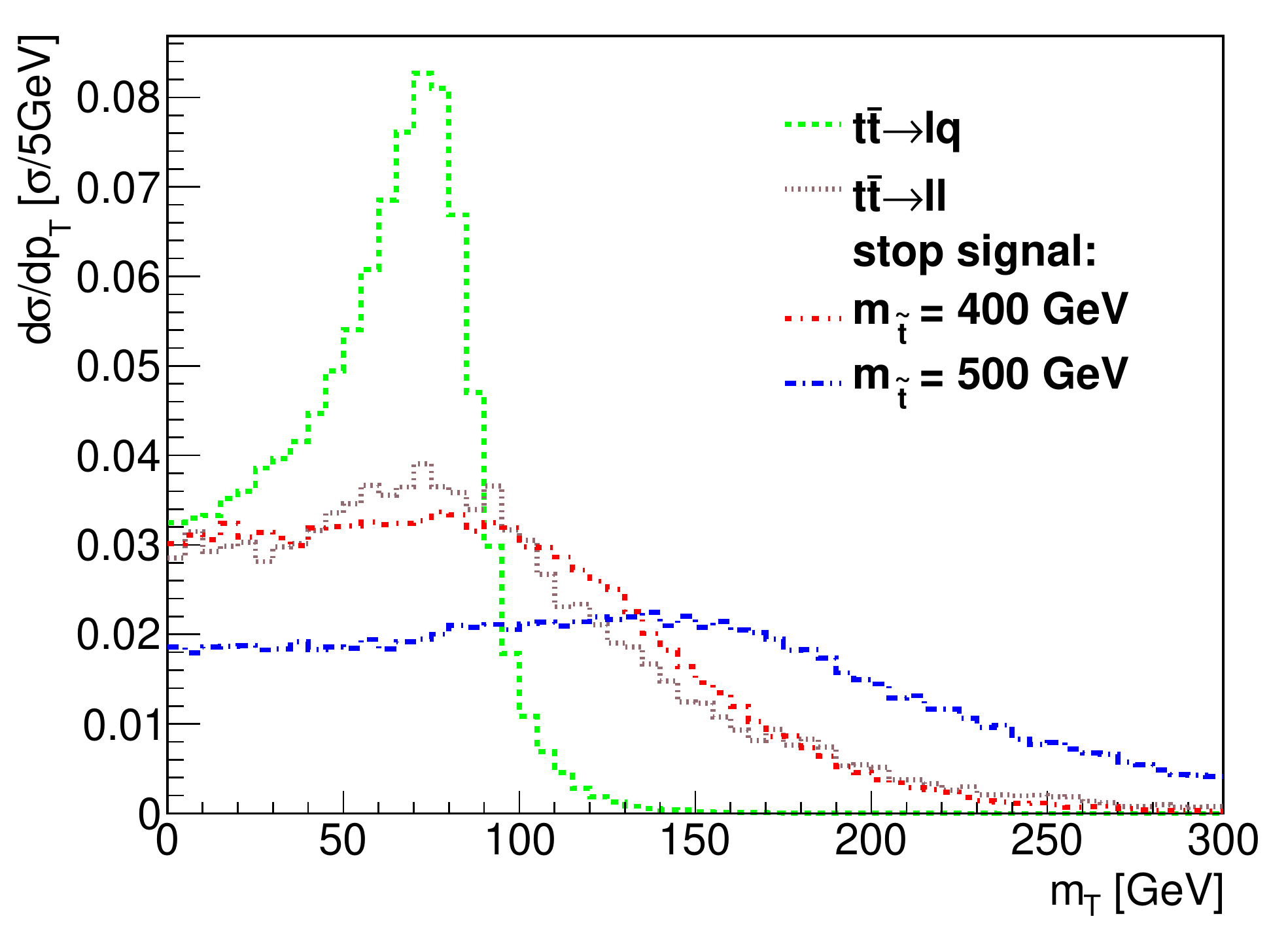}}
\caption{Distribution of \subref{subfig:etmiss} $\etmiss$ and \subref{subfig:mtw} $m_T$ for
top pair production and two different stop signal masses after event selection,
all normalized to have the same area, for events passing selection cuts. The 
neutralino mass is set to 200~GeV. (Color online)}
\label{fig:transverse}
\end{figure}

The missing transverse energy is shown in Fig.~\ref{subfig:etmiss}  for events
passing the selection cuts (Section~\ref{sec:selection}) for
top quark pair production samples and for different SUSY samples that each contain
two non-detected dark matter particles in addition to the top quark pair. 
The distribution in lepton+jets top pair events peaks at the lowest $\etmiss$, with dilepton
top pair events having higher $\etmiss$. The SUSY signals peak at higher $\etmiss$ values, but
there is significant overlap of distributions, and $\etmiss$ alone is not a powerful
discriminator.
The transverse mass $m_T$ from Eq.~\ref{eq:genmt} is shown in Fig.~\ref{subfig:mtw}. 
The distribution in lepton+jets top pair events cuts off at around 100~GeV, slightly above the $W$~boson mass, as expected. 
Dilepton top pair events have an additional contribution to the $\etmiss$ vector from the additional neutrino,
allowing the $m_{T}$ distribution to extend further above the kinematic limit set by the $W$ mass ($M_{W}$).
Similarly, the SUSY signal has additional
particles contributing to $\etmiss$ and its distribution
extends much higher. Note that for a compressed spectrum scenario ($m_{\tilde{t}}=400$~GeV, 
$m_\chi=200$~GeV), the transverse mass variable does
not extend as high and this variable therefore loses its power. This effect limits current
experimental analyses.

\begin{figure}[!h!tbp] 
\centering
\subfigure[]{\label{subfig:Q}\includegraphics[width=0.475\textwidth]{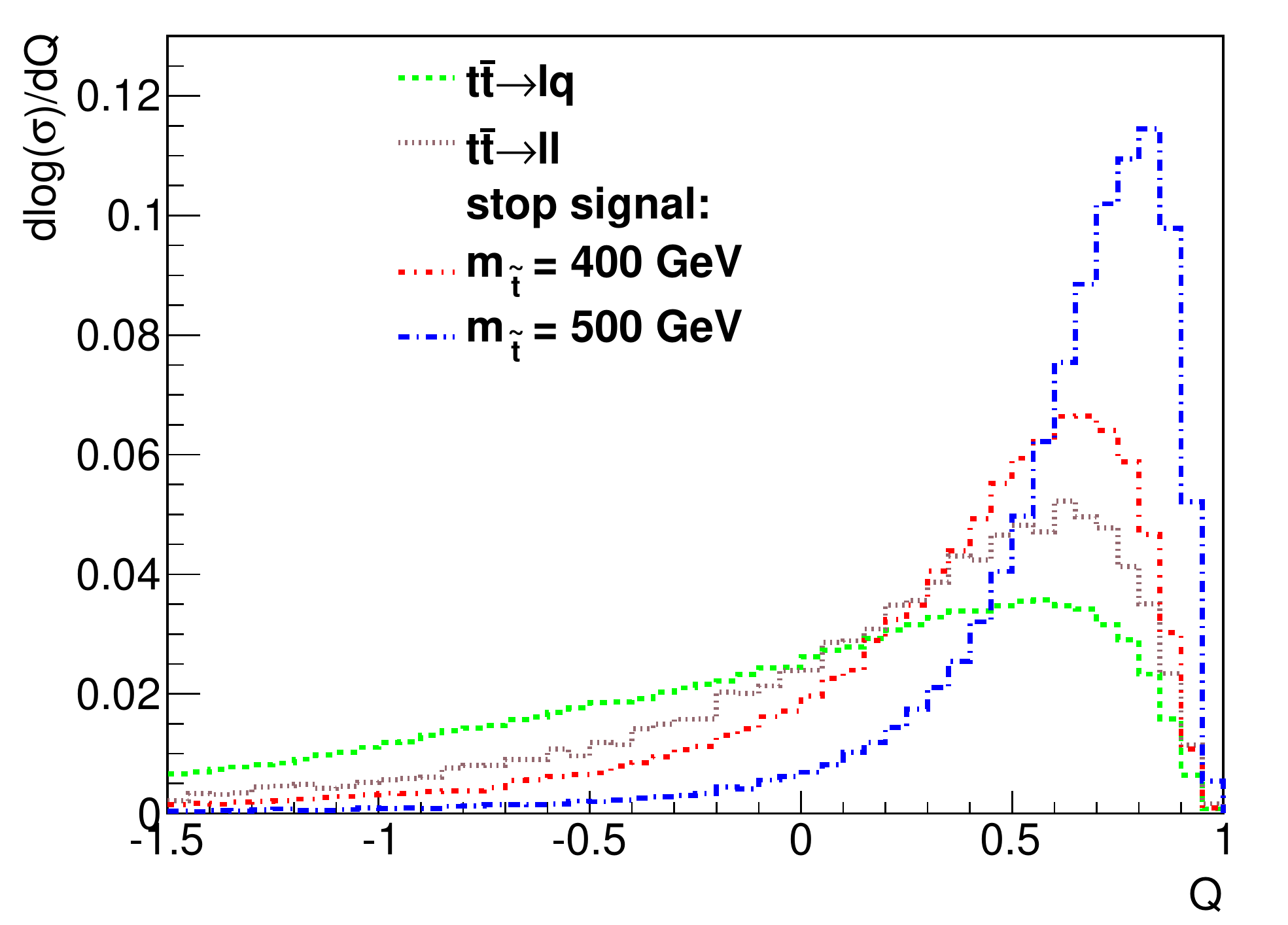} }
\subfigure[]{\label{subfig:cosphi}\includegraphics[width=0.475\textwidth]{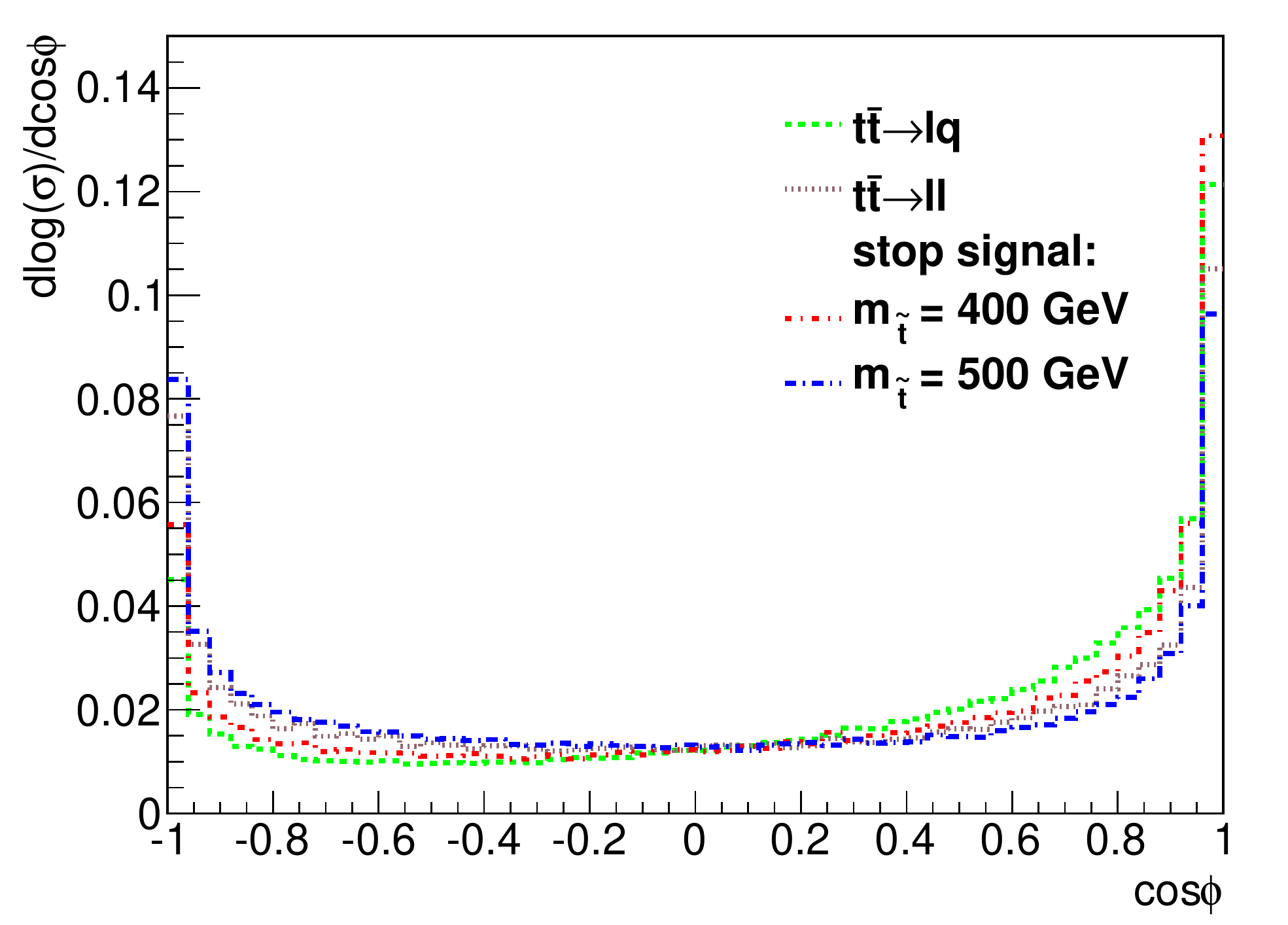}}
\caption{Distribution of \subref{subfig:Q} $Q$ and \subref{subfig:cosphi} $\cphi$ for top
pair production and two different stop
signal masses after event selection, all normalized to have the same area. The 
neutralino mass is set to 200~GeV. (Color online)}
\label{fig:Q1d}
\end{figure}

The distribution of $Q$ is shown in Fig.~\ref{subfig:Q}. 
Top pair events generally have a lower value of $Q$ than the SUSY signals, with $\ttbar$ dilepton events peaking higher than $\ttbar$ lepton+jet events due to the presence of two neutrinos.
We will see that top pair events with one hadronically decaying $\tau$ contribute significantly at higher $Q$, in direct competition with SUSY signals.

The distribution of $\cphi$ is shown in Fig.~\ref{subfig:cosphi}. Signals and backgrounds peak at $\cphi=1$, with a smaller peak at $\cphi=-1$. The usefulness of this angular correlation will become apparent only when plotting it in two dimensions vs $Q$ and vs $\etmiss$.

\subsubsection{Neutrino $p_{\nu\,L}$  Reconstruction}

The parameters $Q$ and $\cphi$ also have a meaning in the reconstruction of the neutrino momentum along the beam direction ($p_{\nu\,L}$) in SM lepton+jets $\ttbar$ events.
This longitudinal neutrino momentum cannot be measured directly
and instead must be inferred, typically using a $W$~boson mass constraint. This leads
to a quadratic equation, which has two solutions,
\begin{eqnarray}
p_{\nu L}^{\pm}  =\frac{1}{{2\, p_{l \,T}^2}}
 \left( {A\, p_{l\, L} \pm E_l \sqrt{A^2  - 4\,{p}^{\,2}_{l\, T}\,p_{\nu\,T}^2}} \right)\,, 
\label{eq:pvl}
\end{eqnarray}
where $A = M_W^2 + 2 \,\vec{p}_{l\, T} \cdot \,\vec{p}_{\nu\,T}$, $M_W=80.4$~GeV is the input
SM mass of
the $W$~boson~\cite{Group:2012gb}, $p_{l\,L}$ is the longitudinal lepton momentum, and 
$E_l=p_l$ is the lepton energy. 

This neutrino momentum calculation breaks down and the neutrino longitudinal momentum becomes unphysical when 
\begin{equation}
M_W^2 < 2\bigl(\, p_{l\, T} p_{\nu\,T} - \vec{p}_{l\, T} \cdot \,\vec{p}_{\nu\,T} \, \bigr)\;, 
\label{eq:bound1}
\end{equation}
i.e.~when the reconstructed transverse mass exceeds the SM mass, $m_{l\nu\,T}>M_W$.
That can occur in SM events because the $W$~boson has a natural width or because the neutrino transverse 
momentum was not reconstructed correctly due to the presence of additional neutrinos
or detector effects. The unphysical region in Eq.~\ref{eq:bound1} can easily be interpreted in terms of $Q$ and $\cphi$.
With $m_T^0=M_{W}$, the unphysical region is given by
\begin{equation}
Q \left( m_T^0=M_{W} \right) > \cphi\;.
\label{eq:qgtcosphi}
\end{equation}

\subsubsection{Correlation}
The variables we introduce are correlated;
cuts on one variable can affect the distribution of another variable. This is
intentional as the correlations are what improves the separation of signals from
backgrounds.
As can be seen from Eq.~\ref{eq:bound1}, $Q=\cphi$ if the longitudinal momenta of both lepton and neutrino are zero. In general, the value of $Q$ is bounded from above at one and asymptotes to this value for events with high $p_{T}$ $W$~bosons.
In principle there is no lower bound on $Q$, though experimental requirements on
lepton $p_T$ and $\etmiss$ effectively limit $Q$ to be greater than
${\cal{O}} \left( -10 \right)$ for $m_T^0 \sim M_W$.
For a $W$~boson decaying at rest, the lepton and neutrino will be back-to-back ($\cphi=-1$) and
the momenta of the lepton and neutrino will be half of $M_W$ each. Hence in this case $Q\le-1$.

\begin{figure}[!h!tbp]
\begin{center}
\vspace{-0.5cm}
\subfigure[~$\ttbar\rightarrow lq$]{\label{subfig:Qcosphi_ttbarlq}\includegraphics[width=0.8\textwidth]{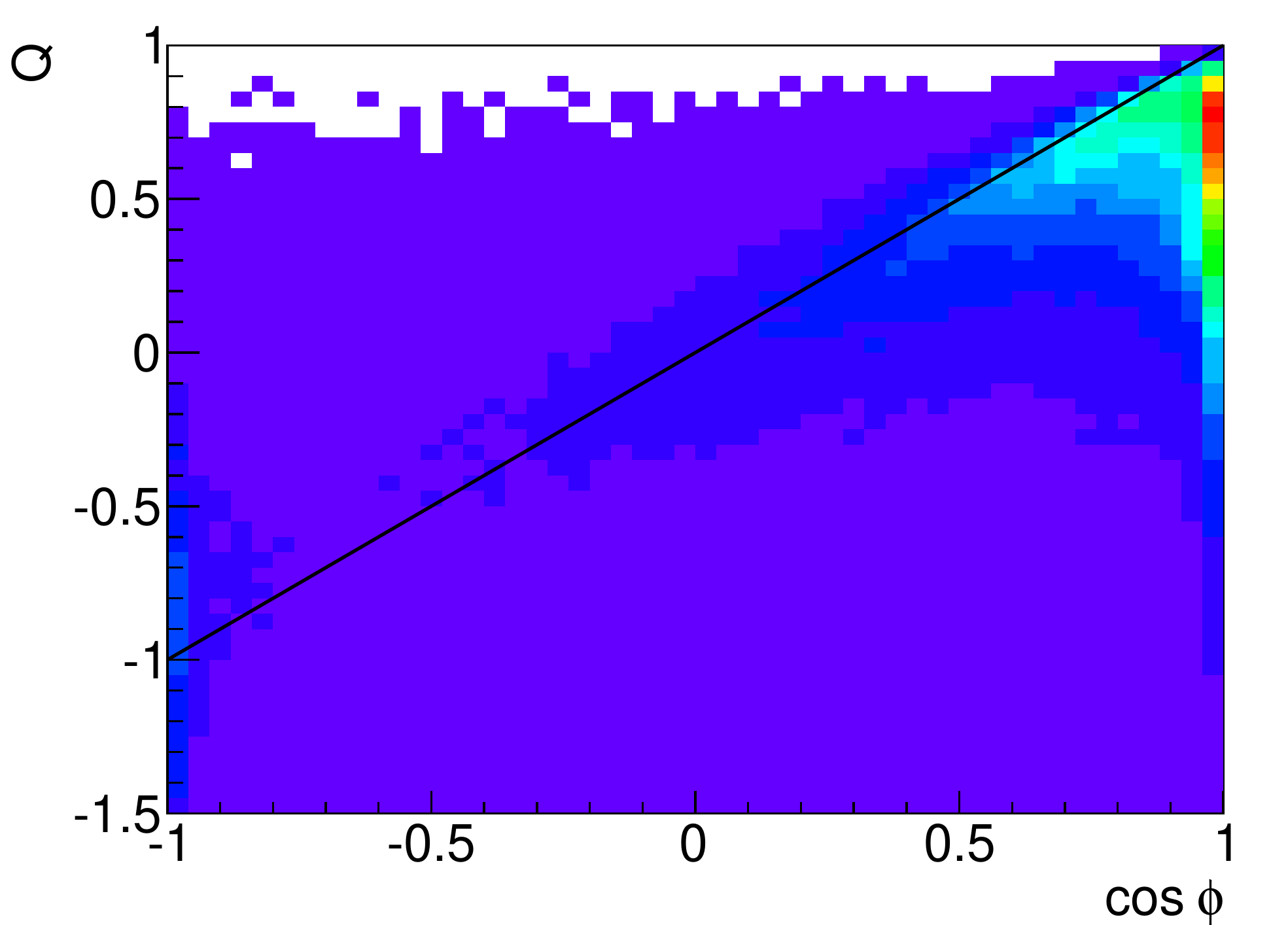}}
\vspace{-0.5cm}
\subfigure[~$\ttbar \rightarrow ll$]{\label{subfig:Qcosphi_ttbarll}\includegraphics[width=0.8\textwidth]{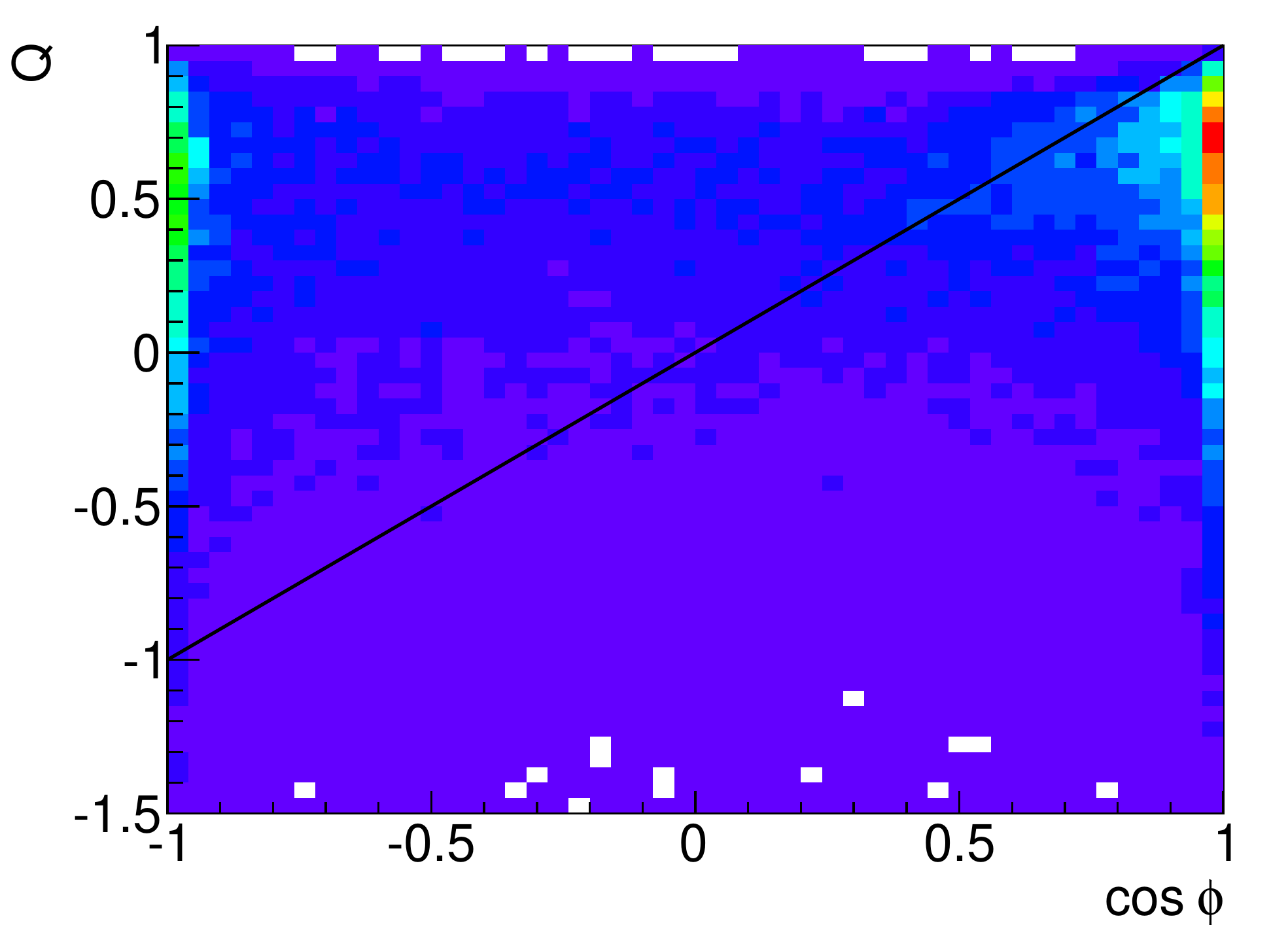}}
\end{center}
\caption{Deconstructed transverse mass for top quark pair production for events
passing selection cuts in the
\subref{subfig:Qcosphi_ttbarlq} lepton+jets decay mode,
\subref{subfig:Qcosphi_ttbarll} di-lepton decay mode. 
The black diagonal line is defined by $Q=\cphi$, i.e. $m_{T} = 80.4$ GeV.  The region above the black line corresponds to $m_T > 80.4$~GeV. The event count per bin follows rainbow colors in linear scale. (Color online)}
\label{fig:ttbarQcos}
\end{figure}

\begin{figure}[!h!tbp]
\begin{center}
\vspace{-0.5cm}
\subfigure[~$\tilde{t}\tilde{t}^*,\;m_{\tilde{t}} = 400$ GeV, $m_\chi=200$~GeV]{\label{subfig:Qcosphi_400_200}\includegraphics[width=0.8\textwidth]{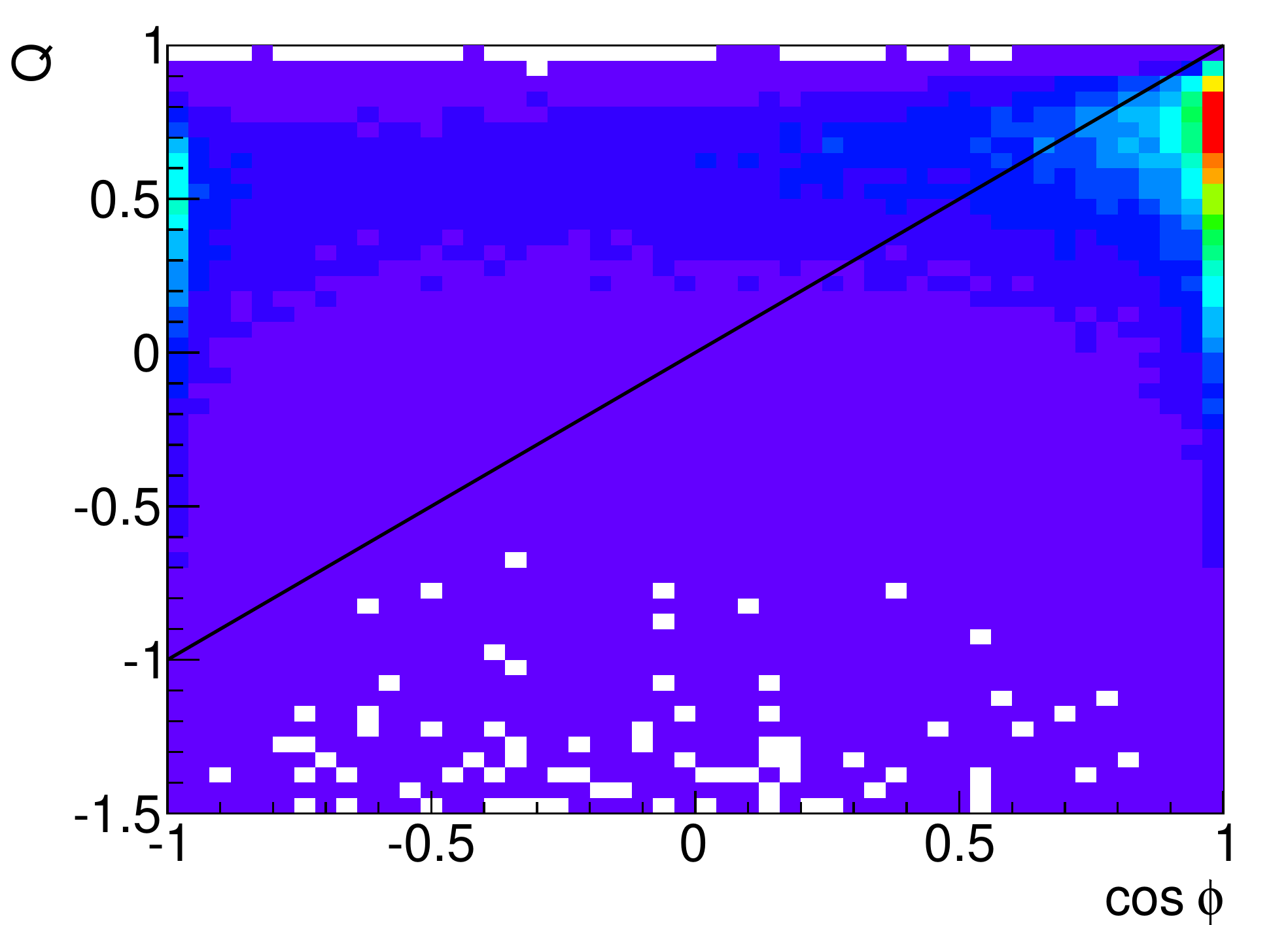}}
\vspace{-0.5cm}
\subfigure[~$\tilde{t}\tilde{t}^*,\;m_{\tilde{t}} = 500$ GeV, $m_\chi=200$~GeV]{\label{subfig:Qcosphi_500_200}\includegraphics[width=0.8\textwidth]{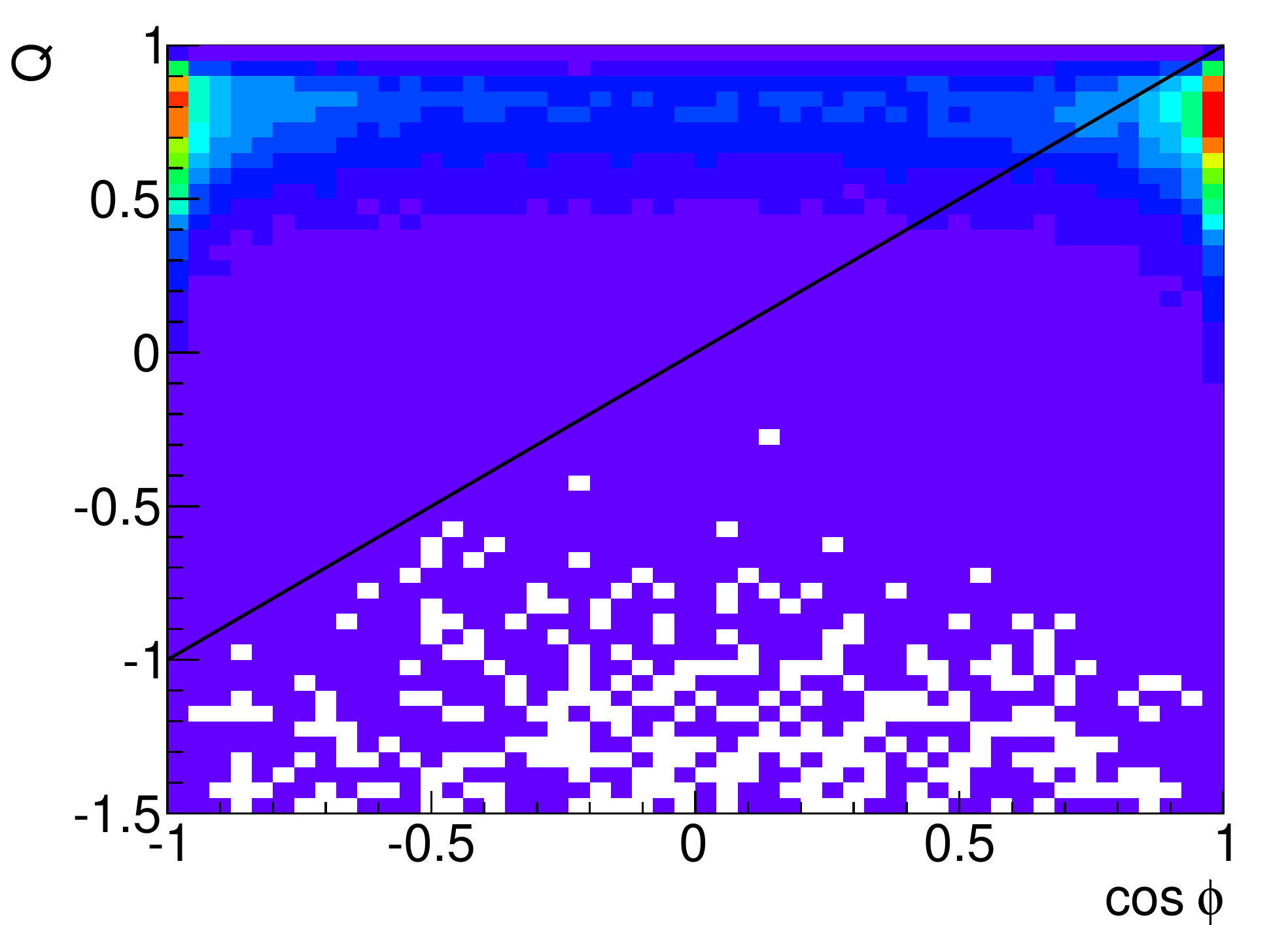}}
\end{center}
\caption{Deconstructed transverse mass for stop pair production for events passing
selection cuts, for stop and neutralino masses of
\subref{subfig:Qcosphi_400_200} $m_{\tilde{t}}=400$~GeV, $m_\chi=200$~GeV,
\subref{subfig:Qcosphi_500_200} $m_{\tilde{t}}=500$~GeV, $m_\chi=200$~GeV.
The black diagonal line is defined by $Q=\cphi$, i.e. $m_{T} = 80.4$ GeV.  The region above the black line corresponds to $m_T > 80.4$~GeV. The event count per bin follows rainbow colors in linear scale. (Color online)
} 
\label{fig:stopQcos}
\end{figure}

Figure~\ref{fig:ttbarQcos} shows the correlation between $Q$ and $\cphi$ for top quark pair events in the lepton+jets and dilepton channels. The majority of events cluster near $\cphi =1$ and $Q=0.6$, corresponding to events where the lepton and neutrino point roughly in the same direction, and where the longitudinal neutrino momentum is small. Entries below the black line given by $Q=\cphi$ with $M_T^0=M_{W}=80.4$~GeV have physical solutions for the longitudinal neutrino momentum in Eq.~\ref{eq:pvl}, while entries above the line do not. Figure~\ref{subfig:Qcosphi_ttbarlq} shows the expected correlation along this diagonal line for lepton+jet events. For dilepton events shown in Fig.~\ref{subfig:Qcosphi_ttbarll}, that correlation is absent. The presence of a second neutrino in dilepton events also results in a narrow band close to  $\cphi =-1$, i.e. where lepton and $\etmiss$ are back-to-back.
Figure~\ref{fig:stopQcos} shows the same correlation for SUSY stop pair lepton+jets events at two different stop masses. In these SUSY events, where the top quark pair is produced together with additional non-interacting particles,
these additional particles are summed into $\etmiss$, which modify both the
magnitude and the direction of the $\etmiss$ vector.
There is little correlation between $Q$
and $\cphi$ along the diagonal line, as expected from the presence of additional sources of $\etmiss$. Instead, there are two peaks, close to $\cphi =1$ and $\cphi =-1$, with the peak at $\cphi =-1$ getting more pronounced as $m_{\tilde{t}}$ increases. It is the presence and location and shape of this additional peak that provides enhanced separation of stop events from the $\ttbar$ background, and this is the basis for our event selection.

Traditionally, dark matter searches have focused on the increase in the magnitude of $\etmiss$
and on removing SM lepton+jets backgrounds through a cut on $m_T$.
A high $\etmiss$ cut is adequate as long as the dark matter particle is produced with a large
transverse momentum. However, such a cut removes much of the phase space of interest here where
the dark matter is not significantly boosted in the transverse direction.
Equation~\ref{eq:bound1} tell us we can do better:  Consider a region of phase space where
$\etmiss$ is relatively small but the missing transverse energy vector is back-to-back with the
transverse lepton momentum.  In this case, Eq.~\ref{eq:bound1} becomes  positive definite
\begin{equation}
M_W^2 \lesssim 4\,p_{l\, T} \etmiss
\end{equation}
which is easily satisfied for sufficiently large $p_{l\, T}$. 

The transverse mass tends to be a powerful variable in searches for new physics with top events
because the presence of the neutralino dark matter
particles increases $m_T$ and moves the signal away from the large backgrounds where $m_T$ comes
from a $W$~boson decay~\cite{Han:2008gy}. A cut on $m_T$ was used in previous searches for stop
pair production~\cite{Aad:2012xqa,Aad:2011wc,Chatrchyan:2012uea}. The power of such a cut can be understood from Figs.~\ref{fig:ttbarQcos} and~\ref{fig:stopQcos}: A cut on $m_T$ corresponds to selecting events above the diagonal line in the $Q-\cphi$ plane starting in the
upper right hand corner at $Q=1$, $\cphi=1$. The black lines in Fig.~\ref{fig:ttbarQcos}
correspond to a cut $m_T \ge 80.4$~GeV. 
In general, the cut $m_{T} \ge m_{C}$ corresponds to the events above the line defined by
\begin{equation}
Q \left( m_T \right) = \frac{m_T^2}{m_C^2} \cphi + 1 - \frac{m_T^2}{m_C^2}\,.
\end{equation}

Such a $m_T$ cut removes the background peak at the upper right, while preserving the signal peak in the upper left. However, the shape of the signal and the shape of the $\ttbar$ dilepton background are quite different from the diagonal behavior. Their separation can be improved through appropriate contour cuts in $Q$ vs. $\cphi$.

An additional correlation can be exploited to provide additional separation between signal and background for events in the upper left-hand corner of Figs.~\ref{fig:ttbarQcos} and~\ref{fig:stopQcos}. By looking at the correlation of $\cphi$ not only with $Q$ but also with $\etmiss$, the kinematic separation of signal and background becomes more obvious.

\begin{figure}[!h!tbp]
\begin{center}
\subfigure[~$\ttbar$]{\includegraphics[width=0.475\textwidth,clip=true]{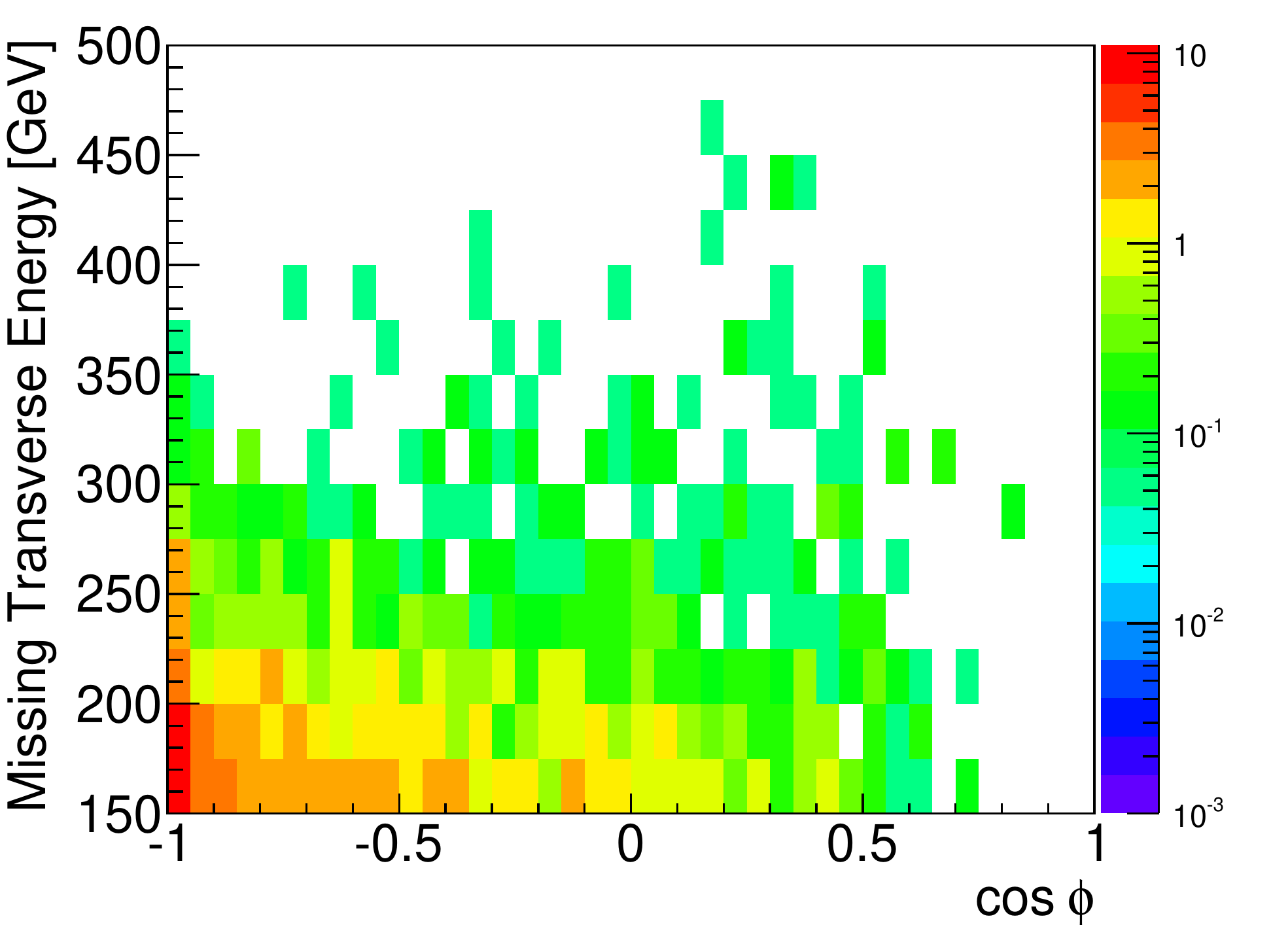}}
\subfigure[~$\tilde{t}\tilde{t}^*,\;m_{\tilde{t}} = 400$ GeV, $m_\chi=200$~GeV]{\includegraphics[width=0.475\textwidth,clip=true]{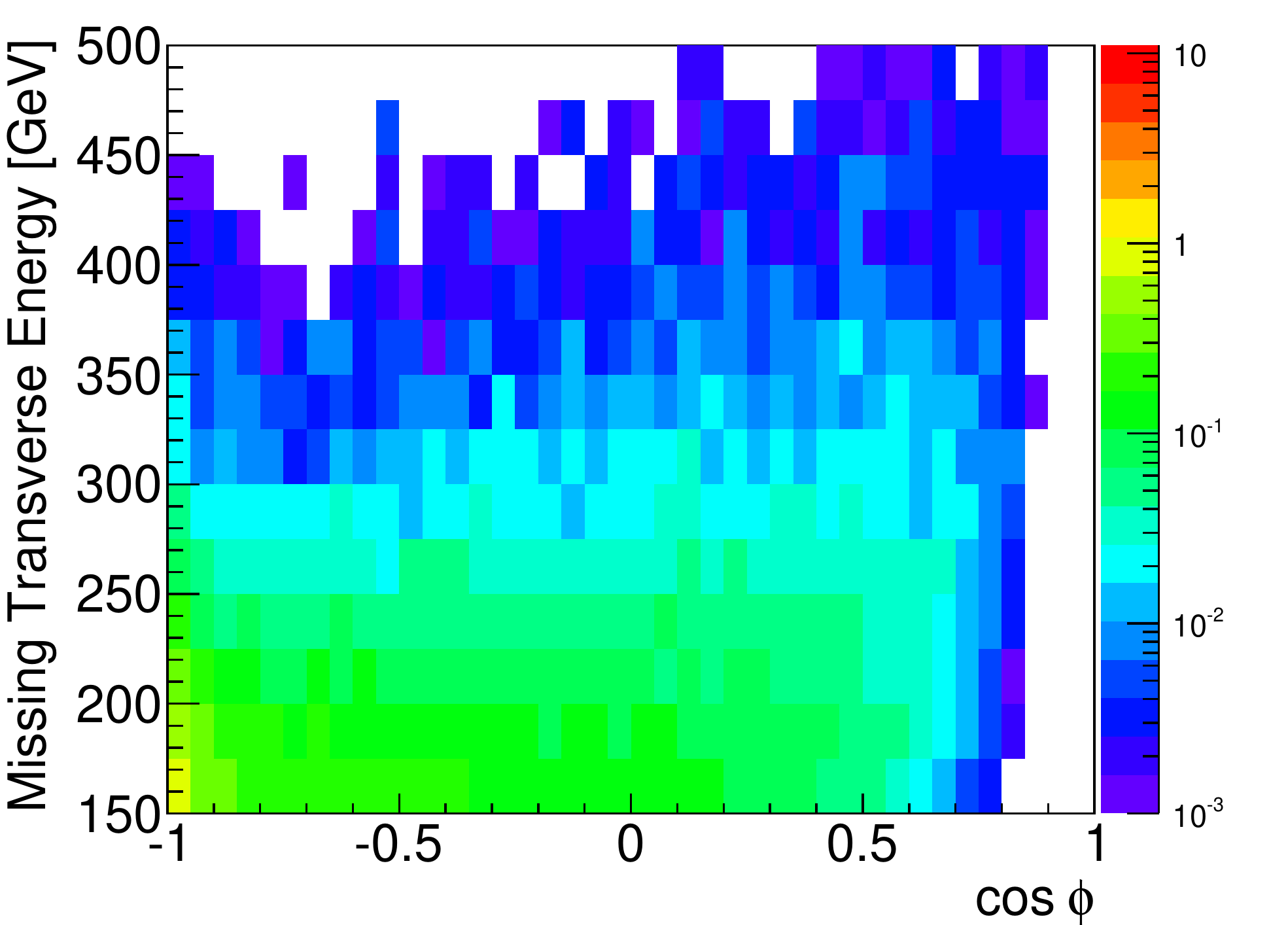}}
\end{center}
\caption{$\etmiss$ vs $\cphi$ distribution for events that pass selection cuts and satisfy 140~GeV$<m_T<250$~GeV and a top mass window cut. The rainbow color log scale indicates the number of events per bin, which is identical in the two distributions. (Color online)}
\label{fig:met_cosphi_sr0}
\end{figure}

Figure~\ref{fig:met_cosphi_sr0} shows the distribution of $\etmiss$ vs $\cphi$ for events with 140~GeV$<m_T<250$~GeV and passing a top mass window cut (see Sec.~\ref{sec:sr0}), i.e. removing the large peak near $\cphi=1$ in the $Q$ vs. $\cphi$ distributions. This $m_T$ window cut reveals that the signal has events not only near $\cphi=-1$, but filling almost the full $\cphi$ range. This is also visible as a horizontal band in distributions in Fig.~\ref{fig:stopQcos}. The $\ttbar$ background by contrast is mostly centered at low $\etmiss$ and near $\cphi=-1$. Thus, the best separation of stop signal from $\ttbar$ background is achieved by exploring the correlation amongst $Q$, $\etmiss$ and $\cphi$.

\section{Stop Pair Analysis}
\label{sec:analysis}

We demonstrate the power of the deconstructed transverse mass in stop pair production for
stop/$\chi$ mass combinations of $m_{\tilde{t}}=500$~GeV, $m_\chi=200$~GeV;
$m_{\tilde{t}}=400$~GeV, $m_\chi=200$~GeV and $m_{\tilde{t}}=350$~GeV, $m_\chi=200$~GeV.
Event yields are computed for 20~fb$^{-1}$ of proton-proton collision data at the 8~TeV LHC.

\subsection{Hadronic Top Mass}

\begin{figure}[!h!tbp] 
\centering
\includegraphics[width=0.6\textwidth,clip=true]{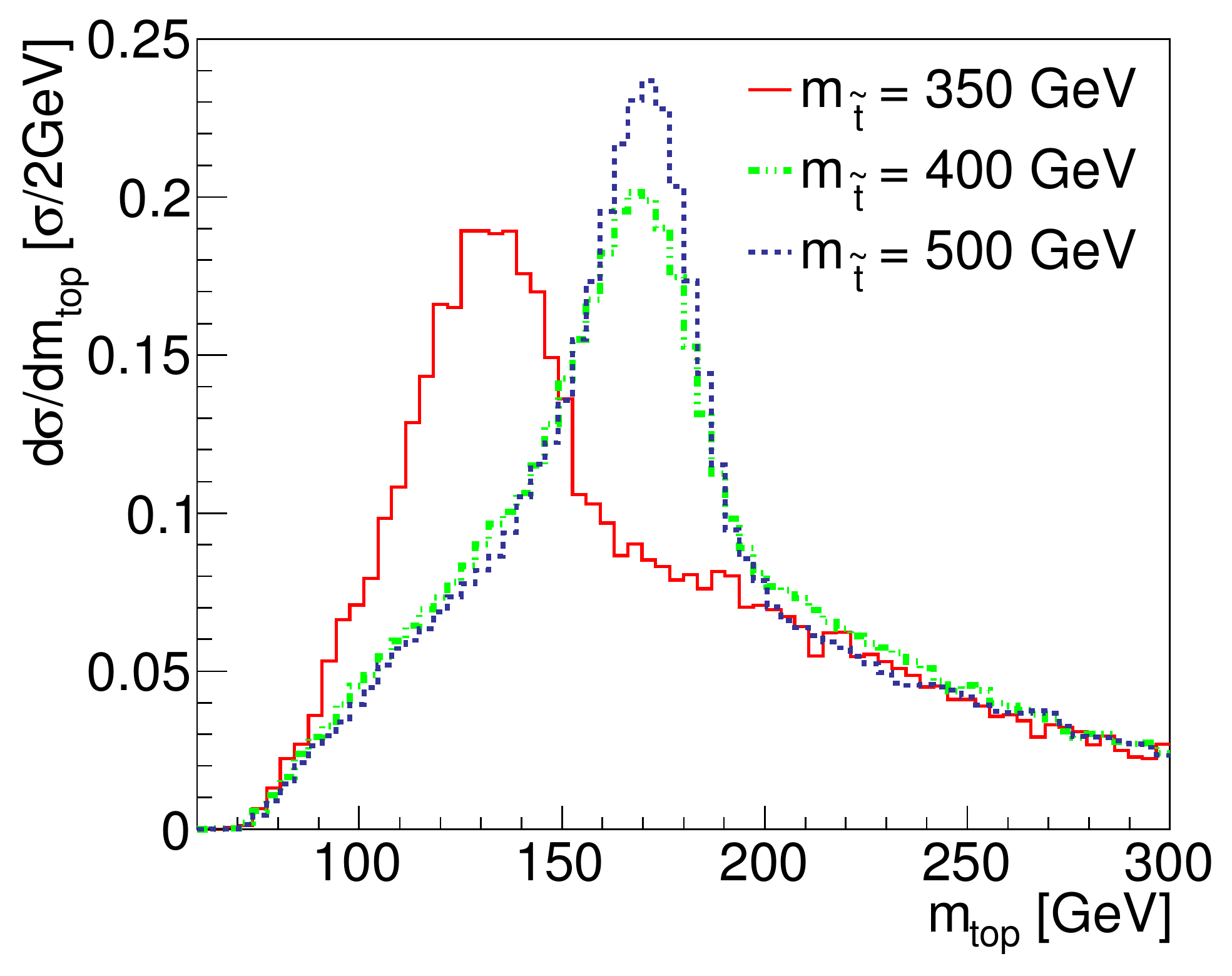}
\caption{Reconstructed 3-jet mass
for stop pair production at the 8~TeV LHC, for $m_{\chi}=200$~GeV and different stop masses.
(Color online) }
\label{fig:mjjj}
\end{figure}

In the same manner as in ~\cite{ATLAS-CONF-2013-037},
we form the three-jet mass of the hadronic top candidate $m_{jjj}$ by positing the hadronic $W$ candidate as the closest pair of jets with an invariant mass $ \ge 60 $~GeV.
The closest jet to the $W$ candidate is used to form the invariant mass $m_{jjj}$.
The $m_{jjj}$ distributions for the three signal masses are shown in Fig.~\ref{fig:mjjj}.
For $m_{\tilde{t}} - m_{\chi} \ge \mtop$, the distributions are consistent with a peak near $\mtop$.
For $m_{\tilde{t}} - m_{\chi} \le \mtop$, 
e.g. $m_{\tilde{t}}=350$~GeV and $m_\chi=200$~GeV, the distribution peaks significantly below $\mtop$. This signal sample is still accessible despite the low reconstructed top mass as will be discussed in Section~\ref{sec:sr2}.

\subsection{Tau Lepton Rejection}
\label{sec:tau_rejection}

\begin{figure}[!h!tbp]
\begin{center}
\subfigure[~$\ttbar \rightarrow ll'$, where $l=e,\,\mu,\,\tau$ and~$l'=\tau\to$~hadrons]{\label{subfig:Qcosphi_ttbar_leptonic_keep_tau}\includegraphics[width=0.475\textwidth]{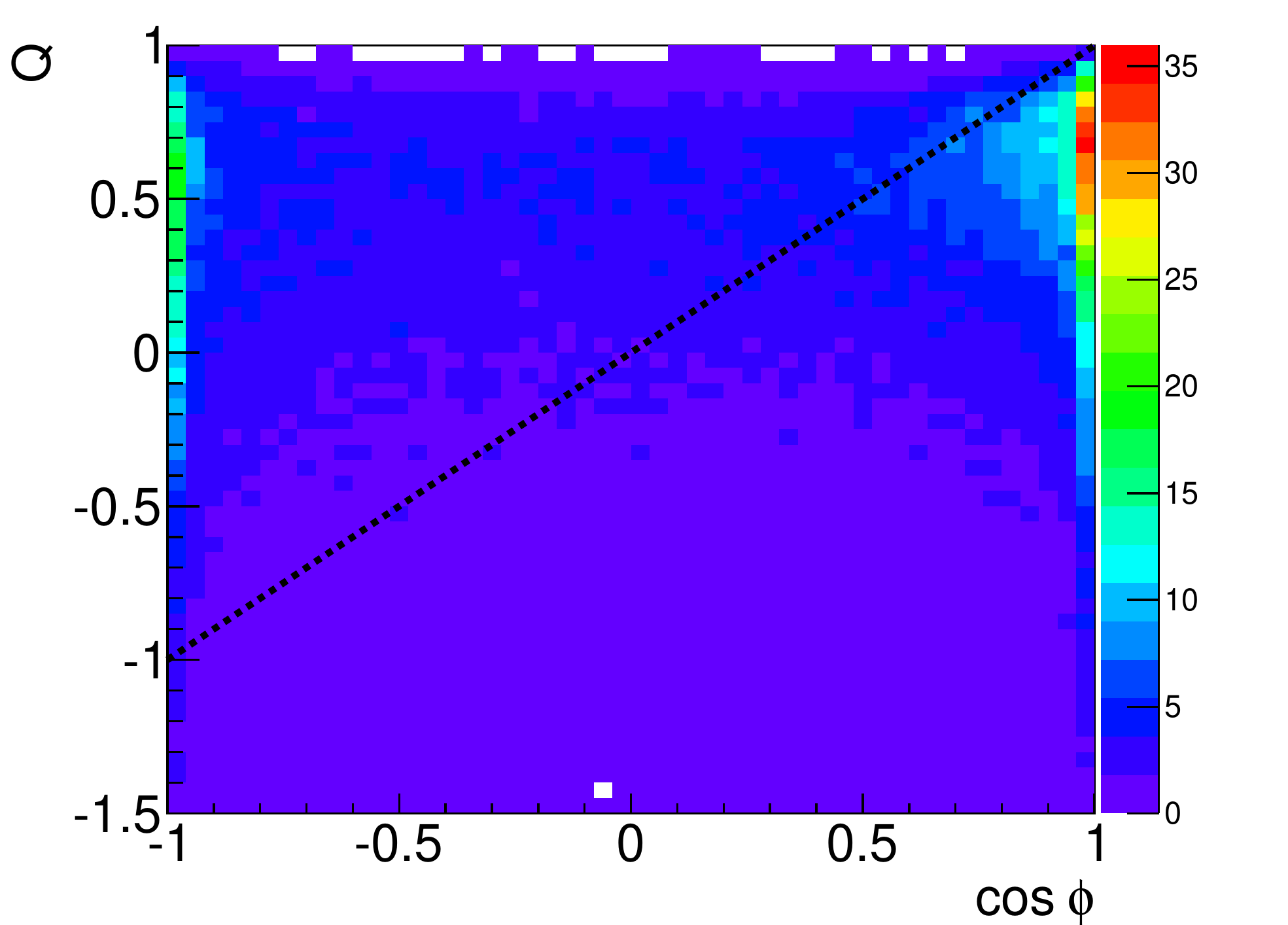}}
\subfigure[~$\ttbar \rightarrow ll'$, where $l=e,\,\mu$ and $l'=e,\,\mu$]{\label{subfig:Qcosphi_ttbar_leptonic_veto_tau}\includegraphics[width=0.475\textwidth]{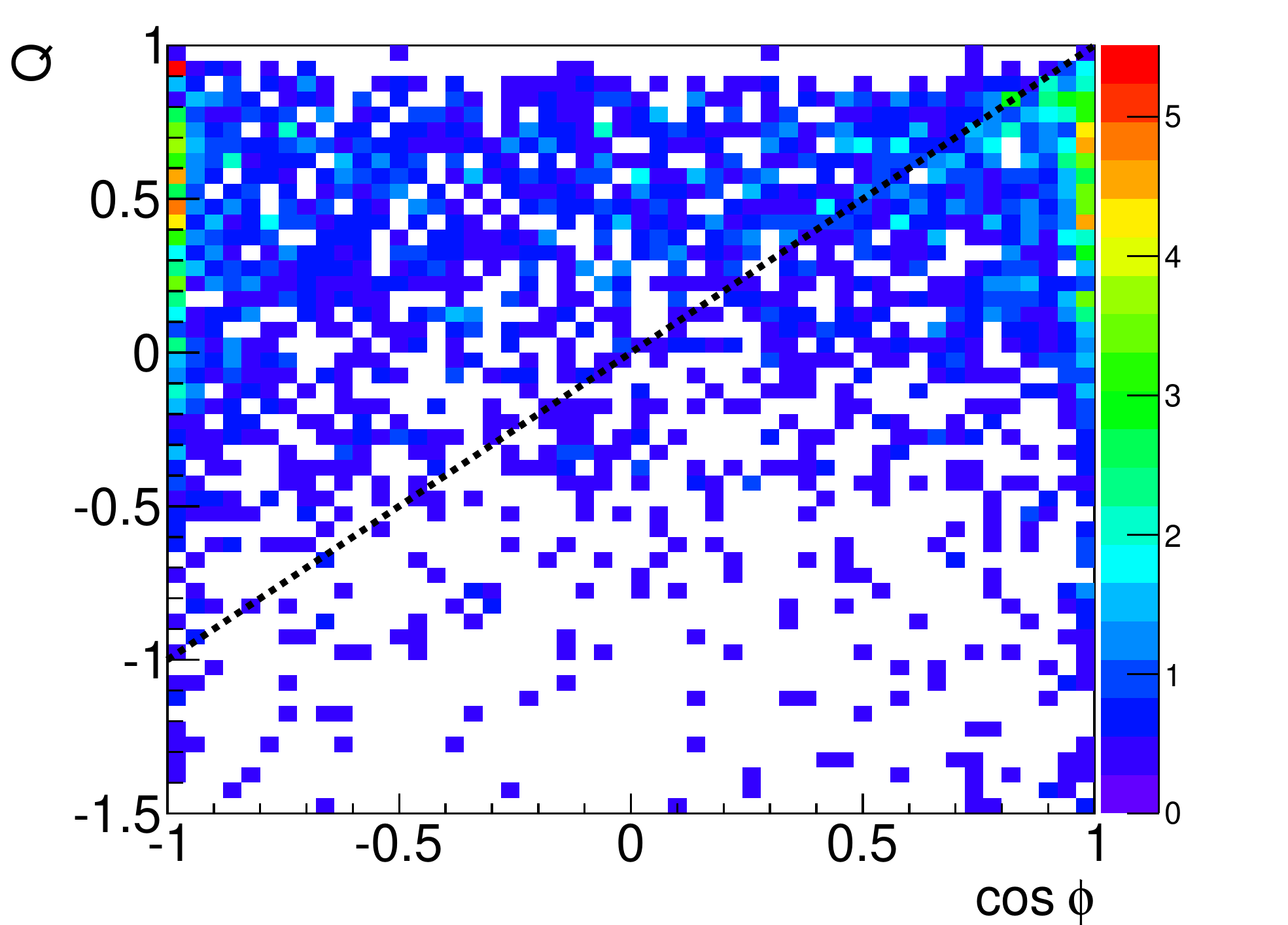}}
\end{center}
\caption{Deconstructed transverse mass for top pair dilepton events that pass selection cuts, for dilepton events that \subref{subfig:Qcosphi_ttbar_leptonic_keep_tau} contain at least one tau lepton and \subref{subfig:Qcosphi_ttbar_leptonic_veto_tau} contain no taus. 
The black diagonal line is defined by $Q=\cphi$. The event count per bin follows rainbow colors in linear scale. (Color online)} 
\label{fig:Qcosphi_ttbar_leptonic}
\end{figure}

Top quark decays involving tau leptons are an important background in this analysis that can be
addressed effectively through a tau ID veto. A large fraction of the $\ttbar$ background in the stop signal region consists of events where a $W$~boson decays to a tau lepton which subsequently decays hadronically. 
Figure~\ref{subfig:Qcosphi_ttbarll} shows the $Q-\cphi$ distribution for top pair dilepton events;
Figure~\ref{subfig:Qcosphi_ttbar_leptonic_keep_tau} shows dilepton events with at least one $\tau$ present in the $W$ decay chain. Figure~\ref{subfig:Qcosphi_ttbar_leptonic_veto_tau} shows the complement where $W \rightarrow \tau \, \nu$ have been removed.

Events where one top quark decays to a light lepton (electron or muon) and the other one
decays to a tau have a similar distribution in $Q-\cphi$ as the signal shown in Fig.~\ref{fig:stopQcos}. Most of these events contain taus which subsequently decays hadronically. The presence of three neutrinos in these events give them kinematic properties similar to the signal, making $\ttbar$ tau events difficult to distinguish from the signal through cuts alone. However,
these events can be effectively rejected by simply looking for a hadronically reconstructed tau.
In dedicated tau analyses for ATLAS and CMS, the efficiency to correctly identify the presence of a tau through its hadronic decay is approximately 80\%, with a hadronic jet rejection of better than a factor of ten~\cite{Aad:2009wy,Ball:2007zza,ATLAS-CONF-2013-064}. We turn this around and apply a hadronic tau veto, reducing events containing a hadronic tau decay by a factor five.
In this paper, we implement hadronic $\tau$ rejection in the $\ttbar$ background by scaling down events containing a hadronic $\tau$ decays by a factor 0.2. The signal and non-$\tau$ events are also reduced to account for the hadronic jet mis-identification. These non-$\tau$ samples are weighted by 0.8, appropriate for the four-jet events in this analysis.

This tau rejection is effective at reducing the background component that is most difficult to otherwise reject. As the color scale in Fig.~\ref{fig:Qcosphi_ttbar_leptonic} indicates, $\tau$ dilepton $\ttbar$ events outnumber non-$\tau$ dilepton $\ttbar$ events by about a factor seven. The hadronic $\tau$-veto reduces this difference to where hadronic $\tau$ and other dilepton events have about the same yield.

\subsection{Signal Region SR0}
\label{sec:sr0}

We demonstrate the efficacy of the deconstructed variable approach by comparing it to a nominal LHC-experiment-like analysis setup. We define the signal region SR0 that has a minimal set of cuts in addition the basic event selection.
This SR is similar to the signal region ``SRtN1'' from~\cite{ATLAS-CONF-2013-037} and the ``tn.diag'' region from~\cite{Aad:2014kra} and serves as a baseline for comparisons. Two additional selection criteria are applied:
\begin{itemize}
\item Hadronic top mass cut: 130~GeV$\le m_{jjj} \le 205$~GeV.
\item Transverse mass cut:  140~GeV$ \le m_{T} \le 250$~GeV.
\end{itemize}

\begin{figure}[!h!tbp]
\begin{center}
\subfigure[~$\ttbar$]{\label{subfig:Qc_ttbar_sr0}\includegraphics[width=0.49\textwidth,clip=true]{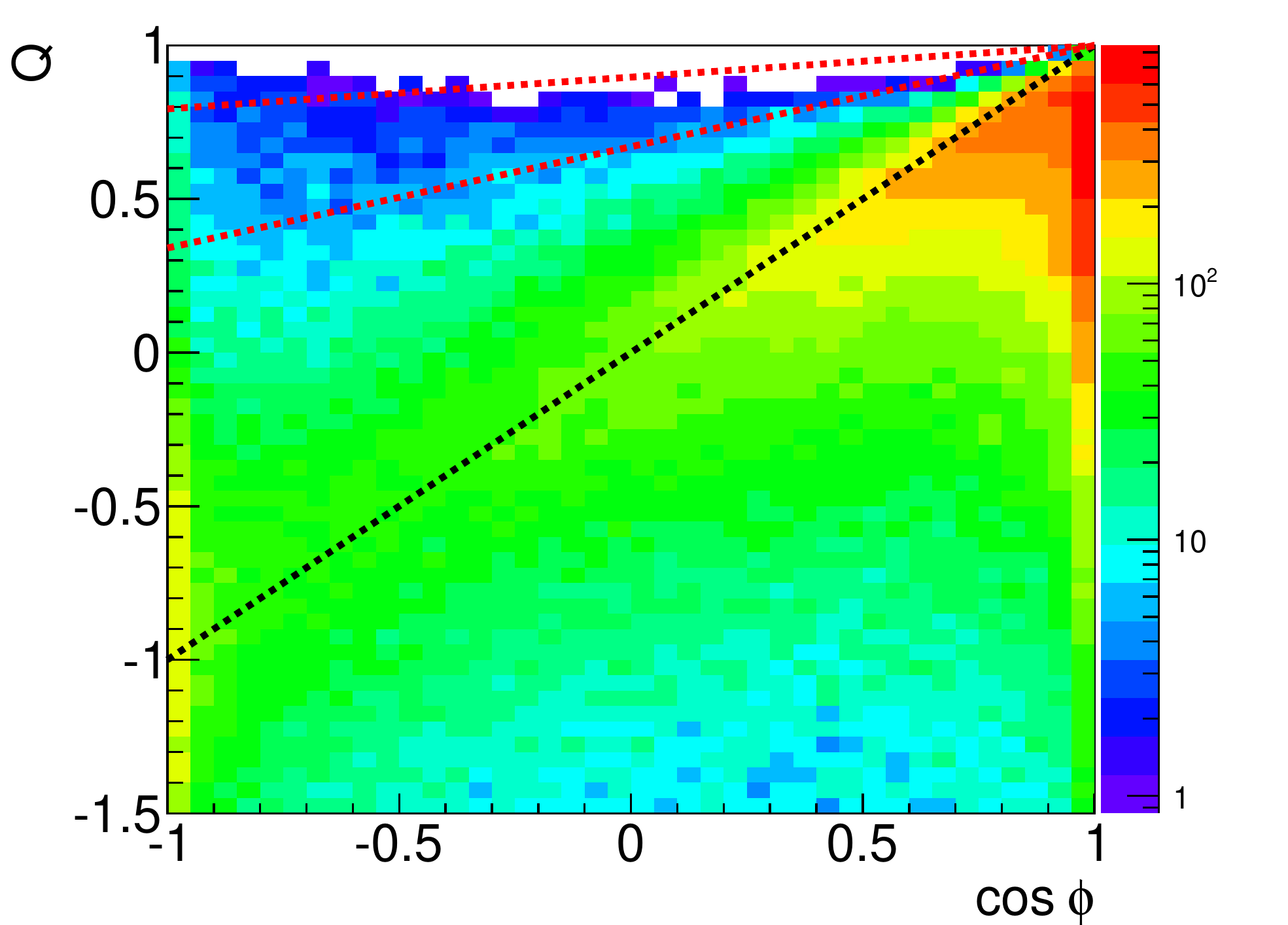}}
\subfigure[~$\tilde{t}\tilde{t}^*,\;m_{\tilde{t}} = 350$ GeV, $m_\chi=200$~GeV]{\label{subfig:Qc_350_sr0}\includegraphics[width=0.49\textwidth,clip=true]{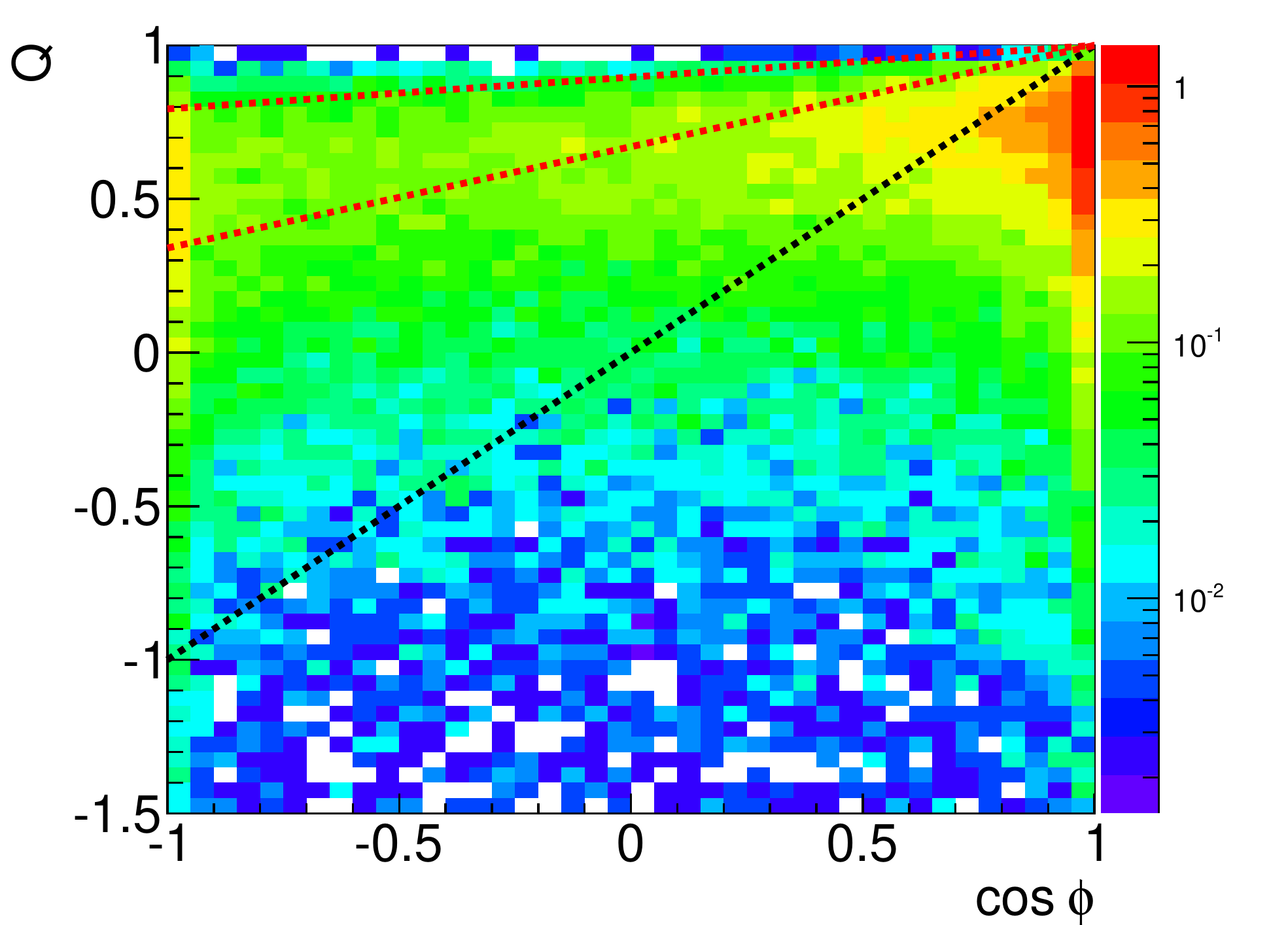}}
\subfigure[~$\tilde{t}\tilde{t}^*,\;m_{\tilde{t}} = 400$ GeV, $m_\chi=200$~GeV]{\label{subfig:Qc_400_sr0}\includegraphics[width=0.49\textwidth,clip=true]{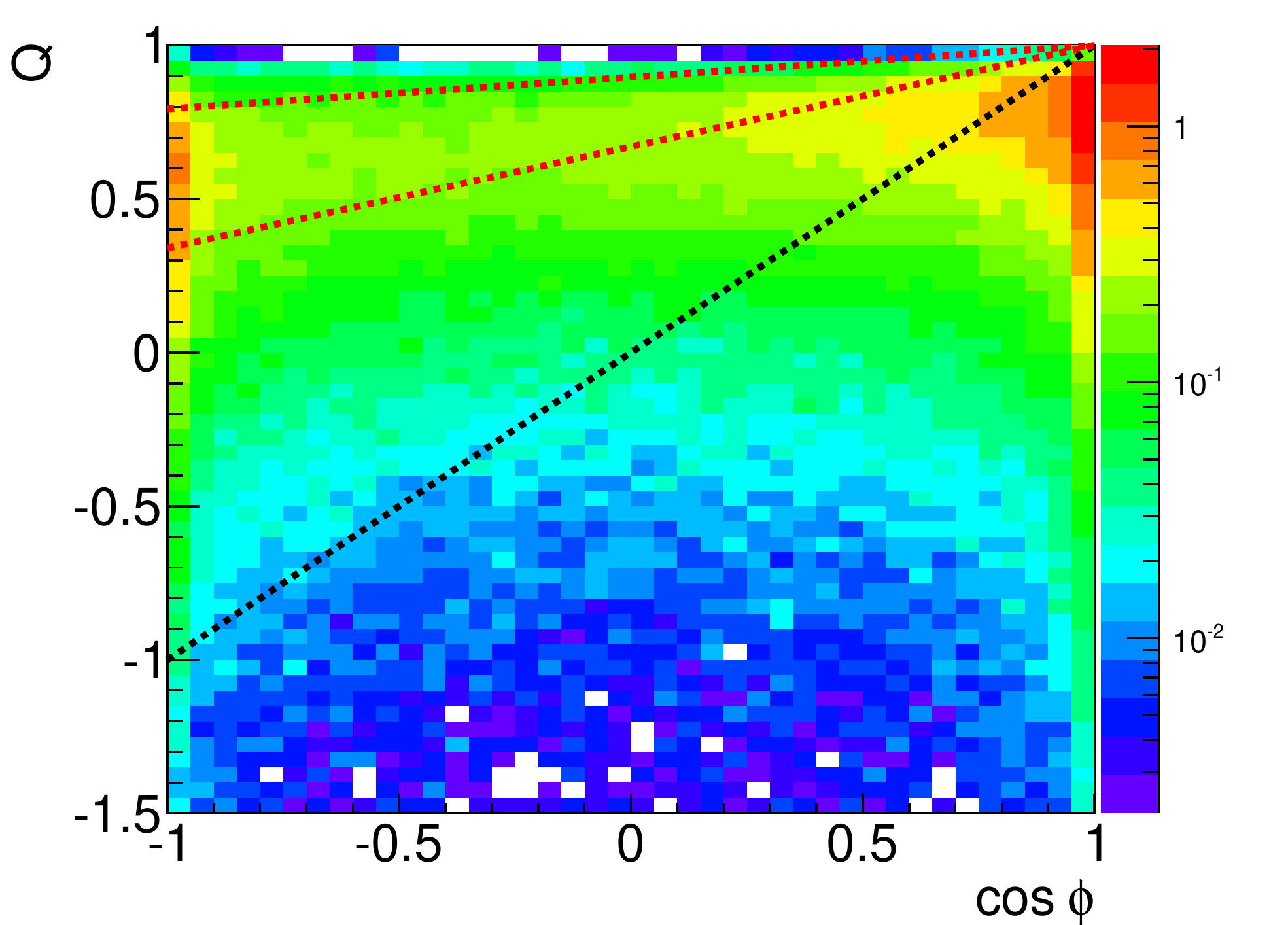}}
\subfigure[~$\tilde{t}\tilde{t}^*,\;m_{\tilde{t}} = 500$ GeV, $m_\chi=200$~GeV]{\label{subfig:Qc_500_sr0}\includegraphics[width=0.49\textwidth,clip=true]{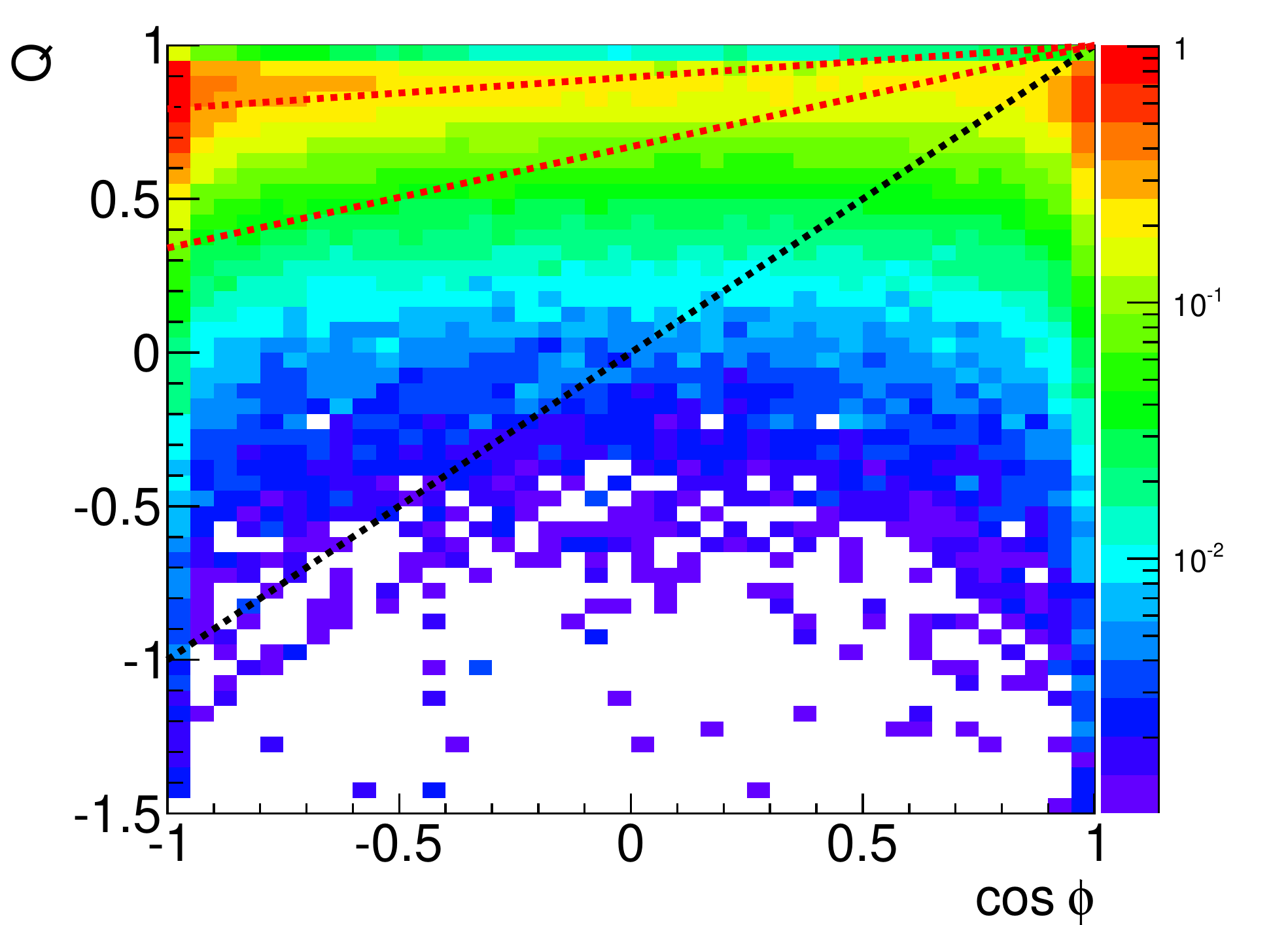}}
\end{center}
\caption{$Q$ vs $\cphi$ distribution for events passing basic selection cuts and a hadronic
top mass window cut. The black diagonal line is defined by $Q=\cphi$.
Events inside the contour bounded by the dashed red lines are selected by the SR0 transverse
mass cut. The event count per bin follows rainbow colors in log scale. (Color online)}
\label{fig:q_cosphi_sr0}
\end{figure}

Fig.~\ref{fig:q_cosphi_sr0} shows the distribution of $Q$ vs $\cphi$ for events passing the hadronic top mass cut. The region between the red lines corresponds to the transverse mass cut. The effectiveness of this cut in reducing the $\ttbar$ background can be clearly seen. However, the limits of the transverse mass cut can also be seen as there remains background in the region close to $\cphi=-1$. Moreover, while the $m_T$ window is appropriate for $m_{\tilde{t}}=400$~GeV, the upper cut is too low for $m_{\tilde{t}}=500$~GeV.
Event yields for signal and backgrounds passing both cuts are tabulated in Table~\ref{table:sr_yields}.

\subsection{Signal Region SR1}

We demonstrate the improvement coming from the deconstructed variables as well as hadronic tau rejection through two additional sets of cuts, one for resolved, on-shell top quarks (SR1), and one for off-shell top quarks (SR2). Signal region SR1 targets $m_{\tilde{t}} - m_{\chi} \ge \mtop$, i.e. signal mass combinations $m_{\tilde{t}} = 400$~GeV, $m_\chi=200$~GeV and $m_{\tilde{t}} = 500$~GeV, $m_\chi=200$~GeV.

\begin{figure}[!h!tbp]
\begin{center}
\subfigure[~$\ttbar$]{\label{subfig:Qc_ttbar_sr1}\includegraphics[width=0.49\textwidth,clip=true]{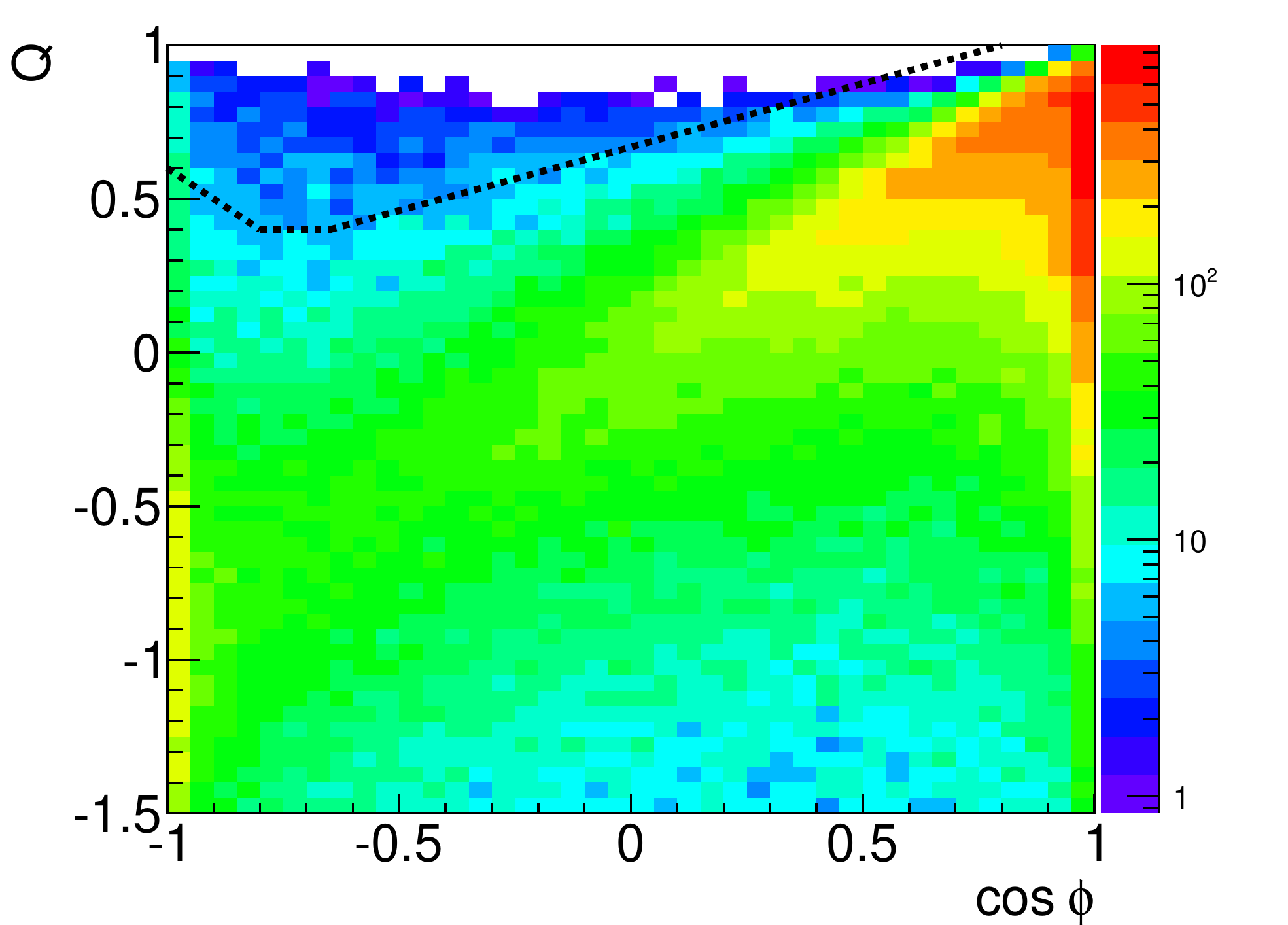}}
\subfigure[~$\tilde{t}\tilde{t}^*,\;m_{\tilde{t}} = 400$ GeV, $m_\chi=200$~GeV]{\label{subfig:Qc_400_sr1}\includegraphics[width=0.49\textwidth,clip=true]{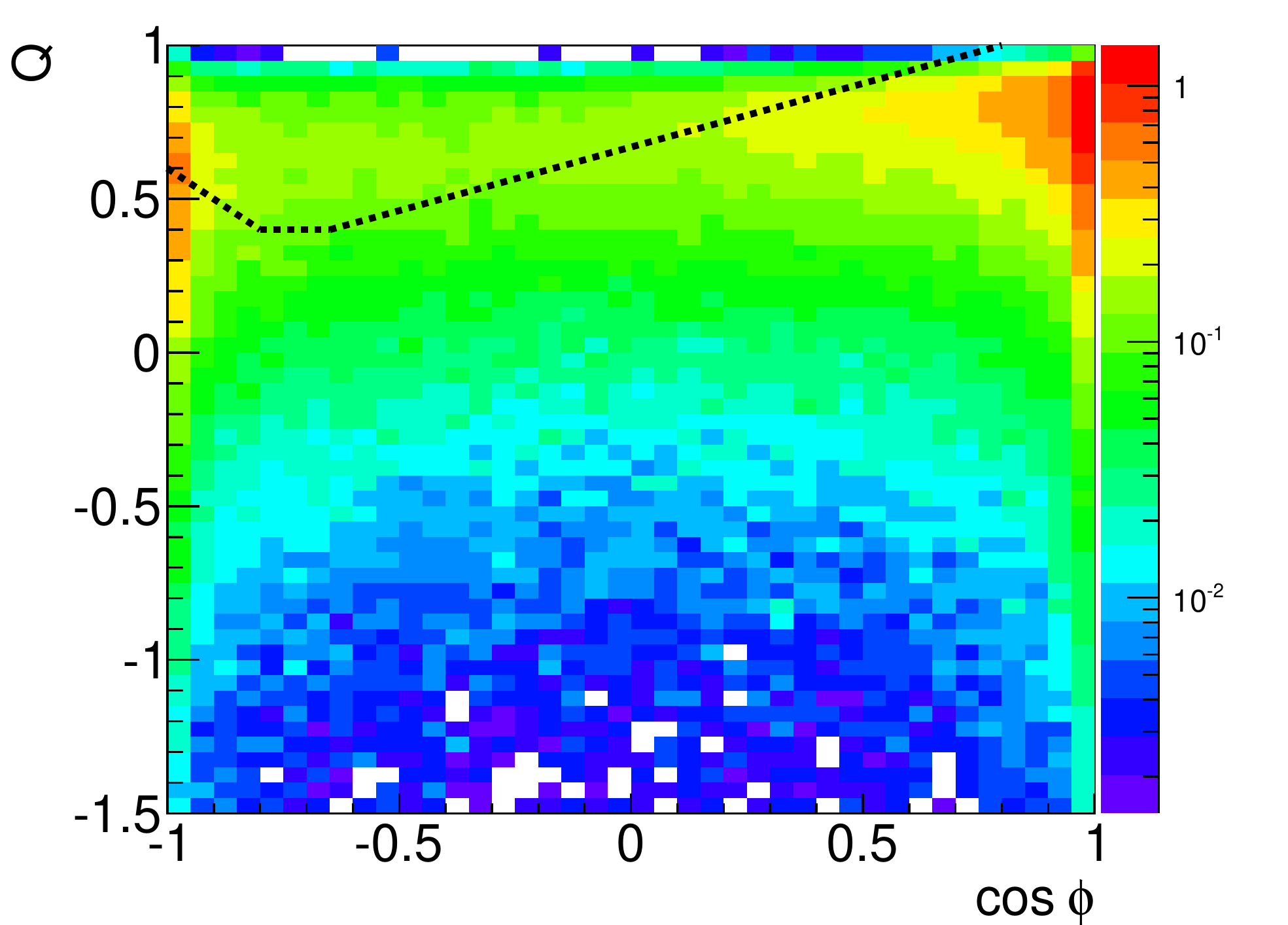}}
\subfigure[~$\tilde{t}\tilde{t}^*,\;m_{\tilde{t}} = 500$ GeV, $m_\chi=200$~GeV]{\label{subfig:Qc_500_sr1}\includegraphics[width=0.49\textwidth,clip=true]{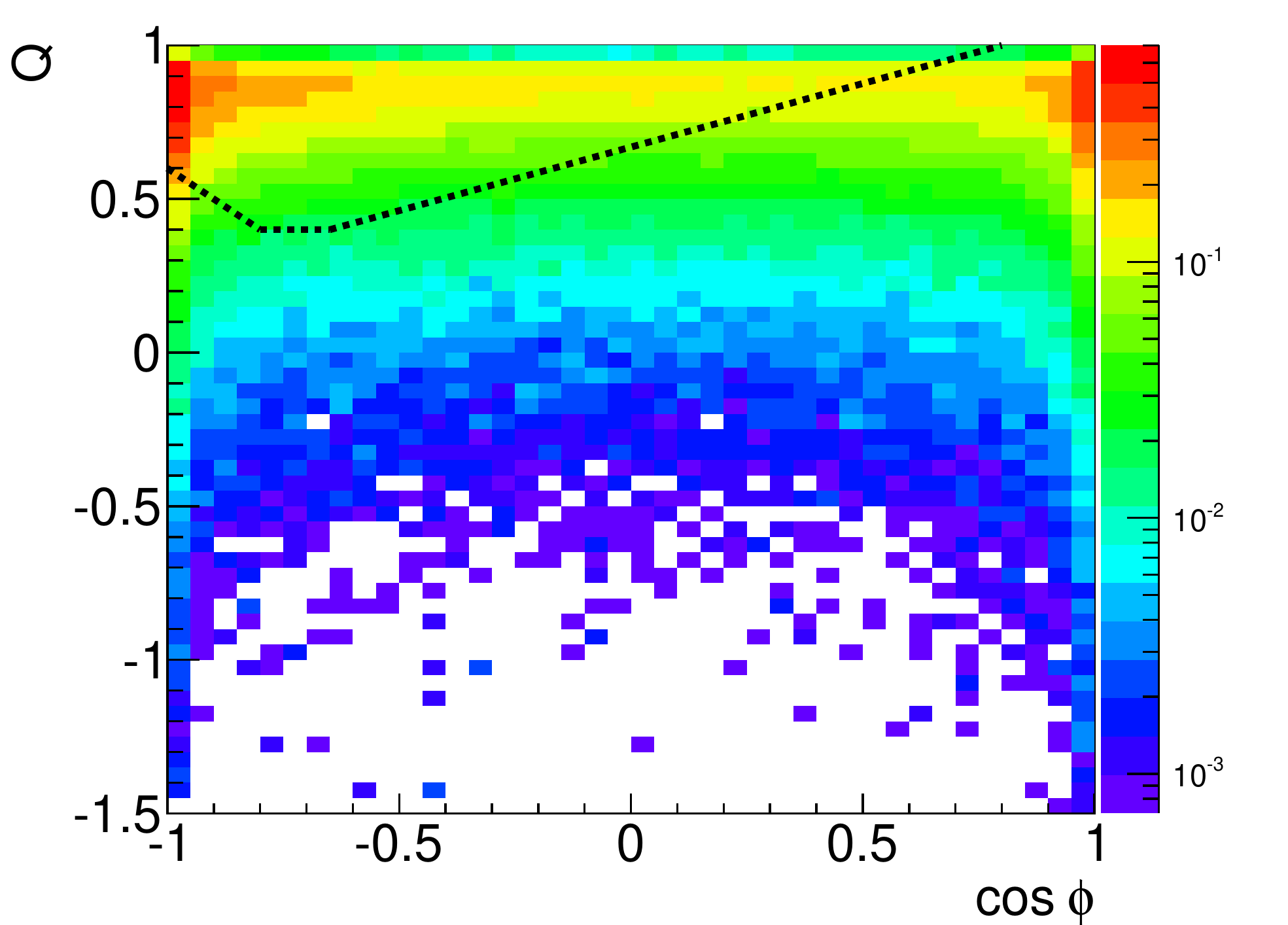}}
\end{center}
\caption{$Q$ vs $\cphi$ distribution for events with $130$ GeV $\le m_{jjj} \le 205$ GeV and tau veto. The event count per bin follows rainbow colors in log scale. (Color online)}
\label{fig:q_cosphi_sr1}
\end{figure}

Since there is sufficient mass difference between $m_{\tilde{t}}$ and $m_\chi$ to produce on-shell top quarks, we retain the SR0 top mass cut 130~GeV~$\le m_{jjj} \le 205$~GeV (c.f. Fig.~\ref{fig:mjjj}). We veto hadronic tau decays (c.f. Section~\ref{sec:tau_rejection}).
Fig.~\ref{fig:q_cosphi_sr1} shows the distribution of $Q$ vs $\cphi$ for these events. 

The $\ttbar$ background shape and event yields are different from Fig.~\ref{subfig:Qc_ttbar_sr0} mainly due to the tau veto. The signal distributions have the same shape as in Fig.~\ref{fig:q_cosphi_sr0} except for a small reduction in yields. The black line indicates the cut in the $Q-\cphi$ plane, which is no longer a simple straight line. Between $\cphi=-0.6$ and $\cphi=1$, the line corresponds to a transverse mass cut $m_T>140$~GeV, but the crucial region close to $\cphi=-1$ has improved background rejection. 

\begin{figure}[!h!tbp]
\begin{center}
\subfigure[~$\ttbar$]{\label{subfig:Ec_ttbar_sr1}\includegraphics[width=0.475\textwidth,clip=true]{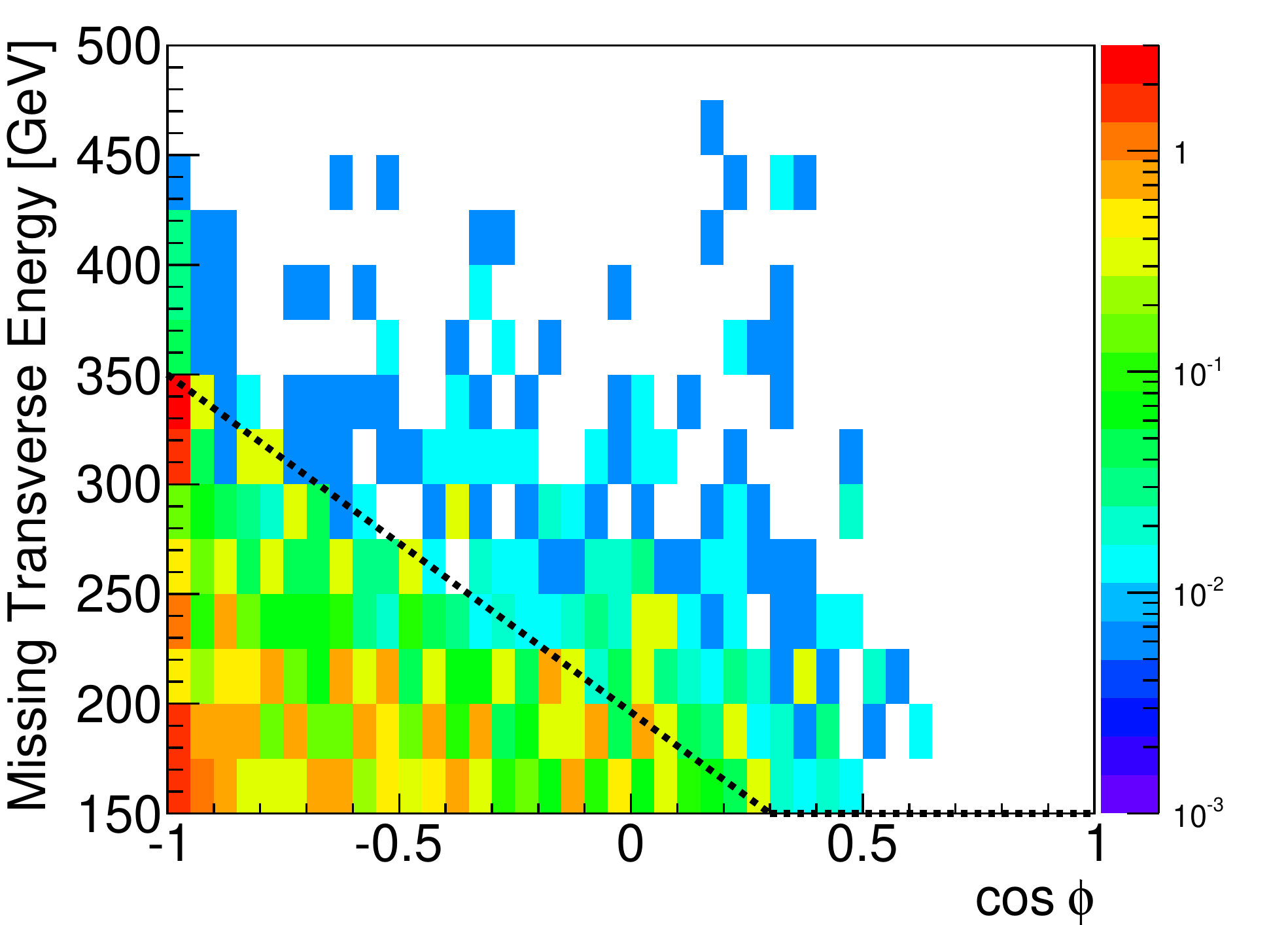}}
\subfigure[~$\tilde{t}\tilde{t}^*,\;m_{\tilde{t}}$ = 400 GeV, $m_{\chi}$ = 200 GeV]{\label{subfig:Ec_400_sr1}\includegraphics[width=0.475\textwidth,clip=true]{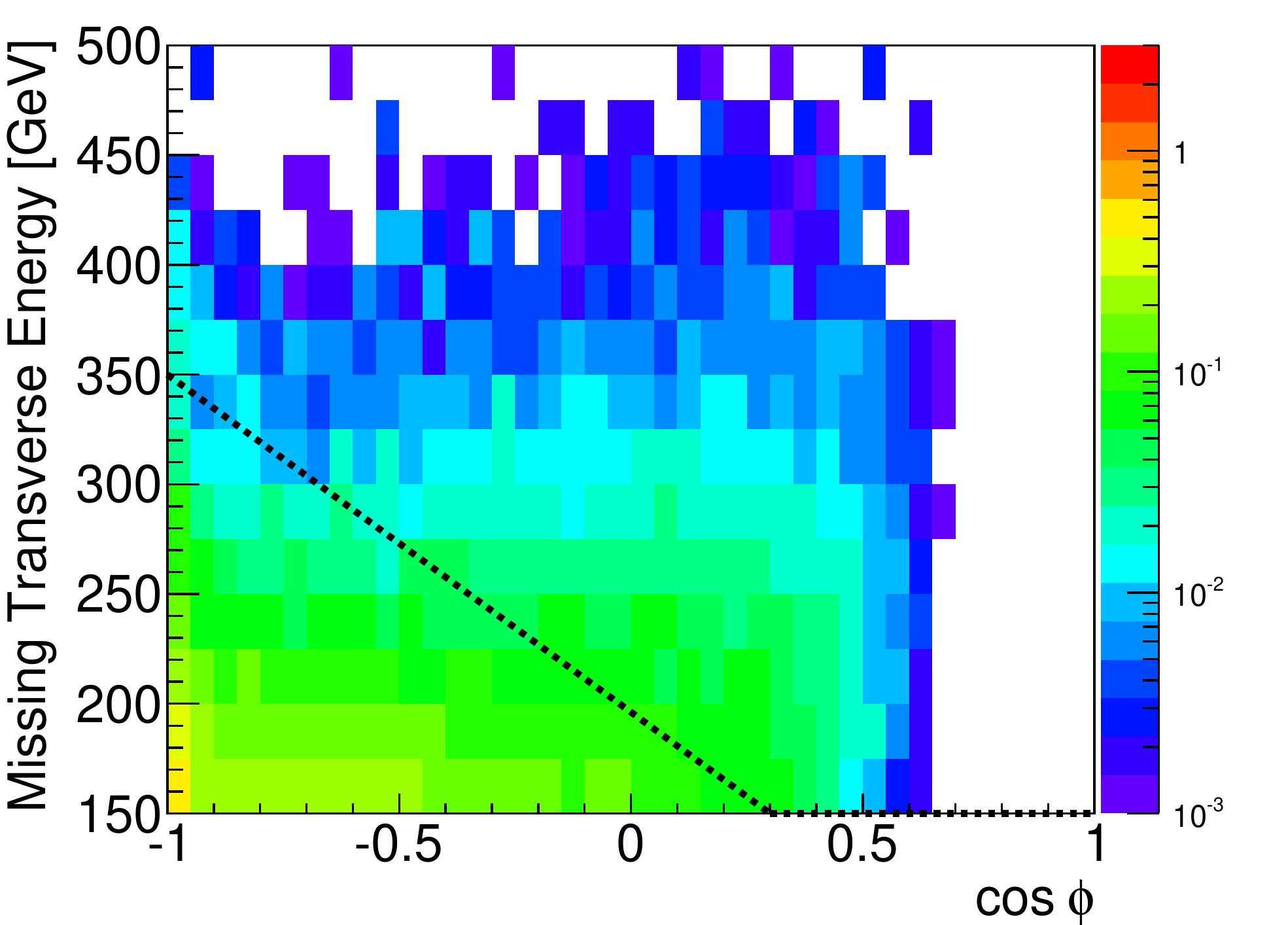}}
\subfigure[~$\tilde{t}\tilde{t}^*,\;m_{\tilde{t}}$ = 500 GeV, $m_{\chi}$ = 200 GeV]{\label{subfig:Ec_500_sr1}\includegraphics[width=0.475\textwidth,clip=true]{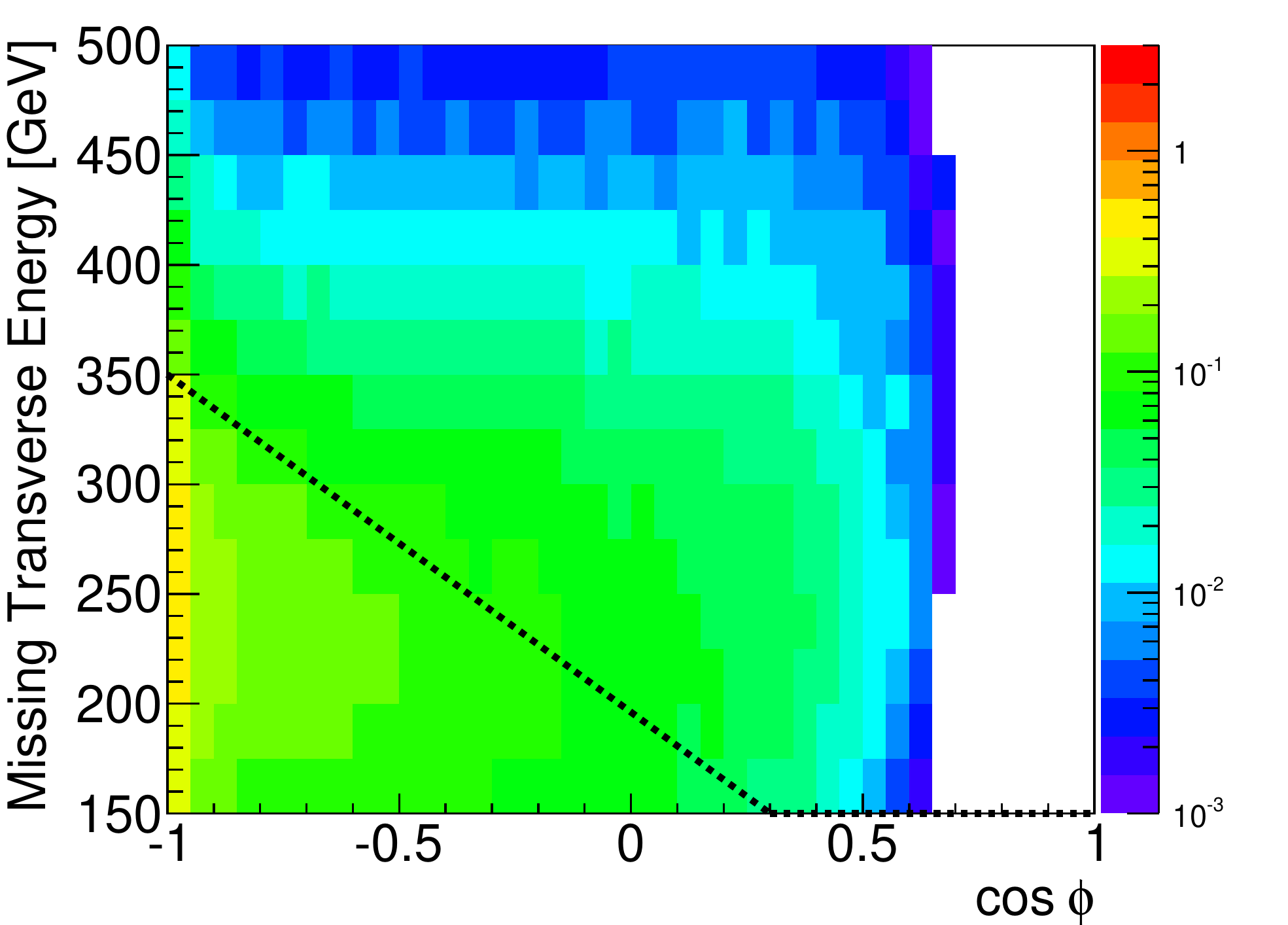}}
\end{center}
\caption{$\etmiss$ vs $\cphi$ distribution for events passing the $Q-\cphi$ contour cut from Fig.~\ref{fig:q_cosphi_sr1}. The black line indicates the $\etmiss-\cphi$ selection cut. The event count per bin follows rainbow colors in log scale. (Color online)}
\label{fig:met_cosphi_sr1t}
\end{figure}

The events in the region of the $Q$-$\cphi$ plane above the black line in Fig.~\ref{fig:q_cosphi_sr1} are selected to populate the $\etmiss$-$\cphi$ distribution in Fig.~\ref{fig:met_cosphi_sr1t}.

The $\etmiss$-$\cphi$ distribution reveals the full power of deconstruction. Most of the signal and background events overlap in the signal region of the $Q-\cphi$ distribution in Fig.~\ref{fig:q_cosphi_sr1}, but are clearly separated when switching the vertical axis from $Q$ to $\etmiss$. The top pair background in Fig.~\ref{subfig:Ec_ttbar_sr1} extends to high $\etmiss$ values predominantly at $\cphi=-1$ and is reduced for higher $\cphi$ values. The signal by contrast has a much flatter distribution in $\cphi$ and extends to very high $\etmiss$ values. Note that the color scale is the same for both plots in Fig.~\ref{fig:met_cosphi_sr1t}. This distribution is the first that shows regions of phase space where the signal is larger than the background. Events above the black line are selected and the final yields are tabulated in Table~\ref{table:sr_yields}.

Compared to SR0, the $\ttbar$ background is reduced by a factor of ${\cal O} \left( 40 \right)$,
while the signal only goes down by a factor of ${\cal O} \left( 2 \right)$. About a factor of ten of this additional background rejection comes from the deconstructed variables. Another factor four rejection is due to the hadronic tau veto, implying that most of the $\ttbar$ events in the signal region indeed contain a hadronically decaying tau.

This analysis of the deconstruction variables shows that large gains can be obtained in signal-background separation by fully exploiting the correlation of the missing energy with other objects in the event. Analyses that use multivariate analysis tools rather than simple cuts to isolate the stop signal~\cite{Chatrchyan:2013xna} will inherently take advantage of some of these correlations. However, only including all three of $Q$, $\etmiss$ and $\cphi$ will unleash the full power of deconstruction.

\subsection{Signal Region SR2}
\label{sec:sr2}

Our treatment so far as well as all experimental analyses select events with on-shell top quarks produced in the decay of the stop. However, in the compressed region where $m_{\tilde{t}} - m_{\chi} \le \mtop$, this is not the case anymore and the analysis procedure needs to be adjusted. The production cross section for stop pair production is the same whether $m_{\chi}\le m_{\tilde{t}} - \mtop$ or $m_{\chi}\ge m_{\tilde{t}} - \mtop$. Moreover, the stop decay branching to $bW\chi$ is 100\% in both cases for the models considered here. 

\begin{figure}[!h!tbp]
\begin{center}
\subfigure[~$\ttbar$]{\label{subfig:Qc_ttbar_sr2}\includegraphics[width=0.6\textwidth,clip=true]{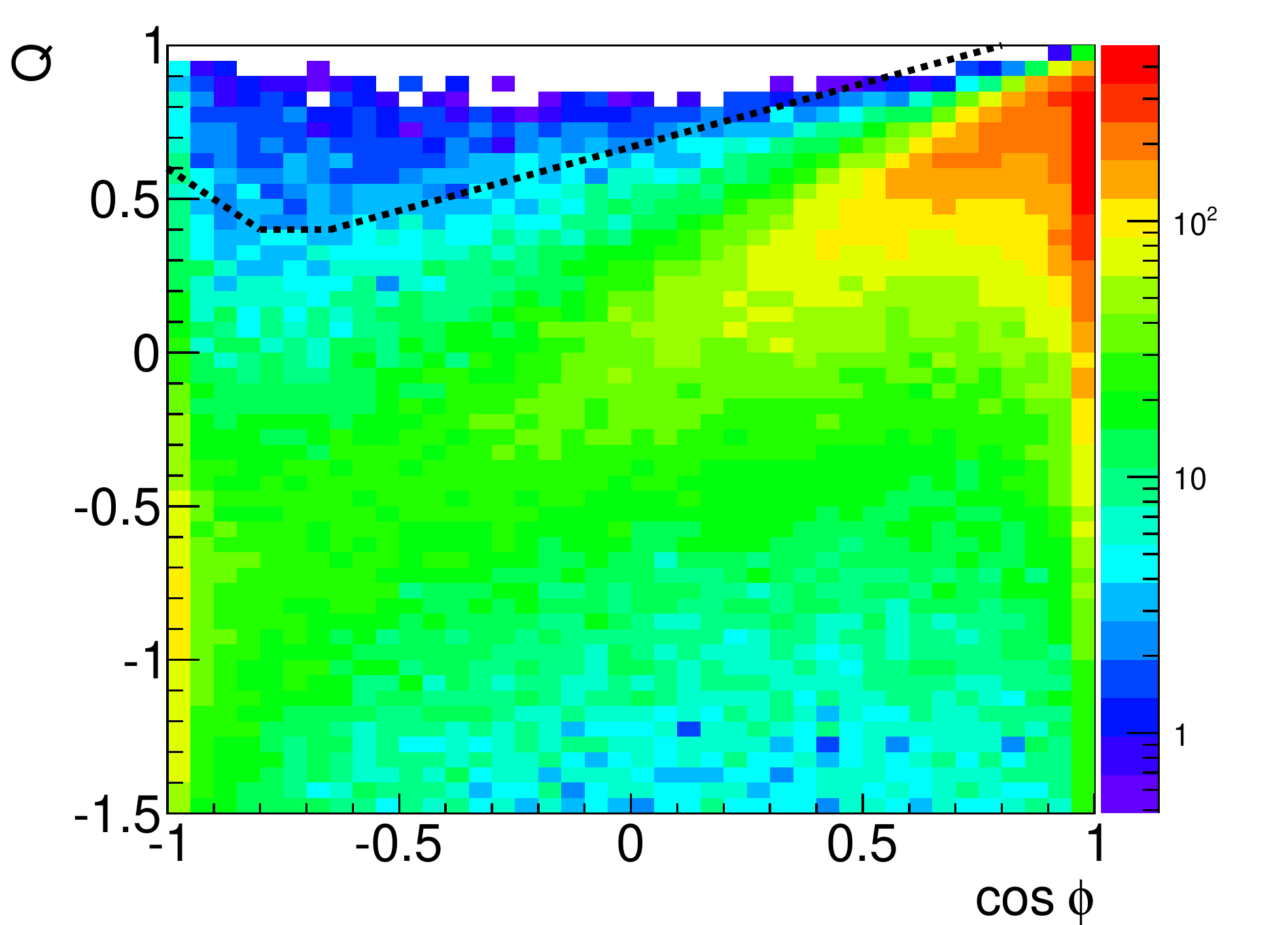}}
\subfigure[~$\tilde{t}\tilde{t}^*,\;m_{\tilde{t}}$ = 350 GeV, $m_{\chi}$ = 200 GeV]{\label{subfig:Qc_350_sr2}\includegraphics[width=0.6\textwidth,clip=true]{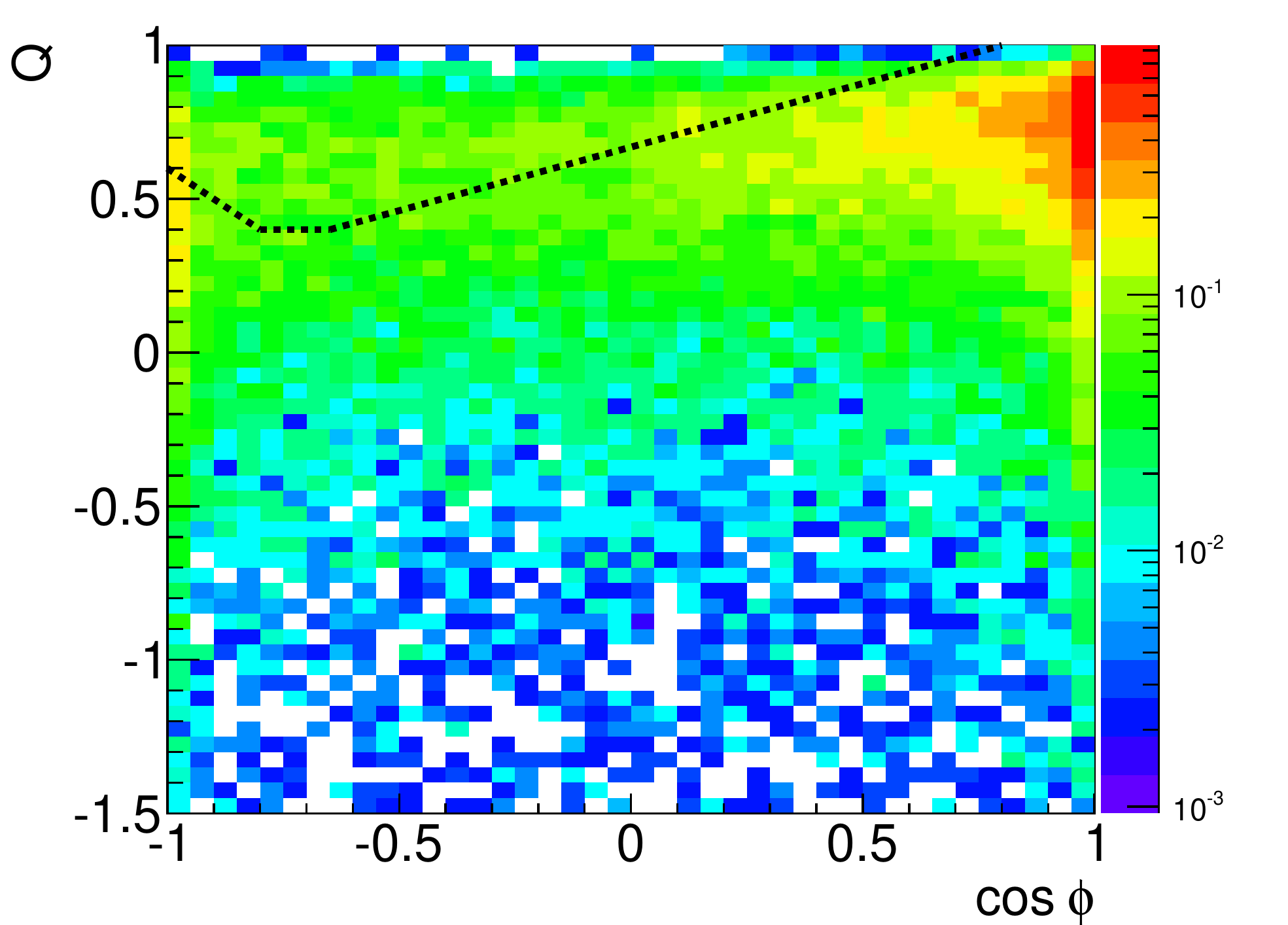}}
\end{center}
\caption{$Q$ vs $\cphi$ distribution for events with 100~GeV~$\le m_{jjj} \le 170$~GeV and passing the tau veto. The event count per bin follows rainbow colors in log scale. (Color online)}
\label{fig:q_cosphi_sr2}
\end{figure}

Here we present the compressed signal region SR2 with the example of a stop and neutralino mass combination $m_{\tilde{t}}=350$~GeV, $m_{\chi}=200$~GeV. For this pairing the top quark will always be off-shell and the top mass window cut is adjusted to 100~GeV~$\le m_{jjj} \le 170$~GeV. This reduces the $\ttbar$ background somewhat, but the largest effect is that the stop signal is increased compared to the higher top mass window cut from SR1 as can be seen in Fig.~\ref{fig:mjjj}. We continue to require a hadronic tau veto.
Fig.~\ref{fig:q_cosphi_sr2} shows the distribution of $Q$ vs $\cphi$ for $\ttbar$ background and stop signal events.

\begin{figure}[!h!tbp]
\begin{center}
\subfigure[~$\ttbar$]{\label{subfig:Ec_ttbar_sr2}\includegraphics[width=0.475\textwidth,clip=true]{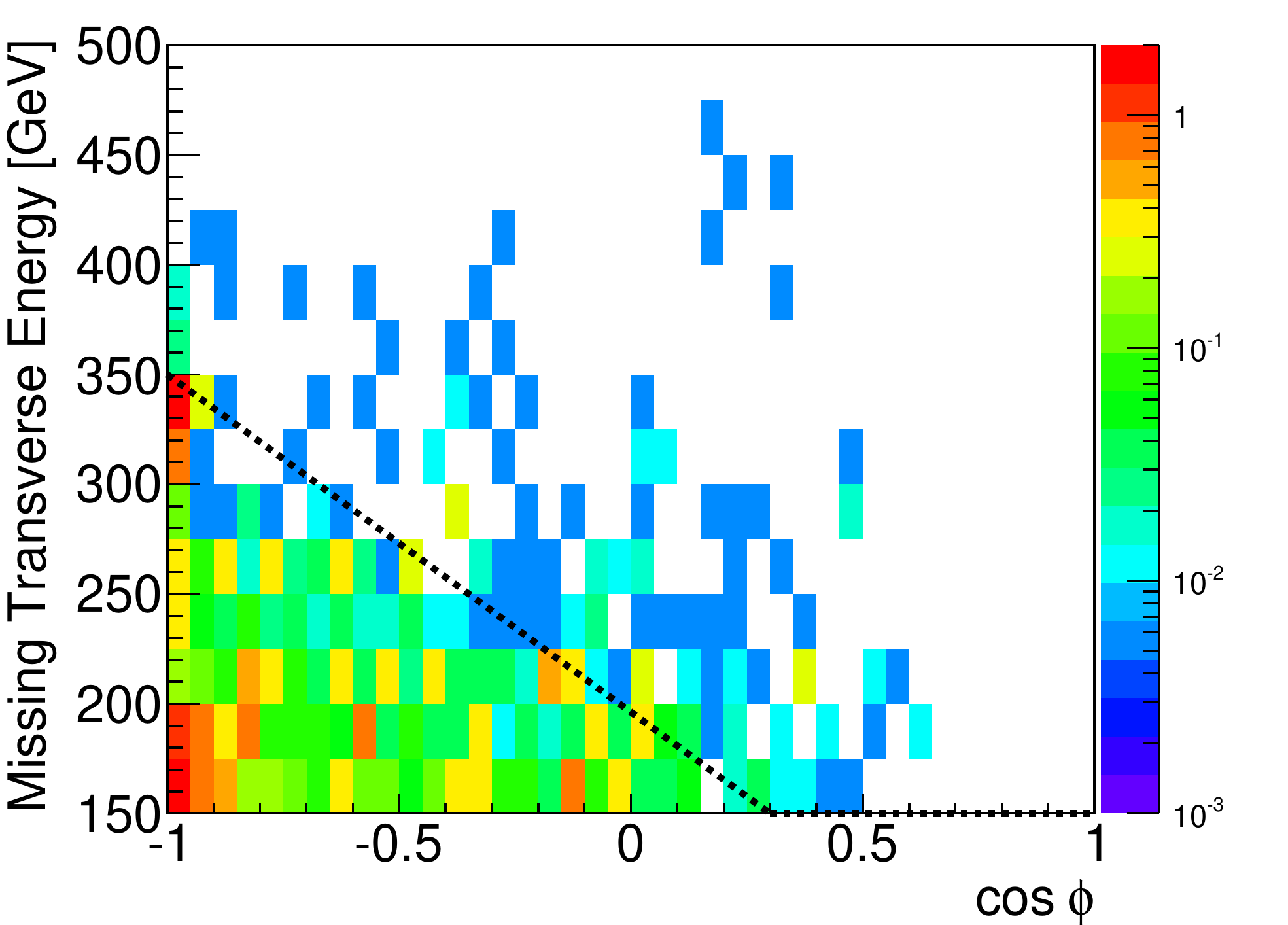}}
\subfigure[~$\tilde{t}\tilde{t}^*,\;m_{\tilde{t}}$ = 350 GeV, $m_{\chi}$ = 200 GeV]{\label{subfig:Ec_350_sr2}\includegraphics[width=0.475\textwidth,clip=true]{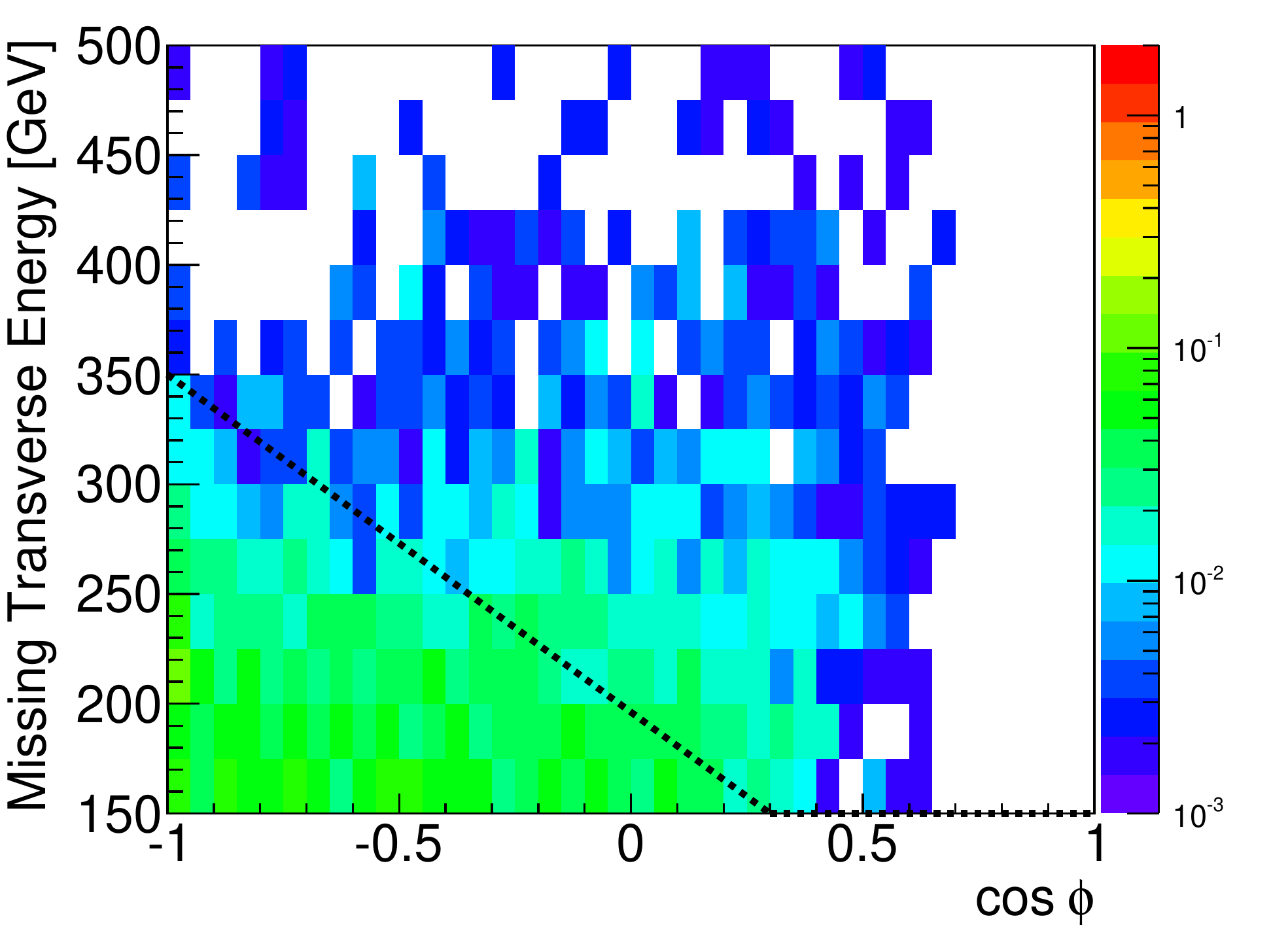}}
\end{center}
\caption{$\etmiss$ vs $\cphi$ distribution for events passing the $Q-\cphi$ contour cut from Fig.~\ref{fig:q_cosphi_sr2}. The black line indicates the $\etmiss-\cphi$ selection cut. The event count per bin follows rainbow colors in log scale. (Color online)}
\label{fig:met_cosphi_sr2t}
\end{figure}

The shape of the $\ttbar$ distribution is similar to Fig.~\ref{fig:q_cosphi_sr1}. The stop signal has a less pronounced peak around $\cphi=-1$ than the higher-mass signals, though still a large number of events in the region.
The events in the region of the $Q$-$\cphi$ plane above the black contour are selected
to populate the $\etmiss-\cphi$ plane in Fig.~\ref{fig:met_cosphi_sr2t}.

Similar to the Fig.~\ref{fig:met_cosphi_sr1t} for SR1, the $\ttbar$ background again peaks close to $\cphi=-1$ and low $\etmiss$, while the signal extends out to higher $\cphi$ and higher $\etmiss$. However, comparing Fig.~\ref{subfig:Ec_350_sr2} to Fig.~\ref{subfig:Ec_ttbar_sr1}, it is clear that the low stop mass leads to lower $\etmiss$ and thus less separation from the background. Note, however, that the color scale is the same for both plots in Fig.~\ref{fig:met_cosphi_sr2t}, and even at this stop mass, there are regions of phase space where the signal is larger than the $\ttbar$ background. Events above the black line are selected and the final yields are tabulated in Table~\ref{table:sr_yields}.

\subsection{Pair Production Summary}
\label{sec:pairsum}

Event yields for the different regions are tabulated in Table~\ref{table:sr_yields}.

\begin{table}[!h!tbp]
\begin{center}
\caption{Event yields for different selection cuts for $\ttbar$ background and stop signals at the 8~TeV LHC with 20~fb$^{-1}$ of collisions.}
\begin{tabular}{|l|c|c|c|}
\hline
Sample & \multicolumn{3}{c|}{Event yield} \\
            &  SR0 & SR1 & SR2 \\
\hline
$\ttbar$ background& \csname ttbarsr0 \endcsname & \csname ttbarsr1t \endcsname & \csname ttbarsr2t \endcsname  \\
Stop signals: & & & \\
~~$m_{\tilde{t}}=350$~GeV, $m_{\chi}$ = 200 GeV & \csname pesr0 \endcsname & & \csname pesr2t \endcsname \\
~~$m_{\tilde{t}}=400$~GeV, $m_{\chi}$ = 200 GeV & \csname pbsr0 \endcsname & \csname pbsr1t \endcsname  & \\
~~$m_{\tilde{t}}=500$~GeV, $m_{\chi}$ = 200 GeV & \csname pasr0 \endcsname & \csname pasr1t \endcsname & \\
\hline
\end{tabular}
\label{table:sr_yields}
\end{center}
\end{table}

Compared to the default selection SR0, the top pair background is reduced by a factor 30, while the stop signal for $m_{\tilde{t}}=500$~GeV is reduced by only a factor two and that for $m_{\tilde{t}}=400$~GeV by a factor four. These signal event yields are low, but should be sufficient for experiments to exclude these mass points after some adjustments and after adopting the techniques presented in this paper.

The event yield for the compressed region ($m_{\tilde{t}}=350$~GeV) is likely too low to be accessible in the current 8~TeV dataset. However, intermediate mass combinations that are in the middle of the currently uncovered region (like $m_{\tilde{t}}=380$~GeV, $m_{\chi}=200$~GeV) should be accessible even in the existing data. Moreover, the cross section goes up by almost a factor five at the 13~TeV LHC, and larger datasets are expected to be collected. Thus, the compressed region should start to be covered using 8~TeV data and completely filled in with 13~TeV data.

\section{Single Stop Production}
\label{sec:singlestop}

There is another parameter region for stop production that is difficult to access experimentally~\cite{Aad:2014kra,Chatrchyan:2013xna}, a small window around
\begin{align}
m_t \leq m_{\tilde{t}}\lesssim 200\, \mathrm{GeV} && 0 \leq m_\chi \leq 20\, \mathrm{GeV}. \label{eq:singlestopmassrange}
\end{align}

When the neutralino is nearly massless and the stop mass close to the top mass, then stop pair production and decay provides a small correction to the much larger SM top pair process and is difficult to distinguish kinematically. Thus, this region will continue to be difficult to probe in pair production. We propose a search for single stop production and argue for its potential to directly probe this region of parameter space. 
Covering this region of parameter space has also been the aim of recent work~\cite{Czakon:2014fka} which focuses on precision measurements of SM top-pair production to indirectly constrain the stop and neutralino masses.  In that work the extent at which the left-~and right-handed stops (as well as the associated neutralinos) can contaminate the SM top mass measurement is important.  

Here we instead propose a novel, direct probe of this region of parameter space that covers most of the parameter space in equation~\leqn{eq:singlestopmassrange} already with the 8~TeV data set at the LHC.

\subsection{Single Stop Production Signal Processes}

\begin{figure}[!h!tbp]
\begin{center}
\includegraphics[width=0.35\textwidth,clip=true]{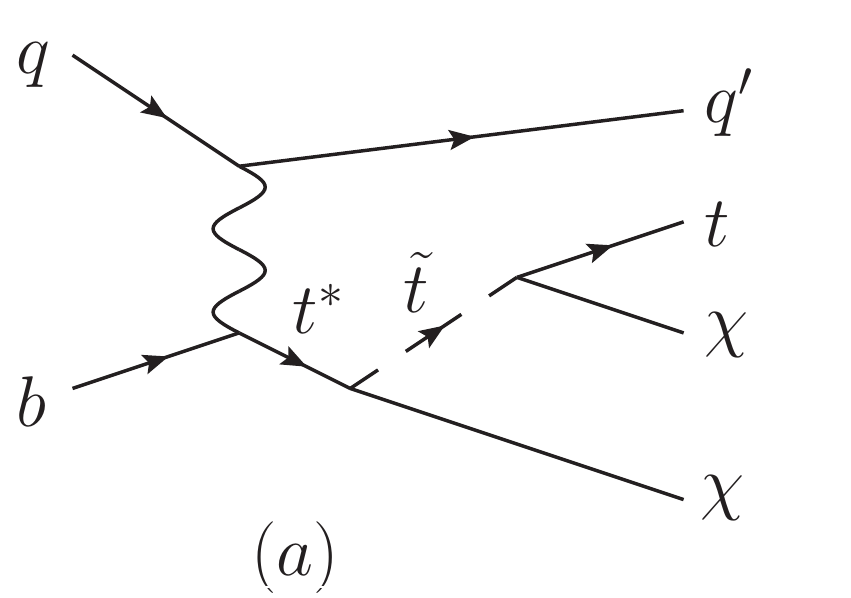}
\includegraphics[width=0.35\textwidth,clip=true]{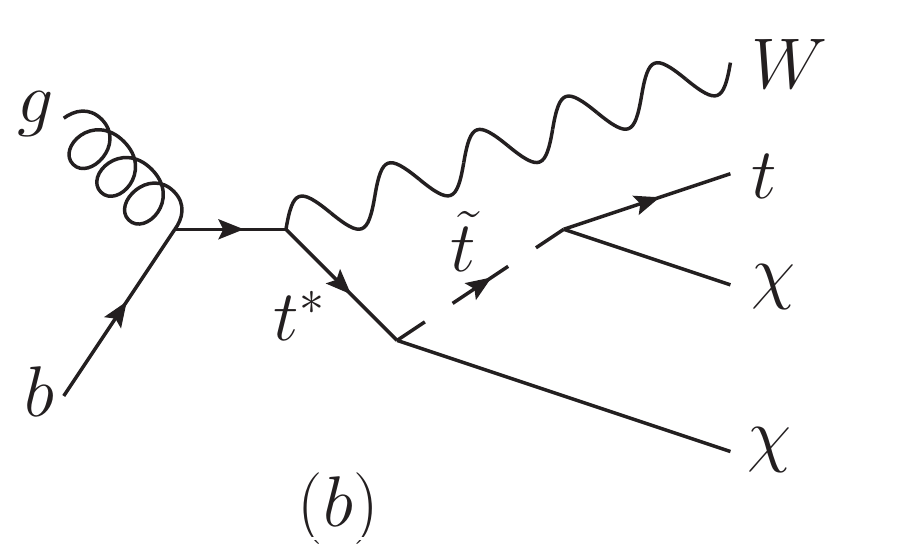}
\end{center}
\caption{Representative Feynman diagrams for production of an off-shell single top quark $t^*$ with decay to a single stop quark, which in turn decays to an on-shell top quark, for (a) the $jt$ mode (Eq.~\ref{eq:tchannel}) and (b) the $Wt$ mode (Eq.~\ref{eq:Wtchannel}). }
\label{fig:sstop_feyn}
\end{figure}

Representative Feynman diagrams for the production of single stop quarks in association with jets and in association with a $W$~boson are shown in Fig.~\ref{fig:sstop_feyn}.

Single stop quarks are produced in the decay of off-shell top quarks from SM single top production,
\begin{align}
&p + p \to j (+j) + t \,\mathrm{(off\textendash shell)}, \label{eq:tchannel}\\ 
&p + p \to W + t  \,\mathrm{(off\textendash shell)},  \label{eq:Wtchannel}
\end{align} 
where the first process includes the large $t$-channel and smaller $s$-channel contributions, and the second process is the associated $Wt$ production process.
Here, $j$ includes gluons and light quarks.  Implicitly, our analysis also includes the analogous processes with top anti-quarks in the above equations.  The off-shell top quarks decay via
\begin{eqnarray}
t\,\mathrm{(off\textendash shell)} &\to& \tilde{t} + \chi \to \chi + \bar{\chi} + t \label{eq:singlestopdecay1} \\
t\,\mathrm{(off\textendash shell)} &\to& \tilde{t} + \chi \to \chi + \bar{\chi} + W + b\,,  \label{eq:singlestopdecay2}
\end{eqnarray}
including on-shell as well as off-shell top quarks in the decay of the $\tilde{t}$ as in Eq.~\ref{eq:signal}. 

Because the stop goes off-shell, the coupling~\cite{Drees:2004jm} plays an important role. The Lagrangian for the top-stop coupling is given by
\begin{eqnarray}
\mathcal{L} = \overline{t}\,\biggl(G_L \,P_L\, + G_R \,P_R\, \biggr) \, \chi\,\tilde{t} +  \,\,\mathrm{h.c.}\;, \label{eq:LHcoupling}
\end{eqnarray}
where $P_{L,R}$ are the projection operators. 
We take the benchmark that the coupling is of order the electroweak coupling, $G_L = G_R = g_\mathrm{ew}$, which produces the decay widths shown in Table.~\ref{tab:sxs}. The decay widths are all much smaller than the SM top decay width of 1300~MeV. This additional top decay mode does not affect the decay of on-shell top quarks due to the limited available phase-space.  In our analysis, we consider the following benchmark (stop, neutralino) mass pairs (in GeV): (175,~1), (190,~10) and (215,~40).

These single stop production processes have a reasonable production cross section as long as the intermediate top quark decaying to stop is not too far off-shell, exactly the situation relevant for the parameter space given in equation~\leqn{eq:singlestopmassrange}. The production cross section for single stops, computed at leading order, using Madgraph~5~\cite{Alwall:2011uj} and the CTEQ6 PDF set~\cite{Pumplin:2002vw}, is shown in Table~\ref{tab:sxs} for the three benchmark stop/neutralino mass pairs. The top quark mass is set to 173~GeV.

\begin{table}[!h!tbp]
\begin{center}
\caption{Off-shell top to stop decay width and single stop production cross sections at LO at a 8~TeV proton-proton collider.}
\begin{tabular}{|c|c|c|c|}
\hline
stop, neutralino mass & off-shell $t\to\tilde{t}\chi$ & \multicolumn{2}{c|}{cross section [pb]} \\
& decay width & $j\tilde{t}$ & $W\tilde{t}$ \\
\hline
$m_{\tilde{t}}$ = 175 GeV, $m_{\chi}$ = 1 GeV &0.67~MeV & 1.08 & 0.20 \\
$m_{\tilde{t}}$ = 190 GeV, $m_{\chi}$ = 10 GeV & 32~MeV & 0.40 & 0.020 \\
$m_{\tilde{t}}$ = 215 GeV, $m_{\chi}$ = 40 GeV & 72~MeV & 0.33 & 0.013 \\
\hline
\end{tabular}
\label{tab:sxs}
\end{center}
\end{table}

In comparison, the NLO with next-to-next-to-leading log corrections cross section for SM production of single top quarks in the $t$-channel is 87~pb~\cite{Kidonakis:2012rm}, and that in the $Wt$-channel is 22~pb~\cite{Kidonakis:2012rm}. The largest contribution to single stop production comes from SM $t$-channel production of an off-shell top quark (Fig.~\ref{fig:sstop_feyn}(a). For the phase space most relevant to Eq.~\ref{eq:singlestopmassrange}, the cross section is sufficiently large to produce over 20,000 events in 20~fb$^{-1}$ of LHC 8~TeV data. 
The single stop production cross section decreases as the sum of stop and neutralino masses moves up and the top quark becomes increasingly virtual. Nevertheless, there is a sufficient number of signal events produced for all three benchmark scenarios to attempt isolating the signal. Moreover, there will be sensitivity to this region of phase space even if the $t\tilde{t}$ coupling is lower than electroweak coupling.

\subsection{Background Processes}

The final state for the single stop signal processes has one hard lepton (electron or muon), missing energy and two or three jets, one of which is from a $b$~quark. The largest backgrounds to this signature come from top quark pair ($\ttbar$), SM single top ($t$-channel) and $W$+jets production. Smaller backgrounds come from SM $Z$+jets, diboson and QCD multijet production, as well as top pair or single top production in association with a $W$~or $Z$~boson. The background from top or $W$~boson production in association with a $Z$~boson decaying to neutrinos also contributes.

Here we consider the three most relevant backgrounds. The $\ttbar$ background is described in Sec.~\ref{sec:process}, and in particular dilepton events also constitute a large background to single stop production. For the $W$+jets background, we include
\begin{align}
 p + p  &\to W + (1-3) j & p + p &\to W + Z\,,  \label{eq:wplusjetsbkg}
\end{align}
which implicitly include $b$~quarks and diboson production with $Z$ boson decays to quarks and neutrinos.
For the SM single top background, we include the processes from Eqs.~\ref{eq:tchannel} and~\ref{eq:Wtchannel}, but producing only on-shell top quarks.

In all, we note that single stop production may outperform stop pair production in this region of parameter space because the dominant SM backgrounds are generated by the weaker electroweak processes. 

\subsection{Event Selection}

All signal and background events are generated at parton level with Madgraph, no hadronization is included. We account for detector effects as described in Sec.~\ref{sec:detector_effects}, except that the smearing is applied to partons rather than jets. 
The basic event selection cuts are similar to those in stop pair production as discussed in Sec.~\ref{sec:selection}, but are adjusted to be similar to ATLAS and CMS SM single top measurements~\cite{Aad:2012ux,Chatrchyan:2011vp, Aad:2014fwa,Khachatryan:2014iya}. We require an isolated, hard lepton (electron or muon) with $p_T>25$~GeV and $|\eta|<2.5$. The missing transverse energy cut is $\etmiss>30$~GeV, lower than in stop pair production searches to retain more of the single stop signal events. We require two or three jets with $p_T>30$~GeV and increase the jet $p_T$ threshold to 35~GeV for forward jets with $|\eta|>2.75$. At least one jet must be $b$-tagged.
The jets and leptons are required to be well separated,
\begin{align}
\Delta R_{lj} > 0.4 &&  \Delta R_{jj} &> 0.4.
\end{align}
which is more conservative than~\cite{Aad:2012ej, Chatrchyan:2011vp, Aad:2014fwa, Khachatryan:2014iya}.  We also make a cut on the transverse mass, $m_T>80$~GeV. The number of signal and background events after these selection cuts is given in Table~\ref{tab:sst_yields}.

\begin{table}[!h!tbp]
\begin{center}
\caption{Event yields for single stop signals and different backgrounds passing basic selection cuts at the 8~TeV LHC with 20~fb$^{-1}$ of collisions.}
\label{tab:sst_yields}
\begin{tabular}{|l|c|}
\hline
\multicolumn{2}{|c|}{Basic event selection} \\
\hline
Sample & Event yield \\
\hline
Signals: & \\
~~$m_{\tilde{t}}=175$~GeV, $m_{\chi}$ = 1 GeV & 592 \\
~~$m_{\tilde{t}}=190$~GeV, $m_{\chi}$ = 10 GeV & 77 \\
~~$m_{\tilde{t}}=215$~GeV, $m_{\chi}$ = 40 GeV & 70 \\
Backgrounds: & \\
~~$\ttbar$ & 3700 \\
~~W+jets   & 10200 \\
~~Single top & 16100 \\
\hline
\end{tabular}
\end{center}
\end{table}

The total background is large, with about 30,000 events. Nevertheless, there are a sufficient number of signal events remaining to isolate the signal further, especially for the lowest stop mass point.

\subsection{Deconstructed Transverse Mass}

Preserving the full information about the missing energy vector is even more important in single stop production than in stop pair production since there are fewer final state objects and larger backgrounds from SM processes. We deconstruct the transverse mass in single stop events according to Sec.~\ref{sec:deconstruct}. 

\begin{figure}[!h!tbp]
\begin{center}
\subfigure[~$\ttbar$]{\label{subfig:Qcst_ttbar}\includegraphics[width=0.45\textwidth,clip=true]{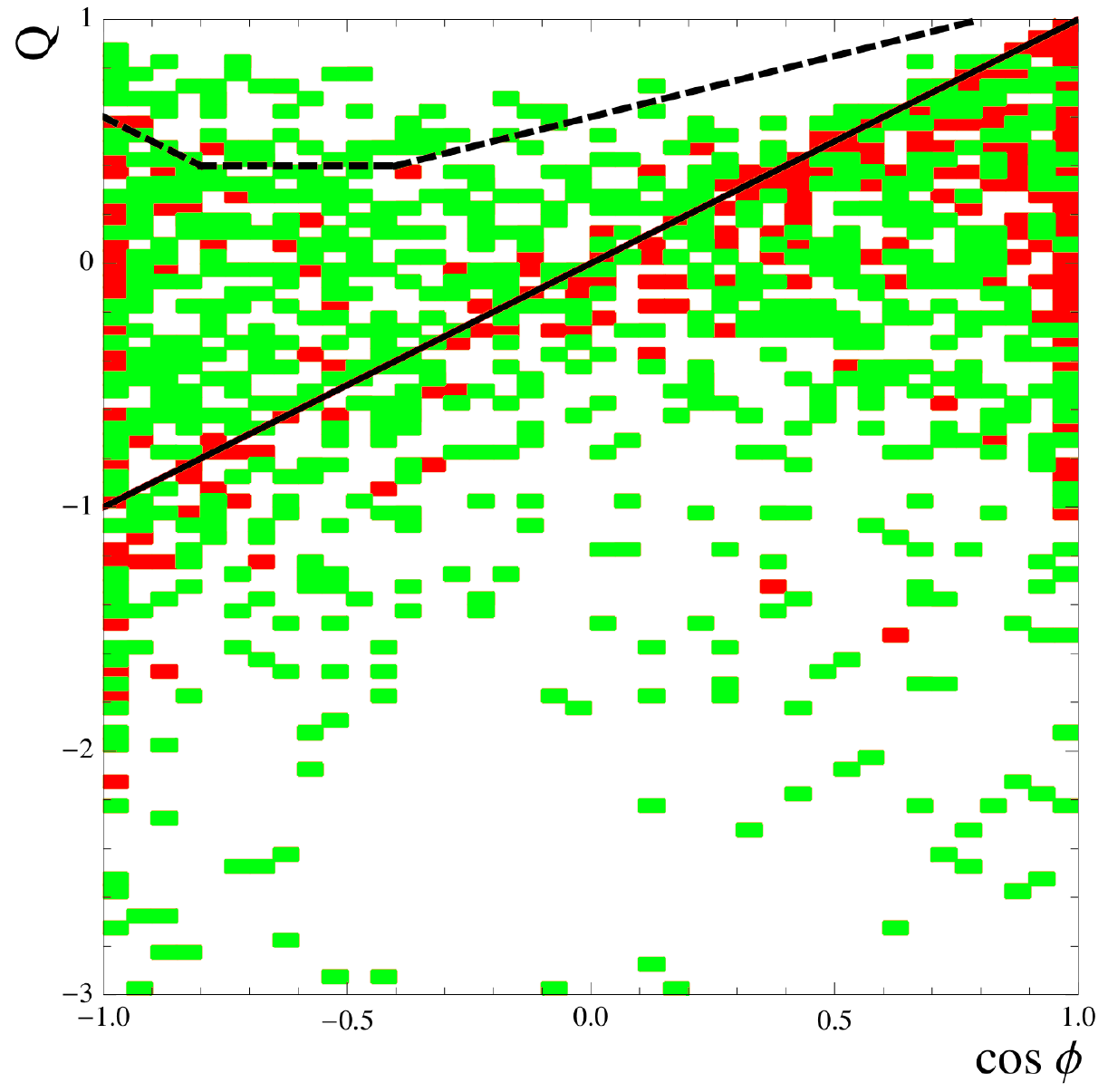}}
\subfigure[~$W$+jets]{\label{subfig:Qcst_Wj}\includegraphics[width=0.45\textwidth,clip=true]{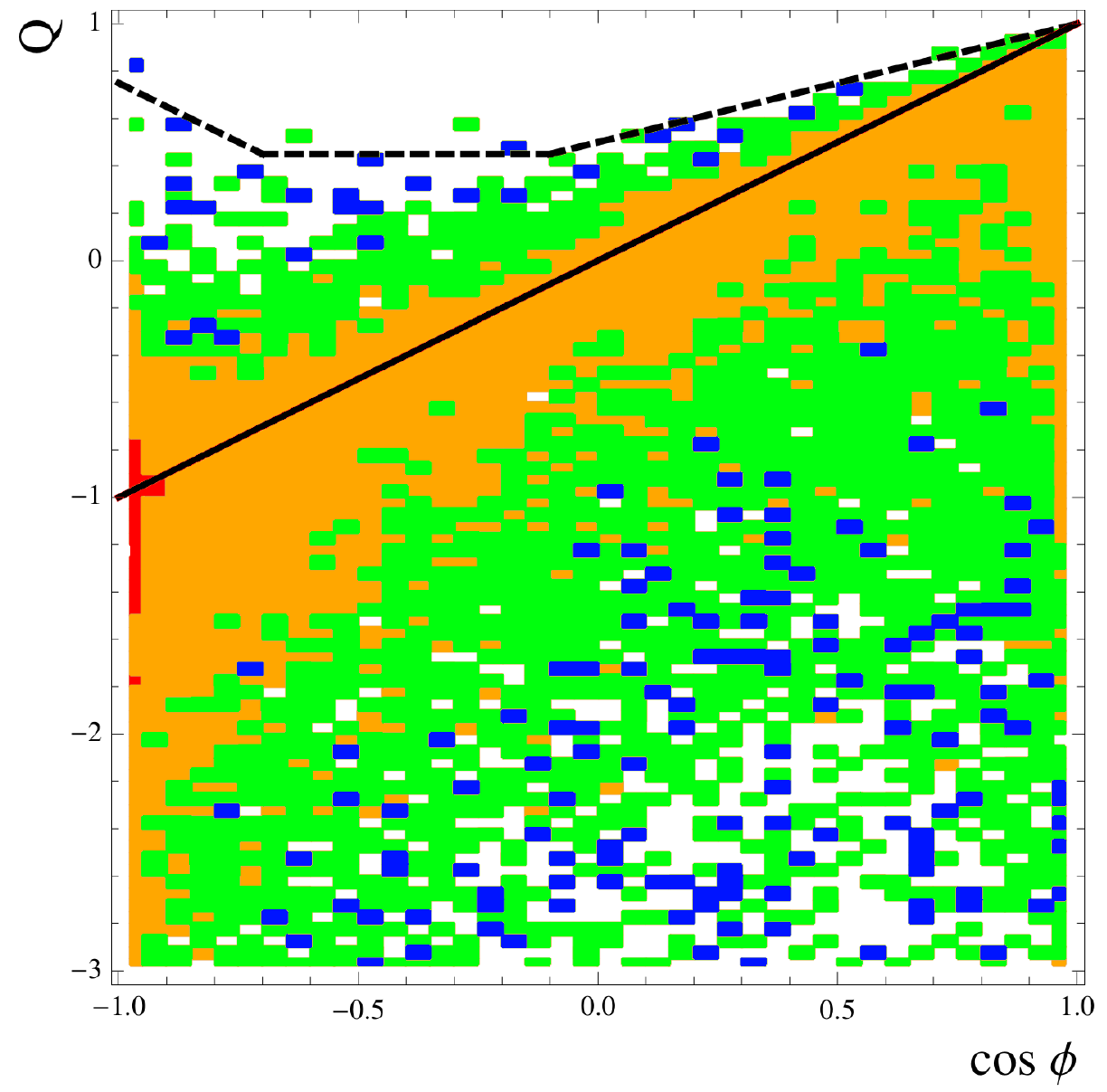}}  \vspace{0.1cm}
\subfigure[~Single top]{\label{subfig:Qcst_st}\includegraphics[width=0.45\textwidth,clip=true]{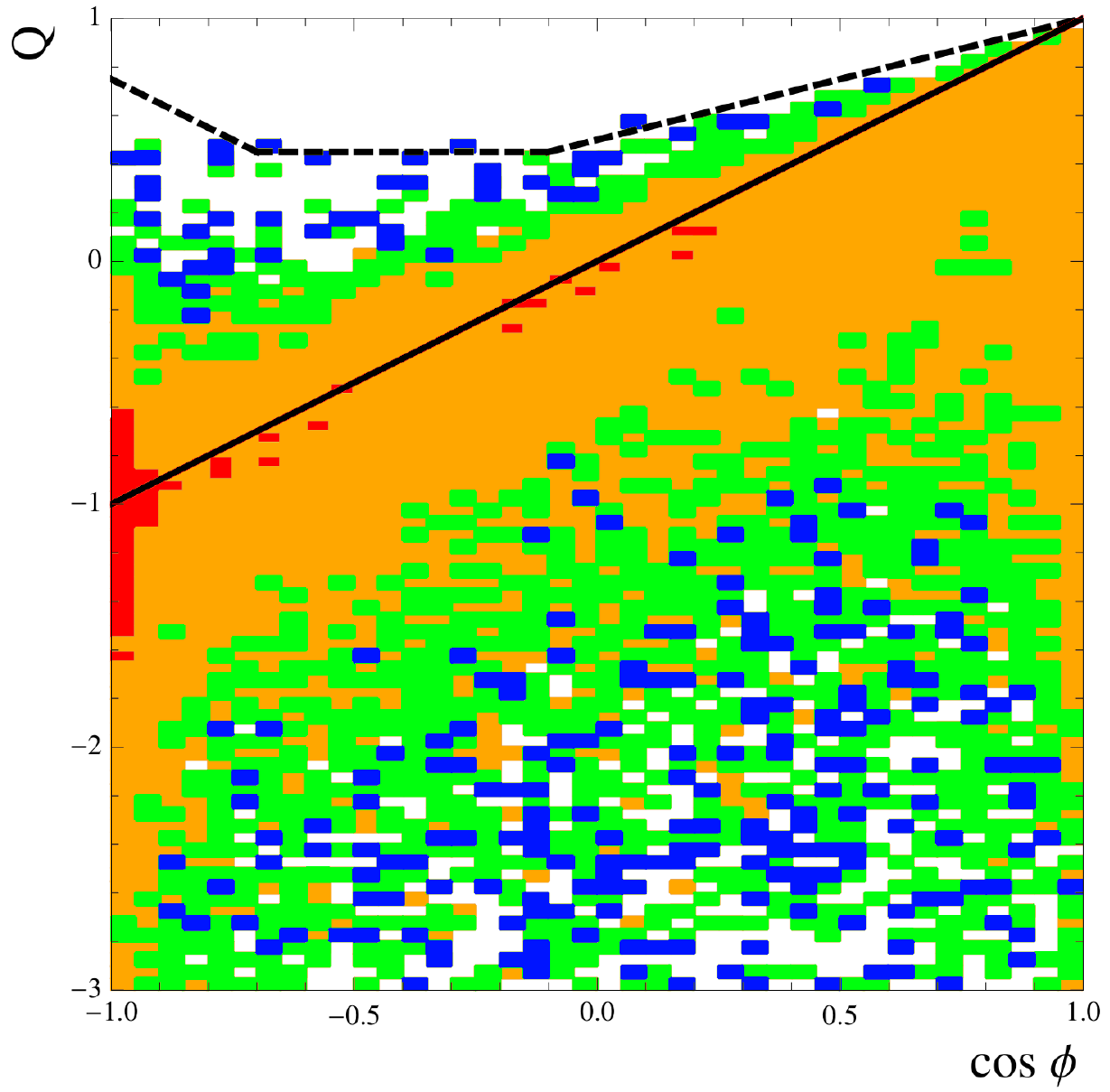}}
\end{center}
\caption{$Q$ vs $\cphi$ distribution for background events passing basic selection, for \subref{subfig:Qcst_ttbar} top quark pair events including both lepton+jets and dilepton decay modes, \subref{subfig:Qcst_Wj} $W$+jets and \subref{subfig:Qcst_st} single top.
The black diagonal line is defined by $Q=\cphi$.
Events inside the contour bounded by the dashed line are selected by the single stop cuts.
The event count per bin follows rainbow colors in log scale. (Color online)}
\label{fig:Qcst_bkg}
\end{figure}

Figure~\ref{fig:Qcst_bkg} shows the $Q-\cphi$ distribution for selected background events (except that the $m_T$ cut has not been applied to show the full range of the distributions).

The $\ttbar$ background has the same features as can be seen in Fig.~\ref{subfig:Qc_ttbar_sr0}, with peaks near $\cphi=1$ and near $\cphi=-1$ and a diagonal distribution consistent with a $W$~boson decay. However, the distribution here is broader, and the peak near $\cphi=1$ is much less pronounced. This is due to the requirement of two or three jets, which reduces the lepton+jets contribution and enhances the dilepton contribution.
The $W$+jets and single top backgrounds show a diagonal trend consistent with the decay of a $W$~boson. This is the same trend also visible in $\ttbar$ lepton+jet events in Fig.~\ref{subfig:Qcosphi_ttbarlq}. Since there are no multiple-neutrino events in SM single top production, the peak near $\cphi=-1$ is absent in single top. It is barely visible in $W$+jets, where it is populated by $WZ$+jets events with $Z\to\nu\nu$ decay.

\begin{figure}[!h!tbp]
\begin{center}
\subfigure[~$\tilde{t},\;m_{\tilde{t}} = 175$ GeV, $m_\chi=1$~GeV]{\label{subfig:Qcst_175}\includegraphics[width=0.45\textwidth, clip=true]{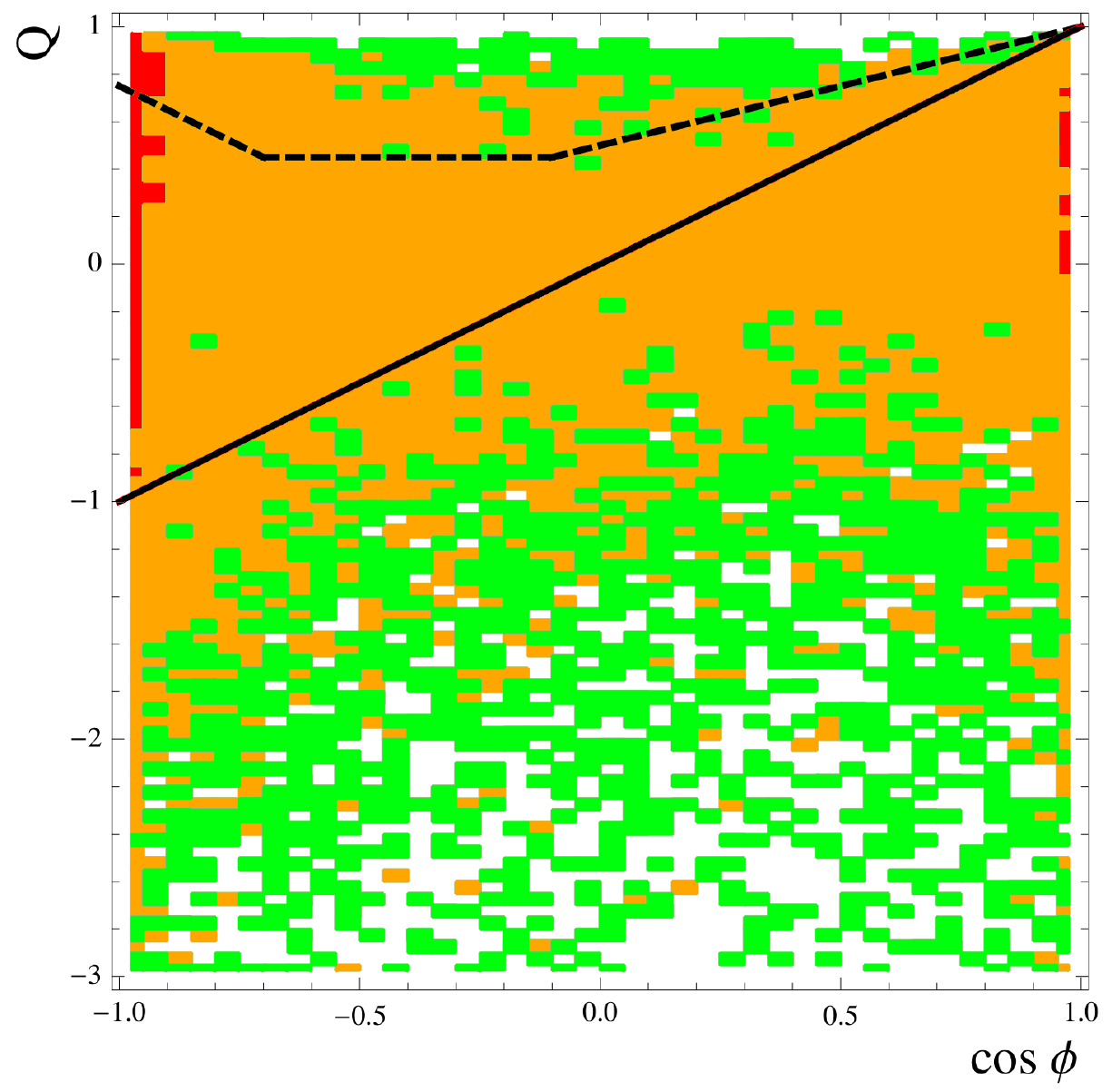}}
\subfigure[~$\tilde{t},\;m_{\tilde{t}} = 190$ GeV, $m_\chi=10$~GeV]{\label{subfig:Qcst_190}\includegraphics[width=0.45\textwidth,clip=true]{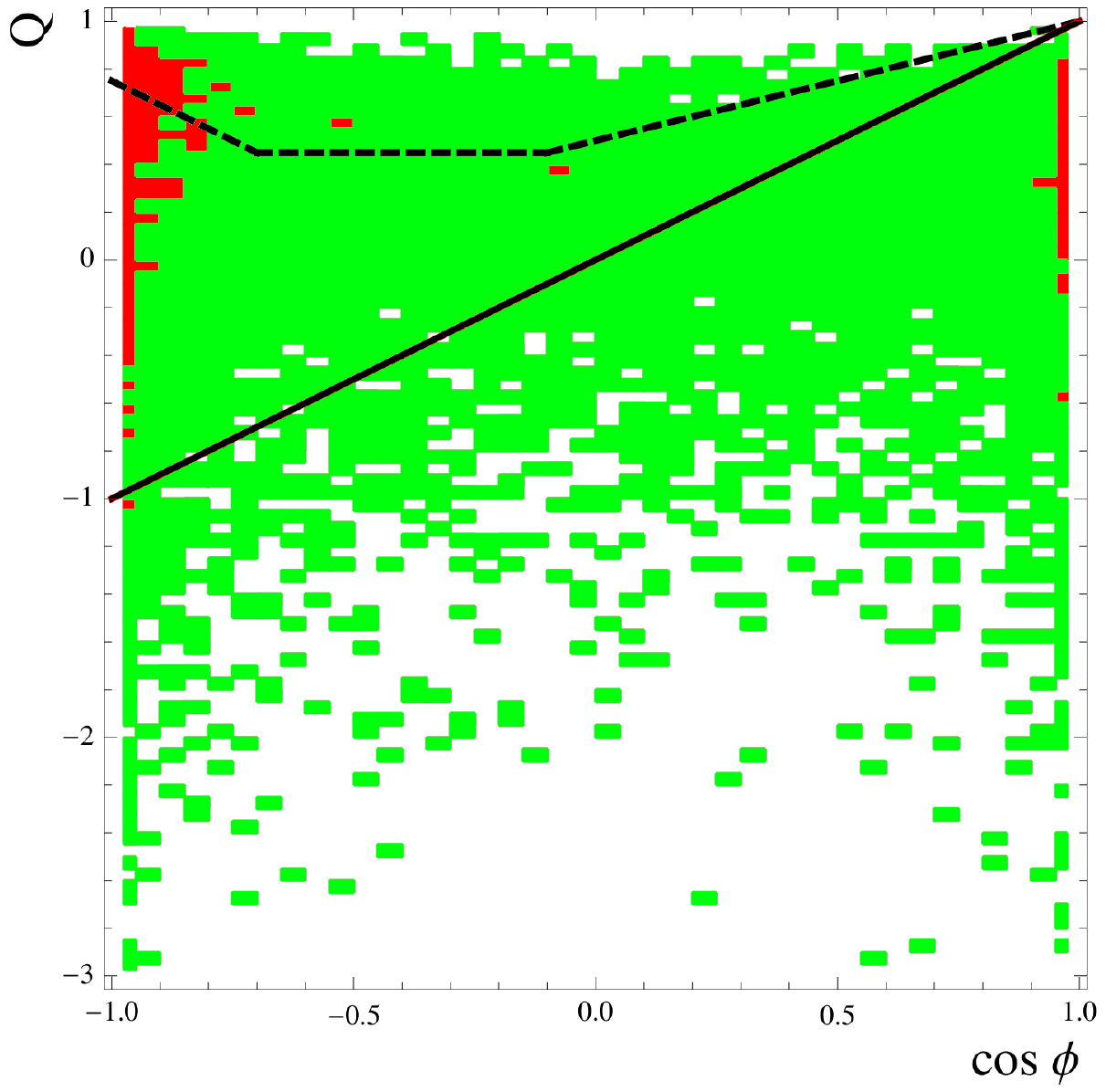}}
\subfigure[~$\tilde{t},\;m_{\tilde{t}} = 215$ GeV, $m_\chi=40$~GeV]{\label{subfig:Qcst_215}\includegraphics[width=0.45\textwidth,clip=true]{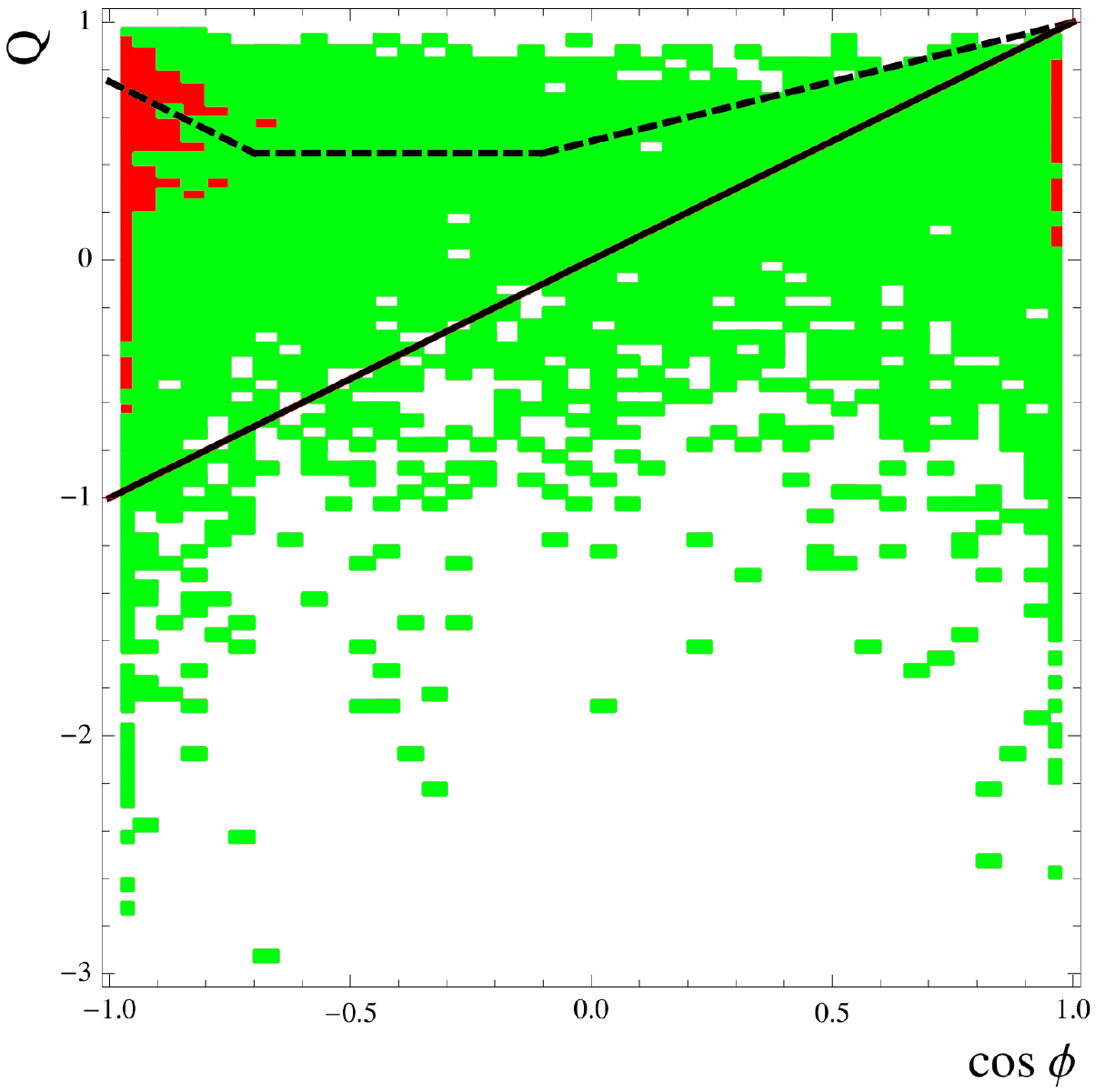}}
\end{center}
\caption{$Q$ vs $\cphi$ distribution for single stop signal events passing basic selection.
The black diagonal line is defined by $Q=\cphi$.
Events inside the contour bounded by the dashed line are selected by the single stop cuts.
The event count per bin follows rainbow colors in log scale. (Color online)}
\label{fig:Qcst_sig}
\end{figure}

Figure~\ref{fig:Qcst_sig} shows the $Q-\cphi$ distribution for the three signal mass pairs. Compared to stop pair production in Fig.~\ref{fig:q_cosphi_sr0}, there is no peak near $\cphi=1$ and the events are instead clustered near $\cphi=-1$. In single stop production, the low mass of the two neutralinos results in a preferred kinematic configuration where they are back-to-back with the lepton from the $W$~boson from the top quark decay. 

\subsection{Single Stop Signal Region}

We employ the deconstructed variables to separate the single stop signals from the large backgrounds. Since there is only one stop quark, hadronic top reconstruction is not possible and a tau veto is not necessary and not required. Events above the dashed line in Figs.~\ref{fig:Qcst_bkg} and~\ref{fig:Qcst_sig} form the single stop signal region. The resulting number of signal and background events are listed in Table~\ref{tab:sst_sr}.

\begin{table}[!h!tbp]
\begin{center}
\caption{Event yields in the single stop signal region for single stop signals and different backgrounds at the 8~TeV LHC with 20~fb$^{-1}$ of collisions.}
\label{tab:sst_sr}
\begin{tabular}{|l|c|}
\hline
\multicolumn{2}{|c|}{Signal region} \\
\hline
Sample & Event yield \\
\hline
Signals: & \\
~~$m_{\tilde{t}}=175$~GeV, $m_{\chi}$ = 1 GeV  & 129 \\
~~$m_{\tilde{t}}=190$~GeV, $m_{\chi}$ = 10 GeV & 21  \\
~~$m_{\tilde{t}}=215$~GeV, $m_{\chi}$ = 40 GeV & 22  \\
Backgrounds: & \\
~~$\ttbar$  & 302 \\
~~W+jets    & 22 \\
~~Single top& 4.3 \\
\hline
\end{tabular}
\end{center}
\end{table}

The background from $W$+jets and single top is reduced by two orders of magnitude compared to Tab.~\ref{tab:sst_yields}, while the signal only goes down by a factor of four. The $\ttbar$ background is dominant, though even at this stage, there is already sensitivity to the 175~GeV mass point.
The 190~GeV mass point signal yield is comparable to the 215~GeV yield despite the larger cross section because the on-shell top quark in the decay chain in equation~\leqn{eq:singlestopdecay1} better recoils off of the heavier neutralino masses.  Thus, the 215~GeV signal is more concentrated at $\cphi = -1$ and therefore easier to separate from background.  
Additional separation can be obtained, in particular for the two higher masses, from additional cuts, for example exploiting the $\etmiss-\cphi$ correlation. Already with the straightforward cuts included here, the stealth stop region of stop mass close to the top mass and low neutralino mass can be accessed directly with the currently available LHC data.

\section{Additional Applications of Deconstruction}
\label{sec:additionalapps}

Beyond searching for stop pair production in the compressed limit, deconstructed transverse mass variables may be useful in a variety of searches for new physics.
Thus far we have focused on the production of stops in processes with a lepton (electron or muon) in the final state and in regions of parameter space that is hard to access.  In this section we expand the application of deconstruction to other examples. In general, deconstruction is useful any time there is information encoded in the magnidude and direction of two objects in an event, in particular when the underlying particles forming those two objects are different between signal and background.
 For events containing a leptonically decaying $W$~boson, these two objects are the lepton and the $\etmiss$ vector, and deconstruction extricates the signal from the background through correlations in two two-dimensional planes. But the two objects could also be two jets or more complex objects.

\subsection{Deconstructed All-Hadronic Final States}
\label{sec:hadronic}
Thus far we have focused on the production of stops in processes with a $W$~boson decaying to a lepton and neutrino. In this section, we describe deconstruction can also apply to all-hadronic multi-jet final states. The signal signature contains multiple jets and one or two undetected dark matter candidates, resulting in large $\etmiss$. The background to this signature is large, but does not contain undetected high-$p_T$ particles. Any $\etmiss$ is generated only from detector mis-reconstruction or mis-identification or from undetected low-$p_T$ particles.  Thus, $\etmiss$ is much lower in the background and constitutes a powerful discriminant between signal and background. Here we present the basic idea of how deconstruction can help separate stop pair signal events from $\ttbar$ all-hadronic backgrounds. A full analysis will be developed in~\cite{devinhst}.

We focus on the hadronic decays stops and consider the signal process
\begin{equation}
p + p \to \tilde{t} + \tilde{t}^* \to {\rm jets} + \chi + \chi \label{eq:hadronicsignal}
\end{equation}
and assume each stop decays solely to a top and neutralino. The final state contains six hard partons, though experiments typically require at least four jets to maximize the signal acceptance~\cite{Aad:2014bva,Chatrchyan:2013lya}.
 
These signal events do not contain a lepton, thus a $W$~boson transverse mass reconstruction seems pointless. However, an analogy to the deconstructed variables in Section~\ref{sec:deconstruct} can be constructed.  When a top quark plays the role of the lepton, then the decay $\tilde{t} \to t + \chi$ is analogous to $W \to l + \nu$, thus we define
\begin{equation}
Q = 1 - {{m_T^0}^2 \over 2\, E_{T\,\mathrm{top}} \,\etmiss} ,
\end{equation}
where $E_{T\,\mathrm{top}} = m_\mathrm{top}^2 + p_{T\,top}^2$, i.e. the mass of the top quark can not be neglected.  A variant of this stop transverse mass variable was also used in~\cite{Aad:2014bva}. 
We consider a stop/neutralino mass combination of 400/200~GeV as an example. We therefore set $M_0 = 400$ GeV.  The top quark is chosen as the one with the largest $p_T$. Thus, $\cos\phi$ measures the transverse angle between the hardest top quark and the missing momentum.  

Stop pair signal and $\ttbar$ background events are generated at parton level with Madgraph, no hadronization is included. We account for detector effects as described in Sec.~\ref{sec:detector_effects}, except that the smearing is applied to partons rather than jets. 
We require six jets with $p_T>40$~GeV, at least one of which must be $b$-tagged.
The jets are clustered into exactly two ``mega-jets'' (representing the two top quarks) by requiring that the invariant mass of each mega-jet is close to the top mass~\cite{Aad:2014bva,Chatrchyan:2013lya}. One jets in each mega-jet must have a  $p_T > 80$ GeV.  

The major backgrounds to this signal is from $\ttbar$, $W$+jets and multi-jet events. Typically, a large missing transverse energy cut is applied to reduce these backgrounds. We do not apply a cut on $\etmiss$ and instead demonstrate the difference in shape of the $Q-\cphi$ contour between signal and $\ttbar$ background. 

\begin{figure}[!h!tbp]
\begin{center}
\subfigure[~$m_{\tilde{t}} = 400$ GeV, $m_\chi=200$~GeV]{\label{subfig:Qcst_had1}\includegraphics[width=0.45\textwidth, clip=true]{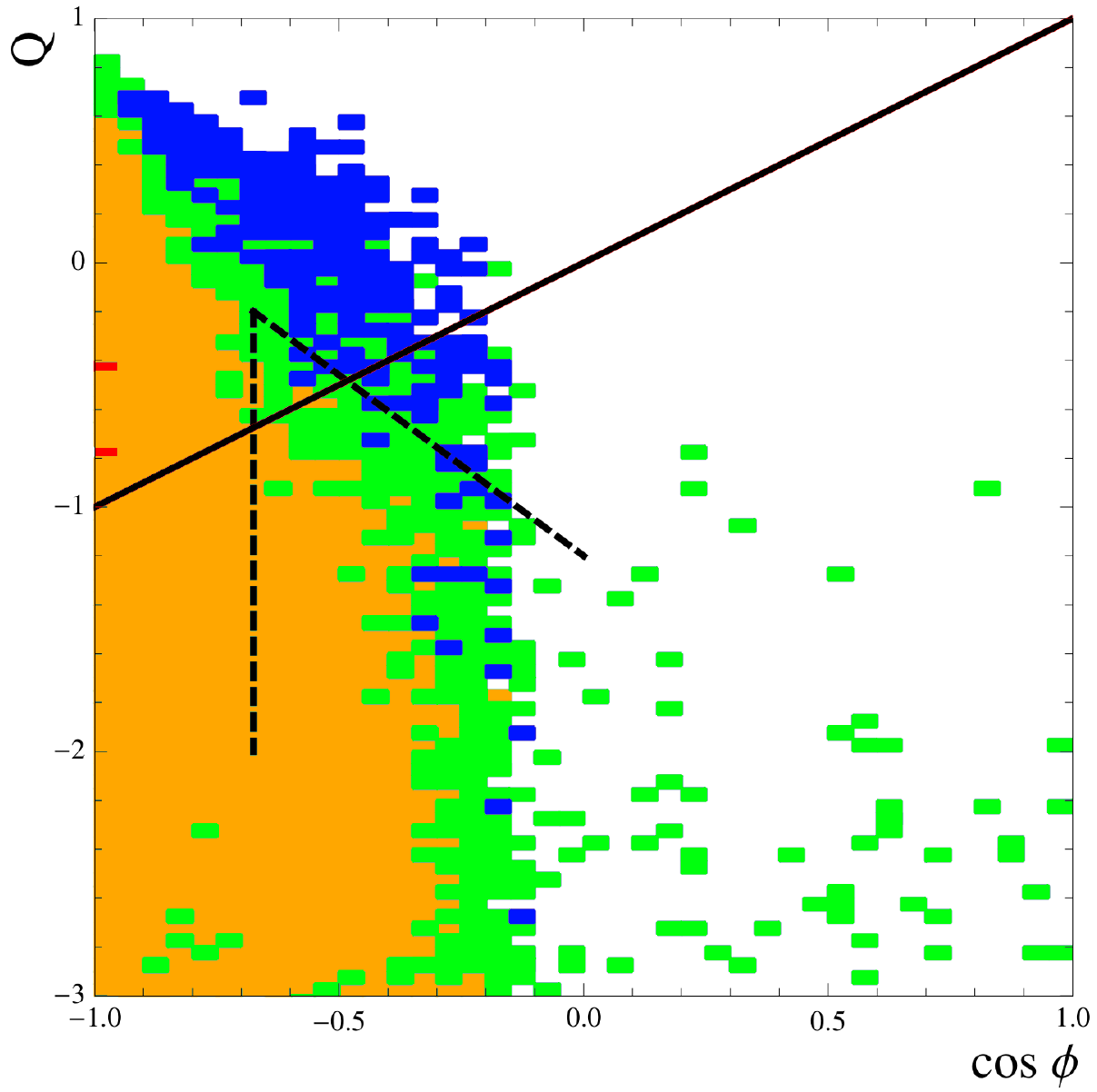}}
\subfigure[~$\ttbar$]{\label{subfig:Qcst_had2}\includegraphics[width=0.45\textwidth,clip=true]{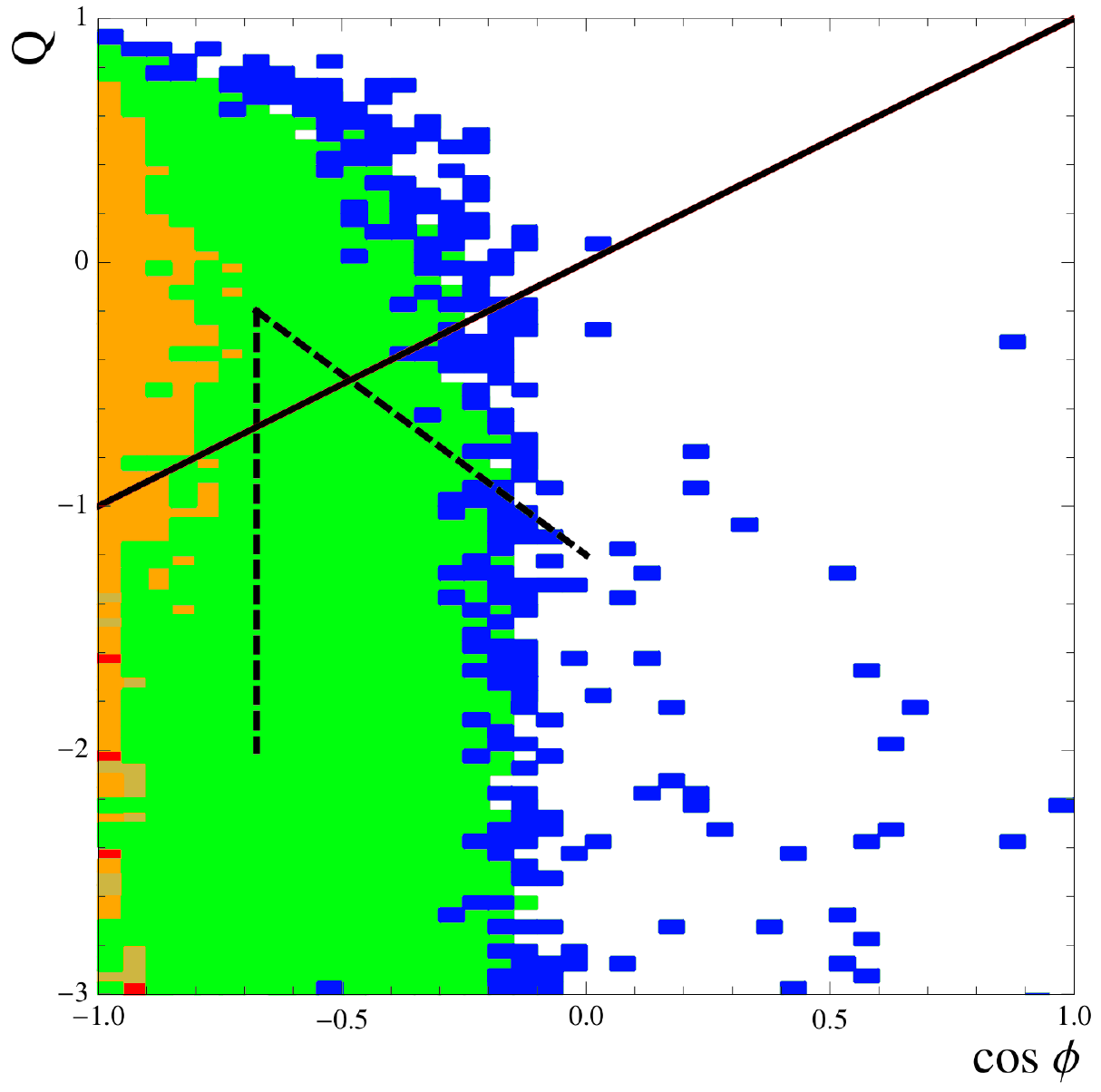}}
\end{center}
\caption{$Q$ vs $\cphi$ distribution for hadronic stop signal and $\ttbar$ background
events passing basic selection. The black diagonal line is defined by $Q=\cphi$.
Selecting events in the triangle defined by the dashed line ensures the greatest separation between signal and background. The event count per bin follows rainbow colors in log scale. (Color online)}
\label{fig:Qcst_had}
\end{figure}

Figure~\ref{fig:Qcst_had} shows the $Q-\cphi$ distribution for the signal and the hadronically decaying $\ttbar$ background. The solid diagonal line corresponds to a transverse mass cut of 400~GeV, which is effective at reducing the background but which also loses a lot of signal. The background peaks near $\cphi=-1$, back-to-back with the leading mega-jet as expected when $\etmiss$ arises mainly from mis-reconstruction and limited resolution. Other backgrounds containing neutrinos, including $\ttbar$ backgrounds where the $W$~boson decays to an electron or muon or tau, will have peaks both near $\cphi=-1$ as well as $\cphi=1$.

The dashed black lines in Fig.~\ref{fig:Qcst_had} enclose the region where the signal shape differs the most from the background shape. This region corresponds to lower $Q$ (and transverse mass) values, but higher $\cphi$ values. This is similar to the lepton+jets final state (c.f. Section~\ref{sec:sr2}), where the best signal-background separation was also obtained for $\cphi$ values away from $+1$ and $-1$.

\subsection{Deconstructed Razor}

Deconstruction language can also be adapted to the formalism of a razor analysis~\cite{Chatrchyan:2014goa,Rogan:2010kb}, which also uses mega-jets in an all-hadronic final state.
Razor analyses are commonly used for events with n-jet events and large amounts of missing energy.  Again, the analysis defines mega-jets in order to form a basic di-jet topology.  The following kinematic variables are then defined from the two mega-jets $j1$ and $j2$:
\begin{eqnarray}
M_R^2 &\equiv&(p_{j_1} + p_{j_2})^2 - (p_z^{j_1} + p_z^{j_2} )^2 \\  
(M_T^R)^2 &\equiv& \bigl(\etmiss\, (p_T^{j_1} + p_T^{j_2}) - \vec{\etmiss} \cdot (\vec{p}_T^{\,j_1} + \vec{p}_T^{\,j_2})\bigr)\bigl/2.
\end{eqnarray}
Here $M_T^R $ is a transverse mass variable.  These variables are used to construct the ratio,
\begin{equation}
R^2 = \biggl({M_T^R \over M_R}\biggr)^2.
\end{equation}

The simplest way to deconstruct razor is to define
\begin{align}
\Bigl(M_{T_S}^{R}\Bigr)^2 =  {1\over 2}\,\etmiss\, (p_T^{j_1} + p_T^{j_2})  && \Bigl(M_{T_V}^R\Bigr)^2 = -{1\over 2}\,\vec{\etmiss} \cdot (\vec{p}_T^{\,j_1} + \vec{p}_T^{\,j_2}),
\end{align}
and write two new ratios 
\begin{align}
R_V^2 = {(M^R_{T_V})^2 \over M_R^2} && R_S^2 = {(M^2_{T_S})^2 \over M_R^2} \label{eq:deconrazor}
\end{align}
where $R^2 = R_V^2 + R_S^2$.  
The combination of $M_{T_S}^{R}$, $M_{T_V}^R$, $R_V$ and $R_S$ provides information on both the magnitude and the direction of the missing energy. Two-dimensional distributions of these variables will provide additional signal-background separation in a razor analysis.

\subsection{Deconstructed Top Quark}
\label{eq:topchi}

In addition to a modification of the reconstructed $W$~boson transverse mass, the presence
of dark matter particles also modifies the reconstruction of the leptonically decaying top
quark. We form a top mass discriminant analogously to Eq.~\ref{eq:bound1},
\begin{equation}
\chi_t = p_{bl\,L}^2A_t + \left( E_{bl}^2-p_{bl\,L}^2\right)\left(A_t - 4E_{bl}\etmiss\right)\,,
\chi_t = 1 - \left(1-E_{bl}^2/p_{bl\,L}^2\right) \times\left(1-\frac{4E_{bl}^2\etmiss^{\,2}}{A_2}\right)\,,
\end{equation}
where $A_2$ is given by
\begin{equation}
A_2 = \left(m_t^2 - M_{bl}^2 + 2 \vec{p}_{bl\,T}\cdot \vec{\etmiss} \right)^2\;.
\end{equation}
Here, $bl$ is the reconstructed lepton-$b$~quark system and $m_t=173$~GeV is the input top
quark mass~\cite{ATLAS:2014wva}. As before, $T$ and $L$ refer to the transverse and
longitudinal components of the momentum, respectively. 

The discriminant $\chi_t$ encodes information on the magnitude and direction of the missing
transverse energy. It provides some additional information in the analysis beyond the
variables from Sec.~\ref{sec:deconstruct} by also involving a jet (the $b$~quark from the
top decay, and thus has a different sensitivity to mis-reconstructed jet energies.

\begin{figure}[!h!tbp]
\begin{center}
\includegraphics[width=0.475\textwidth,clip=true]{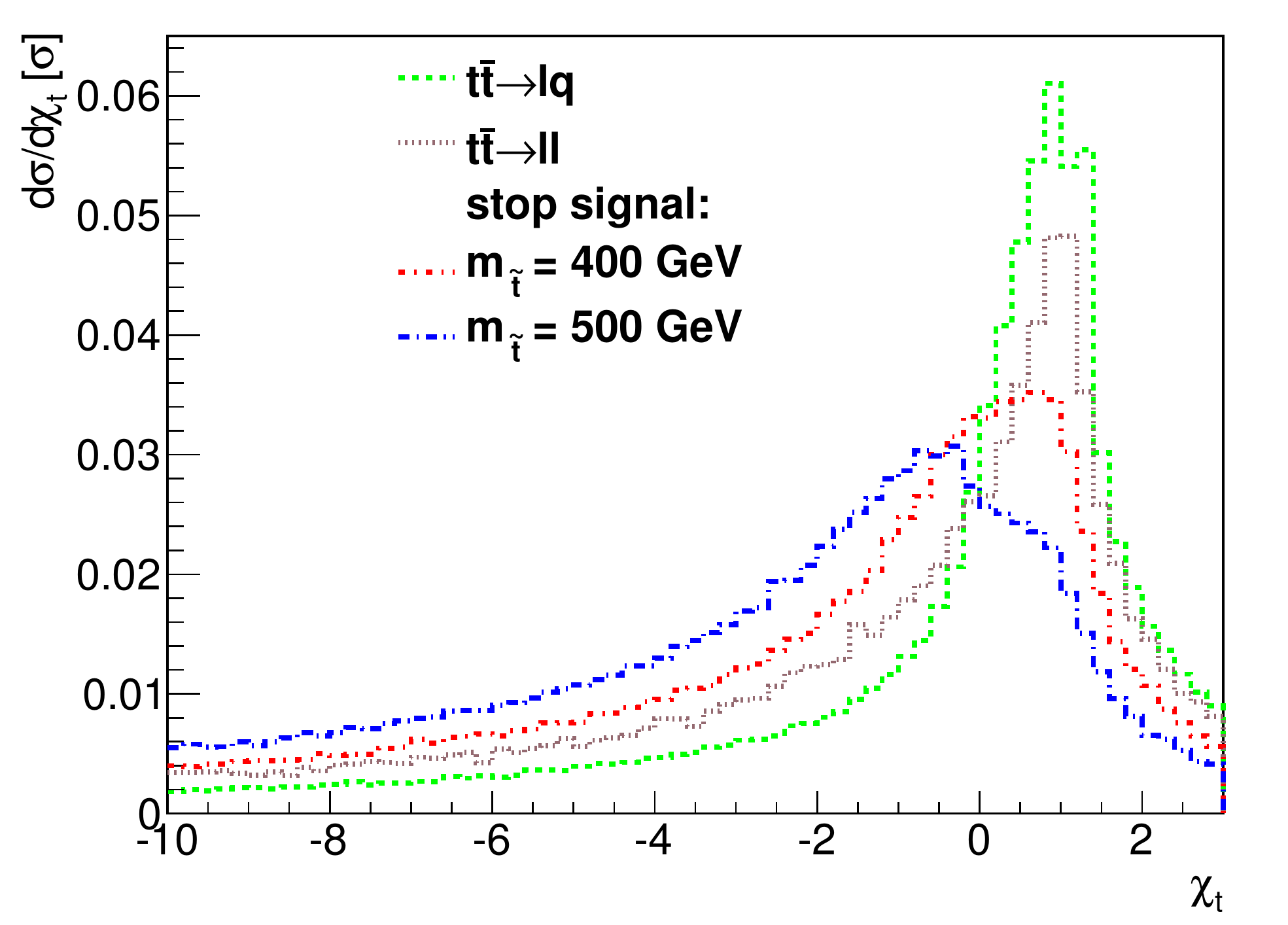}
\end{center}
\caption{Distribution of $\chi_t$ for top pair production and three different stop
signal masses after event selection, all normalized to have the same area, for events
passing selection cuts. (Color online)}
\label{fig:ttbarchi}
\end{figure}

The distribution of $\chi_t$ is shown in Fig.~\ref{fig:ttbarchi} for SM top pair events. It
peaks at one, corresponding to events where the top quark mass is reconstructed properly.
The distribution has both positive and negative tails
due to the effects that detector resolution and mis-reconstruction have on
$\etmiss$ and the $b$-quark jet. There is an additional bump around $\chi_t=1.3$ from the
kinematic threshold of the lepton and $\etmiss$ cuts for events where the $W$~boson and
$b$~quark are back-to-back.

For new physics signals, $\chi_t$ is typically negative, equivalent to Eq.~\ref{eq:bound1}.
This can also be seen in Fig.~\ref{fig:ttbarchi} which shows that the two stop signal lines
also peak at zero, but both distributions have large tails extending to negative values.
Therefore a cut on $\chi_t$ is effective at removing top pair background, preserving the
stop signal. This variable is also useful for regions of parameter space with large
stop masses and small neutralino masses.

\subsection{Other Applications}

All of the deconstructed variables defined here will also be relevant in searches for SUSY at 13~TeV. In particular, with the large datasets expected, tighter selection cuts can be made, giving access to virtual top quarks far away from the on-shell top mass. A significant fraction of the stealth stop and compressed spectra regions can be covered already with the existing 8~TeV data, and should be completely covered using 13~TeV data with the techniques described here. 

High-mass gluino and other searches that have four top quarks in the final state~\cite{Cacciapaglia:2011kz,Gregoire:2011ka,AguilarSaavedra:2011ck}, which will result in multiple sources of missing energy, an ideal playground for deconstruction, which will then also allow for lower $\etmiss$ cuts while still controlling backgrounds. Similarly, vector-like quark decays often have $W$~bosons and multiple neutrinos in the final state~\cite{AguilarSaavedra:2009es}, and deconstruction can improve the signal-background separation in these searches.

Similarly, allowing top quarks and $W$~bosons to be off-shell in order to access kinematic ranges not otherwise accessible also applies to other searches. For example stop decays to charginos don't have to end their sensitivity at the on-shell chargino mass and can instead extend below that value by allowing for off-shell chargino decays.

\section{Conclusion }
\label{sec:conclusion}

We have investigated the pair production of supersymmetric top quark partners,
with subsequent decay to SM top quarks and neutralinos.
In particular, we investigated the lepton+jets decay mode of the resultant SM top quark pairs,
with a single lepton and neutrino (missing energy) in the final state.
The neutralinos (as dark matter candidates) manifest as additional missing energy in any detector. To make contact with existing analyses, we focused on the 8~TeV LHC. We demonstrate that significant gains in sensitivity are available to ATLAS and CMS beyond the already-published 8~TeV results.
We introduced new deconstructed transverse mass variables which exploit the correlations between the amplitudes and directions of the $\vec{\etmiss}$ and lepton.
We showed how these correlations can be used to separate the SM background from the SUSY signal, improving significantly upon existing analyses. 
We have shown that a hadronic $\tau$ veto can be used to reduce the significant $\ttbar$ dilepton background where one lepton is a tau decaying hadronically.
We pointed out that the compressed regions, where the stop mass is less than the sum of top
and dark matter particle mass, should be accessed by shifting the top mass cut window to reflect the off-shell-nature of the top quark in these events. The stealth top region of light neutralinos is accessible through single stop production together with deconstructed transverse mass variables.
Our techniques can be applied to other kinematic variables such as mega-jet kinematics in hadronic final states and will also be important at future hadron collider experiments such as the LHC at 13~TeV.

\section*{Acknowledgements}

We thank S.~Chivukula, L.~Dixon, H.-C.~Fang, S.~El Hedri, J.~Hewett, I.~Hinchliffe, M.~Peskin, T.~Rizzo, E.~Simmons, M.~Shapiro and J.-H. Yu for useful discussions. We thank A.~Schwartzman for his contributions to early drafts of this work. A.I. is supported by the Department of Energy under Grants No. DE-AC02-06CH11357, DE-AC02-76SF00515 and DE-FG02-12ER41811. The work of R.S. is supported in part by the US National Science Foundation under Grant No. PHY-0952729.  D.W. is supported by Department of Energy under Grants No. DE-AC02-76SF00515 and in part by a grant from the Ford Foundation via the National Academies of the Sciences as well as the National Science Foundation under Grants No. NSF-PHY-0705682, the LHC Theory Initiative.

\bibliography{topetmiss_v4}

\end{document}